\documentclass[twocolumn,trackchanges,twocolappendix]{aastex701}

\newcommand{\msun}{\mbox{${\rm M}_\odot$}}
\newcommand{\mstar}{\mbox{${M}_{\rm star}$}}
\usepackage{booktabs}
\usepackage{adjustbox}
\usepackage{graphicx} 
\usepackage{chngcntr}
\usepackage{comment}
\usepackage{float}
\usepackage{threeparttable}

\def\ion#1#2{#1$\;${\small\rm\@{#2}}\relax}
\newcommand{\kms}{\mbox{km\ s${^{-1}}$}}

\newcommand{\cmjjj}{\mbox{${\rm cm^{-3}}$}}

\begin{document}

\title{The Cosmic Ultraviolet Baryon Survey X: Density and Pressure Profiles of Clumpy Circumgalactic Gas at $z=0.4-0.7$}

\author[orcid=0000-0003-4427-4831]{Suyash Kumar}
\affiliation{Department of Astronomy and Astrophysics, The University of Chicago, Chicago, IL 60637, USA}
\email[show]{suyashk@uchicago.edu}

\author[orcid=0000-0001-8813-4182]{Hsiao-Wen Chen}
\affiliation{Department of Astronomy and Astrophysics, The University of Chicago, Chicago, IL 60637, USA}
\affiliation{Kavli Institute for Cosmological Physics, The University of Chicago, Chicago, IL 60637, USA}
\email{hwchen@uchicago.edu}

\author[orcid=0000-0002-8459-5413]{Gwen C.~Rudie}
\affiliation{The Observatories of the Carnegie Institution for Science, 813 Santa Barbara Street, Pasadena, CA 91101, USA}
\email{gwen@carnegiescience.edu}

\author[orcid=0000-0002-2941-646X]{Zhijie Qu}
\affiliation{Department of Astronomy, Tsinghua University, Beijing 100084, People’s Republic of China}
\email{quzhijie@tsinghua.edu.cn}

\author[orcid=0000-0001-9487-8583]{Sean D.~Johnson}
\affiliation{Department of Astronomy, University of Michigan, 1085 S. University, Ann Arbor, MI 48109, USA}
\email{seanjoh@umich.edu}

\author[orcid=0000-0001-7396-3578]{Alexander Beckett}
\affiliation{Aix Marseille Universit\'e, CNRS, CNES, LAM
(Laboratoire d’Astrophysique de Marseille), 13388 Marseille, France}
\email{alexander.beckett@lam.fr}

\author[orcid=0000-0001-5804-1428]{Sebastiano Cantalupo}
\affiliation{Department of Physics, University of Milan Bicocca, Piazza della Scienza 3, I-20126 Milano, Italy}
\affiliation{INAF,  Osservatorio Astronomico di Trieste, Via G. B. Tiepolo 11, 34143 Trieste, Italy}
\email{sebastiano.cantalupo@unimib.it}

\author[orcid=0000-0002-8739-3163]{Mandy C.~Chen}
\affiliation{Cahill Center for Astronomy and Astrophysics, California Institute of Technology, Pasadena, CA 91125, USA}
\affiliation{The Observatories of the Carnegie Institution for Science, 813 Santa Barbara Street, Pasadena, CA 91101, USA}
\email{mandyc@caltech.edu}

\author[0009-0003-8927-2140,gname=David,sname=DePalma]{David DePalma}
\affiliation{Department of Physics, MIT, 77 Massachusetts Avenue, Cambridge, MA 02139, USA}
\affiliation{MIT--Kavli Institute for Astrophysics and Space Research, 77 Massachusetts Avenue, Cambridge, MA 02139, USA}
\email{depalma@space.mit.edu}

\author[orcid=0000-0002-4900-6628]{Claude-André Faucher-Giguère}
\affiliation{Department of Physics \& Astronomy, Center for Interdisciplinary Exploration and Research in Astrophysics (CIERA), Northwestern University, 1800 Sherman Avenue, Evanston, IL 60201, USA}
\email{cgiguere@northwestern.edu}

\author[orcid=0000-0003-0389-0902]{Sebastian Lopez}
\affiliation{Departamento de Astronomía, Universidad de Chile, Camino El Observatorio 1515, Las Condes, Santiago, Chile}
\email{slopez@das.uchile.cl}

\author[orcid=0000-0002-9141-9792]{Nishant Mishra}
\affiliation{Department of Astronomy, University of Michigan, 1085 S. University, Ann Arbor, MI 48109, USA}
\affiliation{The Observatories of the Carnegie Institution for Science, 813 Santa Barbara Street, Pasadena, CA 91101, USA}
\email{mishran@umich.edu}

\author[]{Patrick Petitjean}
\affiliation{Institut d’Astrophysique de Paris, 98bis Boulevard Arago, 75014, Paris, France}
\email{ppetitje@iap.fr}

\author[orcid=0000-0002-9946-4731]{Marc Rafelski}
\affiliation{Space Telescope Science Institute, Baltimore, MD 21218, USA}
\affiliation{Department of Physics \& Astronomy, Johns Hopkins University, Baltimore, MD 21218, USA}
\email{mrafelski@stsci.edu}

\author[orcid=0000-0002-1690-3488]{Michael Rauch}
\affiliation{The Observatories of the Carnegie Institution for Science, 813 Santa Barbara Street, Pasadena, CA 91101, USA}
\email{mr@obs.carnegiescience.edu}

\author[orcid=0000-0002-0668-5560]{Joop Schaye}
\affiliation{Leiden Observatory, Leiden University, PO Box 9513, NL-2300 RA Leiden, the Netherlands}
\email{schaye@strw.leidenuniv.nl}

\author[orcid=0000-0001-7869-2551]{Fakhri S.~Zahedy}
\altaffiliation{Deceased}
\affiliation{Department of Physics, University of North Texas, Denton, TX 76203, USA}
\email{fakhri.zahedy@unt.edu}

\begin{abstract}

The thermodynamic state and turbulent properties of the cool circumgalactic medium (CGM) are investigated for 23 galaxies and galaxy groups at intermediate redshift ($z=0.4$--0.7) in the Cosmic Ultraviolet Baryon Survey (CUBS) midz sample using background QSOs within the virial radius. Kinematically resolved absorption-line measurements from \textit{Hubble Space Telescope}/COS and STIS spectra, combined with photoionization modeling, yield gas densities of $n_\mathrm{H}\approx10^{-4}$--$10^{-2}$ \cmjjj, temperatures of $T\approx(1$--$5)\times10^4\,\mathrm{K}$, and non-thermal line broadening of $b_\mathrm{NT}\approx2$--$30$ \kms\ for individual cool clumps. Combined with published samples at similar redshifts, the results indicate that photoionization equilibrium accurately describes the cool CGM and that turbulent motions are predominantly subsonic. At fixed projected distance, cool-gas density decreases systematically with halo mass, while inferred cool-gas pressures in $M_h\approx10^{12}$--$10^{13}\,\msun$ halos are consistent with hot-gas pressures inferred from stacked X-ray observations.  The analysis uncovers a near constant total cool gas mass across a broad halo mass range, which directly translates to a declining baryon mass fraction in the cool phase from $\approx 6$--19\% in halos of $M_h \approx 10^{12}\, \msun$ to $\approx 0.6$--1.5\% in halos of $M_h \approx 10^{13}\, \msun$.  Finally, several CUBS galaxies exhibit kinematically aligned low- and high-ionization species spanning up to two orders of magnitude in density, reflecting local fluctuations in the halo gas pressure.

\end{abstract}



\section{Introduction} 

\setcounter{footnote}{0}

Building a census of baryons in galaxy halos is critical for connecting feedback on the smallest, star-forming scales to structure formation on cosmological scales. The stellar population and interstellar gas in external galaxies have been well characterized across various cosmic epochs using photometric and spectroscopic observations \citep[see][]{Conroy:2013,Kewley:2019}. Although these data reveal a variety of physical processes driving galaxy formation, stars and the interstellar medium (ISM) constitute less than 25\% of the baryonic matter in galaxy halos. The remaining baryons reside in the circumgalactic medium \citep[CGM, see][]{Tumlinson:2017,Chen:2026}, which is the interface between the star-forming region and the intergalactic medium.

The CGM hosts gas with a vast range of thermodynamic properties. The hot CGM with volume density $n_\mathrm{H} < 10^{-4} \ \mathrm{cm^{-3}}$ and temperature $T\gtrsim 10^6 \ \mathrm{K}$ is the largest reservoir of baryons and traces cosmological processes like halo assembly and intergalactic accretion. While the role of the hot CGM in galaxy evolution is critical, observations remain challenging. The X-ray surface brightness and thermal Sunyaev-Zeldovich (tSZ) signal are probes of the hot gas density and pressure, respectively. However, these measurements are difficult to obtain on galaxy scales. Attempts to study individual galaxies using these methods have been restricted to massive objects in the nearby universe \citep[see e.g.,][]{Hodges-Kluck:2018,Bregman:2022}, although insights about lower-mass halos can be obtained through stacking efforts \citep[see e.g.,][]{Singh:2018,Zhang:2024,Popesso:2025,Siegel:2026}. 

Absorption line spectroscopy of background quasars (QSOs) remains the most common technique for probing gaseous reservoirs in and around individual galaxies. Absorption line measurements of the CGM typically trace the cool- and warm-hot phases ($n_\mathrm{H} \approx 10^{-4}$--$0.01 \ \mathrm{cm}^{-3}, T \approx 10^{4}$--$10^{5} \ \mathrm{K}$), revealing exquisite details about ongoing physical processes. Ambitious absorption-line surveys have uncovered baryons in the CGM at different redshifts and halo masses \citep[see][]{Weymann:1981, Rauch:1998, Chen:2017a,Tumlinson:2017,Muzahid:2021,Dutta:2024,Chen:2026}. 
An important feature of circumgalactic gas is its clumpy nature, which is revealed by the presence of multiple kinematic components observed in high-resolution spectra \citep[see e.g.,][]{Rauch:1996CIV, Crighton:2015, Rudie:2012, Rudie:2019}.

Explaining the formation, growth, and survival of these clouds remains a significant challenge \citep[see e.g.,][]{Mo:1996, Maller:2004, Schaye:2007, Afruni:2019, F-G:2023}. The pressure balance between the cool- and hot-phases is a critical ingredient in theoretical treatments of this problem \citep[see also][]{Faerman:2023}. In addition to thermal motions, the pressure support in cool gas may be attributed to a variety of non-thermal sources, including turbulence \citep[see e.g.,][]{Chen:2023}, cosmic rays \citep[see e.g.,][]{Butsky:2020,Ji:2020}, and magnetic fields \citep[see e.g.,][]{vandeVoort:2021}.

Existing absorption line data can provide unique insights into the pressure support in circumgalactic clouds. The varying absorption strengths of ionic tracers across kinematic components reflect the ionizing conditions encountered along the QSO line of sight. An estimate of the underlying ionization parameter, or equivalently gas density, can be obtained using simple photoionization models that assume ionization equilibrium with the metagalactic radiation background \citep[see e.g.,][]{Khaire:2019,F-G:2020}. Such models have been successful at predicting the observed ionic column densities within individual absorption components and providing quantitative estimates of the underlying gas density \citep[see e.g.,][]{Crighton:2015,Zahedy:2019}. Additionally, the observed line broadening for individual ions parameterized by the Doppler parameter $b$, can be decomposed into thermal and non-thermal contributions as $b_{\rm I}^2=b_{T, \rm{I}}^2+b_{\rm NT}^2$ with $b_{T,\rm{I}}\equiv\sqrt{2k_\mathrm{B} T/m_{\rm I}}$ and $b_{\rm NT}\equiv\sqrt{2}\,\sigma_{v,{\rm NT}}$, where $k_\mathrm{B}$ is the Boltzmann constant, $T$ is the gas temperature, $m_{\rm I}$ is the particle mass, and $\sigma_{v, {\rm NT}}$ is the line-of-sight velocity dispersion due to non-thermal motions. Comparisons of line widths between species with sufficiently different masses enable decomposition of gas temperature and non-thermal velocity \citep[see e.g.,][]{Rauch:1996CIV,Rudie:2019,Qu:2022}. Recovering the gas density, temperature, and non-thermal broadening for individual absorption components provides direct estimates of the total kinetic pressure within the clumpy CGM.

While a photoionization origin is accepted for gas hosting low- and intermediate-ionization species (e.g., Mg$^+$, C$^{2+}$), the mechanism for producing higher-ionization species like O$^{5+}$ is more nebulous. Absorption line surveys find a diversity in the properties of high-ionization absorbers. While broad absorption ($b>30 \ \kms$) is consistent with warmer collisionally ionized gas \citep[see e.g.,][]{Fox:2007,Lehner:2014,Savage:2014,Werk:2016,Oppenheimer:2016}, narrower high-ionization components can be accommodated by low-density photoionized gas \citep[see e.g.,][]{Carswell:2002,Thom:2008,Stern:2018,Zahedy:2021,Cooper:2021}. Accordingly, kinematically aligned low- and high-ionization species may indicate large fluctuations in local gas pressure \citep[see e.g.,][]{Haislmaier:2021,Qu:2022,Kumar:2024,Sameer:2024}. While projection effects can be important \citep[see e.g.,][]{Ticoras:2026}, a larger incidence of components hosting both low- and high-ions in the inner halo suggests that the corresponding gas phases are likely co-spatial \citep[see e.g.,][]{Qu:2023}.
Therefore, investigating the physical conditions underlying kinematically aligned absorption from low- and high-ionization species is critical for understanding the presence of multiphase gas in galaxy halos \citep[see e.g.,][]{Donahue:2022}.

Despite the accuracy with which physical conditions in the absorbing gas can be studied, constructing density and pressure profiles for cool gas in galaxies spanning a range of halo masses remains a significant observational challenge for two reasons. First, sufficiently deep galaxy surveys are required to uncover low-mass galaxies in QSO fields \citep[see e.g.,][]{Chen:2019,Muzahid:2021}. Second, because galaxies tend to reside in overdense cosmological environments, the survey also needs to be wide for uncovering satellite galaxies associated with the host galaxy of an absorber \citep[see e.g.,][]{Qu:2023}. The Cosmic Ultraviolet Baryon Survey \citep[CUBS, see][]{Chen:2020} is uniquely posed for addressing this challenge. With high-quality far- and near-UV spectra of 15 $z \approx 1$ QSOs obtained using the \textit{Hubble Space Telescope} (\textit{HST}) and accompanying deep galaxy surveys extending out to projected distances of $d_\mathrm{proj} \lesssim 1 \ \mathrm{Mpc}$ from the sightline, the CUBS data enable the construction of a galaxy-selected absorber sample for examining the density structure and pressure support in the CGM.

Constraints on the gas density and pressure are presented for absorbing gas around 23 galaxies at $z = 0.4$--$0.7$ identified within the CUBS-midz sample \citep[][]{Qu:2024}. These galaxies have a QSO sightline intersecting the halo within the virial radius, $R_\mathrm{vir}$, aiding the galaxy-absorber association. The redshift range of the survey is optimized for simultaneously covering low-, intermediate-, and high-ionization species in the absorption spectra. Recently, echelle-resolution near-UV data were also acquired for the CUBS QSOs, enabling more robust absorption profile decomposition, subsequently improving constraints on the gas ionization state. Combined with a previously conducted study for passive luminous red galaxies at $z=0.2$--$0.6$ from \cite{Chen:2018,Zahedy:2019}, this effort results in a characterization of the cool gas density and pressure profile around galaxies with dark matter halo masses of $M_h \approx 10^{12} \msun$--$10^{13} \msun$, for which corresponding constraints on the hot CGM are also available. 

The paper is organized as follows. Details of the QSO spectroscopy and accompanying galaxy surveys are presented in \S\,\ref{sec:obs}. The methodology for absorption line fitting and ionization modeling of cool clumps is described in \S\,\ref{sec:methods}. Finally, key results regarding the physical properties of cool gas in galaxy halos are discussed in \S\,\ref{sec:results}. For brevity, details regarding individual systems are deferred to the Appendix. A flat $\Lambda$ cosmology with $\Omega_m = 0.3, \Omega_\Lambda = 0.7, H_0 = 70 \ \mathrm{km \ s^{-1} \ Mpc^{-1}}$ is adopted throughout the paper, and projected distances ($d_{\rm proj}$) are in physical units.

\section{Observations and Data} \label{sec:obs}

Probing the ionizing conditions and thermal state in the diffuse CGM requires both galaxy survey data and quasar absorption spectra.  The available QSO absorption spectra and the accompanying galaxy imaging and spectroscopic data for the study are summarized here.

\subsection{QSO spectroscopy} \label{sec:stis_redn}

The Cosmic Ultraviolet Baryon Survey \citep[CUBS, ][]{Chen:2020} is designed to map diffuse baryons around galaxies at $z \lesssim 1$ by combining high-resolution quasar spectroscopy with accompanying deep galaxy surveys. Fifteen far-UV (FUV) bright QSOs ($\mathrm{FUV}_{AB} \lesssim 18.5 \ \mathrm{mag}$) were targeted using the {Cosmic Origins Spectrograph} \citep[COS; ][]{Green:2012} on board the \textit{HST} as part of a Cycle 25 General Observer Program (GO-CUBS, PID=15163, PI: Chen). The data presented in this work are available on the Mikulski Archive for Space Telescopes: \dataset[doi:10.17909/qasf-4k87]{https://doi.org/10.17909/qasf-4k87}. The COS G130M and G160M spectra cover a wavelength range from $\lambda \approx 1140$ to 1800 \AA\ with a spectral resolution of ${\rm FWHM}\approx 20$ \kms \   (corresponding to a Doppler parameter of $b \approx 12  \ \kms$).  The FUV spectra provide optimal coverage of multiple ionization stages of heavy metals (e.g., carbon, nitrogen, oxygen, sulfur) at $z < 1$, enabling constraints on the ionization state and relative abundances of the gas. Additionally, the Lyman series transitions available for absorption systems at $z>0.2$ enable a robust determination of the neutral hydrogen column density $N(\mathrm{HI})$, and correspondingly the gas metallicity $\mathrm{[O/H]}$. Optical echelle spectra were also obtained for the QSOs using the Magellan Inamori Kyocera Echelle (MIKE) instrument on the Magellan telescopes \citep{Bernstein:2003}, covering transitions from low-ions (e.g., $\mathrm{Mg}^+, \ \mathrm{Fe}^+$), which are particularly critical for constraining the component structure of absorption systems, aiding the broad line spread function for COS \citep[see e.g.,][]{Zahedy:2019,Qu:2022}. The COS and MIKE spectra are co-added using custom software described in \cite{Chen:2018} and \cite{Chen:2014,Zahedy:2016}, respectively, and the details of this procedure for CUBS sightlines can be found in \cite{Chen:2020}.

Recently, near-UV (NUV) echelle spectra were also obtained using the Space Telescope Imaging Spectrograph \citep[STIS;][]{Woodgate:1998} with the E230M grating for all CUBS sightlines under a Cycle 31 General Observer Program (GO-CONTACT, PID=17517, PI: Chen). These new STIS NUV spectra provide additional spectral coverage from $\lambda\approx 2000$ \AA\ to $\lambda\approx 2800$ \AA\ with a spectral resolving power of ${\rm FWHM}\approx 10$ \kms \ (corresponding to a $b$-value of $6 \ \kms$).  A primary motivation of this program was to target the associated $\mathrm{C}^{3+}$ transitions, which are intermediate in ionization potential to low-ions (e.g., $\mathrm{Mg}^+$) and high-ions (e.g., $\mathrm{O}^{5+}$), and therefore can help constrain the ionization structure of gas clouds more robustly \citep[see e.g.,][]{Kumar:2024}. NUV observations are complete for 14 out of the 15 sightlines with a mean signal-to-noise per resolution element of $\mathrm{S/N}\gtrsim 7$, and the data are incorporated in this work. 

All STIS exposures were processed and extracted using \texttt{stistools} (version 1.4.5) with dark current correction omitted. Applying default dark correction consistently leads to negative background counts, while omitting dark correction leads to consistency between the observed fluctuations in the continuum and the error spectrum. A custom software was developed to co-add multiple exposures to achieve optimal spectral resolution and $\mathrm{S/N}$. A finer bin size of $2 \ \kms$ was adopted, compared to the native bin size of $5 \ \kms$. This choice takes advantage of wavelength offsets between different exposures to achieve a finer spectral resolution than the instrument specification, enabling the recovery of
narrow, unsaturated metal lines (e.g., \ion{Si}{II} $\lambda$1526). 
 
For each spectral bin, the effective exposure time, defined as the total contribution from all unique exposure–order pairs, is tracked and subsequently used for flux calibration. Uncertainties are derived from the gross counts using two-sided Poisson confidence intervals corresponding to the 68\% (1\,$\sigma$) confidence level following the exact formulation of \citet{Gehrels:1986} and propagated through the stacking procedure. To perform flux calibration, the combined net count spectrum is compared with the pipeline-generated flux spectrum to obtain a wavelength-dependent scaling factor based on a running median across the full wavelength range. This scaling function is then applied to the net count rates and associated errors to obtain the final flux spectra for each sightline.

\subsection{Galaxy survey} \label{sec:gal_survey}

The CUBS program design includes a deep galaxy survey along all 15 QSO fields to associate gas properties observed along a sightline to the galaxy host. The survey design is summarized in \cite{Chen:2020}. Briefly, the survey includes a shallow and wide component carried out using the IMACS multi-object spectrograph on the Magellan Baade telescope \citep[][]{Dressler:2006} to map the large-scale environment at $d_\mathrm{proj} < 3$--$5 \ \mathrm{Mpc}$ from the QSO, down to $L_*$ galaxies at $z_\mathrm{gal} \lesssim 0.8$. A deep and narrow component performed using the Low Dispersion Survey Spectrograph 3 (LDSS3C) on the Magellan Clay telescope targets galaxies down to $0.1 L_*$ at $z_\mathrm{gal} \approx 1$ at $d_\mathrm{proj} < 300$--$500 \ \mathrm{kpc}$ from the sightline. Finally, the ultradeep component using MUSE on VLT UT4 \citep[][]{Bacon:2010} is designed to uncover galaxies as faint as $0.01 L_*$ at $z_\mathrm{gal} \lesssim 1$ at $d_\mathrm{proj} < 250 \ \mathrm{kpc}$ from the sightline.

In addition to the galaxy spectroscopy campaign, the CUBS design includes deep optical and near-infrared imaging for all 15 fields. The optical imaging performed using LDSS3C and IMACS covers $g, r, i, z$ bands, and the near-infrared portion covers the $H$ band using the FourStar imaging camera on the Magellan Baade telescope \citep{Persson:2013}. The available photometry for spectroscopically confirmed galaxies provides constraints on their spectral energy distributions (SEDs). Specifically, the photometric measurements for individual galaxies are fit using the Bayesian Analysis of Galaxies for Physical Inference and Parameter EStimation code \citep[\texttt{BAGPIPES}; ][]{Carnall:2018} with an adopted initial mass function (IMF) from \cite{Kroupa:2001}. The inferred stellar masses ($M_\mathrm{star}$) for CUBS galaxies have a typical systematic uncertainty of $0.2 \ \mathrm{dex}$ \citep[see][]{DePalma:2025}. The corresponding dark matter halo masses $M_h$ are evaluated using an average between the stellar-mass to halo-mass relations from \cite{Kravtsov:2018} and \cite{Behroozi:2019}, which differ primarily in their attribution of intrahalo light to the central galaxy or satellite mergers. Recently, high-resolution images of the CUBS fields were obtained for the F814W band of Advanced Camera for Surveys \citep[ACS;][]{Ford:1998} on the \textit{HST} as part of a Cycle 32 General Observer Program (GO-DISCS, PID=17815, PI: Beckett). The \textit{HST} galaxy photometry was not included in the SED fitting, but the galaxy images are presented in the Appendix for individual systems.

\startlongtable
\begin{deluxetable*}{rrrrrrrrr}
\tablewidth{\textwidth}
\renewcommand{\arraystretch}{0.95}
\tablecaption{Galaxy properties for the CUBS-midz sample\label{tab:gal_props}}
\tablehead{
\colhead{} & \colhead{} & \colhead{} & \colhead{Spec} & \colhead{} & \colhead{} & \colhead{$d_\mathrm{proj}$} & \colhead{$d_\mathrm{proj}$} & \colhead{$v_\mathrm{abs}-v_\mathrm{gal}$} \\
\colhead{ID} & \colhead{Galaxy} & \colhead{$z_\mathrm{gal}$} & \colhead{Type} & \colhead{$\log(M_\mathrm{star}/\mathrm{M}_\odot)$} & \colhead{$\log(M_h/\mathrm{M}_\odot)$} & \colhead{$(\mathrm{kpc})$} & \colhead{$(R_\mathrm{vir})$} & \colhead{$(\mathrm{km\,s^{-1}})$}
}
\startdata
s1 & J230837.40$-$525844.94 & 0.54248 & SF & 10.6 & 12.1 & 34 & 0.16 & +49 \\
s2 & J213553.27$-$531700.70 & 0.67577 & SF & 8.0 & 10.8 & 35 & 0.49 & +5 \\
s3 & J011139.36$-$031605.19 & 0.57655 & Q & 10.9 & 12.6 & 42 & 0.14 & $-$76 \\
 & J011137.89$-$031605.58 & 0.57545 &  & 9.2 & 11.3 & 131 & 1.20 &  \\
 & J011138.08$-$031551.43 & 0.57745 &  & 9.8 & 11.5 & 167 & 1.27 &  \\
 & J011141.33$-$031602.00 & 0.57620 &  & 9.4 & 11.4 & 220 & 1.90 &  \\
 & J011142.19$-$031557.24 & 0.57610 &  & 10.2 & 11.8 & 310 & 1.96 &  \\
s4 & J015455.18$-$071222.15 & 0.45501 & SF & 8.0 & 10.8 & 43 & 0.55 & +592 \\
s5 & J011035.01$-$164833.32 & 0.47239 & SF & 10.1 & 11.7 & 54 & 0.35 & $-$22 \\
 & J011033.82$-$164828.42 & 0.47299 &  & 9.0 & 11.2 & 144 & 1.34 &  \\
s6 & J233912.33$-$552354.27 & 0.57259 & SF & 8.5 & 11.0 & 55 & 0.61 & +48 \\
s7 & J224500.93$-$493154.29 & 0.53378 & SF & 8.0 & 10.8 & 58 & 0.78 & $-$129 \\
s8 & J011138.57$-$031609.64 & 0.64825 & Q & 10.9 & 12.6 & 63 & 0.22 & $\ldots$ \\
 & J011139.65$-$031540.77 & 0.64835 &  & 8.5 & 11.0 & 214 & 2.49 &  \\
s9 & J213552.98$-$531645.51 & 0.50818 & SF & 8.1 & 10.8 & 65 & 0.83 & +10 \\
s10 & J035723.09$-$481215.82 & 0.43574 & SF & 10.8 & 12.5 & 66 & 0.21 & $-$102 \\
s11 & J035721.44$-$481225.86 & 0.44894 & SF & 8.3 & 10.9 & 67 & 0.78 & +54 \\
s12 & J042052.57$-$565038.97 & 0.58386 & SF & 9.4 & 11.4 & 80 & 0.68 & $-$112 \\
 & J042055.26$-$565025.92 & 0.58396 &  & 10.4 & 11.9 & 140 & 0.82 &  \\
s13 & J045416.38$-$611640.90 & 0.43149 & Q & 10.1 & 11.7 & 82 & 0.50 & $-$73 \\
 & J045412.79$-$611645.81 & 0.43159 &  & 8.7 & 11.1 & 167 & 1.64 &  \\
 & J045422.36$-$611707.50 & 0.43071 &  & 10.9 & 12.8 & 346 & 0.92 &  \\
s14 & J035723.55$-$481215.38 & 0.48074 & SF & 9.9 & 11.6 & 97 & 0.68 & $-$142 \\
 & J035725.26$-$481208.90 & 0.47874 &  & 10.7 & 12.2 & 203 & 0.90 &  \\
s15 & J015453.61$-$071216.79 & 0.46791 & SF & 10.1 & 11.7 & 99 & 0.63 & +78 \\
 & J015453.23$-$071231.92 & 0.47051 &  & 8.7 & 11.1 & 140 & 1.44 &  \\
 & J015452.06$-$071205.48 & 0.46775 &  & 9.7 & 11.5 & 249 & 1.82 &  \\
s16 & J224458.73$-$493158.12 & 0.58627 & SF & 10.3 & 11.8 & 114 & 0.71 & $-$49 \\
s17 & J213552.14$-$531700.49 & 0.62187 &  & 9.0 & 11.2 & 72 & 0.70 &  \\
 & J213552.16$-$531707.08 & 0.62247 &  & 9.7 & 11.5 & 99 & 0.77 &  \\
 & J213553.93$-$531709.54 & 0.62017 &  & 9.4 & 11.4 & 103 & 0.89 &  \\
 & J213553.48$-$531714.50 & 0.62187 & Q & 10.8 & 12.4 & 128 & 0.53 & +124 \\
 & J213554.29$-$531716.16 & 0.62147 &  & 9.7 & 11.5 & 153 & 1.20 &  \\
 & J213550.14$-$531713.39 & 0.62187 &  & 10.6 & 12.1 & 222 & 1.15 &  \\
 & J213550.58$-$531720.31 & 0.62347 &  & 10.7 & 12.3 & 231 & 1.02 &  \\
 & J213549.83$-$531715.08 & 0.61977 &  & 10.3 & 11.9 & 243 & 1.48 &  \\
 & J213550.94$-$531728.07 & 0.61827 &  & 9.0 & 11.2 & 258 & 2.56 &  \\
 & J213551.04$-$531734.52 & 0.62403 &  & 10.4 & 11.9 & 295 & 1.73 &  \\
s18 & J224458.93$-$493129.29 & 0.48928 & SF & 10.1 & 11.7 & 138 & 0.88 & $\ldots$ \\
s19 & J011955.08$-$201007.02 & 0.45109 &  & 8.4 & 10.9 & 122 & 1.37 &  \\
 & J011957.16$-$201003.55 & 0.45169 & SF & 10.6 & 12.0 & 141 & 0.68 & +21 \\
s20 & J045415.97$-$611602.32 & 0.60109 & Q & 10.5 & 12.0 & 162 & 0.87 & $\ldots$ \\
s21 & J224459.54$-$493133.21 & 0.61817 &  & 7.8 & 10.7 & 112 & 1.66 &  \\
 & J224457.67$-$493141.90 & 0.61777 &  & 10.3 & 11.9 & 173 & 1.05 &  \\
 & J224457.76$-$493159.13 & 0.62027 & Q & 10.7 & 12.2 & 177 & 0.81 & $-$309 \\
 & J224456.77$-$493130.73 & 0.61827 &  & 9.2 & 11.3 & 257 & 2.35 &  \\
s22 & J011035.31$-$164753.57 & 0.54119 &  & 9.4 & 11.4 & 218 & 1.84 &  \\
 & J011035.28$-$164753.17 & 0.54075 &  & 8.8 & 11.1 & 220 & 2.26 &  \\
 & J011036.83$-$164740.54 & 0.53987 &  & 10.0 & 11.6 & 323 & 2.23 &  \\
 & J011038.69$-$164804.69 & 0.54132 &  & 9.3 & 11.3 & 325 & 2.81 &  \\
 & J011037.69$-$164915.63 & 0.54043 & Q & 11.3 & 13.6 & 364 & 0.56 & +167 \\
 & J011039.21$-$164900.82 & 0.54242 &  & 8.9 & 11.2 & 398 & 3.91 &  \\
 & J011039.91$-$164834.19 & 0.53960 &  & 10.4 & 11.9 & 403 & 2.29 &  \\
 & J011041.05$-$164749.29 & 0.53945 &  & 8.5 & 11.0 & 560 & 6.19 &  \\
 & J011041.42$-$164752.64 & 0.54214 &  & 9.8 & 11.5 & 584 & 4.34 &  \\
 & J011042.22$-$164757.80 & 0.53973 &  & 10.2 & 11.7 & 640 & 4.11 &  \\
s23 & J045414.73$-$611722.85 & 0.57859 &  & 8.8 & 11.1 & 375 & 3.89 &  \\
 & J045414.37$-$611511.34 & 0.58164 &  & 11.0 & 13.0 & 501 & 1.29 &  \\
 & J045406.52$-$611538.71 & 0.57874 &  & 9.0 & 11.2 & 547 & 5.28 &  \\
 & J045408.65$-$611522.74 & 0.58790 &  & 9.8 & 11.6 & 548 & 4.08 &  \\
 & J045408.60$-$611522.67 & 0.58790 &  & 8.9 & 11.2 & 550 & 5.55 &  \\
 & J045406.34$-$611518.54 & 0.58113 & Q & 11.6 & 14.2 & 640 & 0.65 & +99 \\
 & J045404.65$-$611411.54 & 0.58113 &  & 9.3 & 11.3 & 1039 & 9.26 &  \\
\enddata
\end{deluxetable*}

The CUBS-midz sample at $z_{\rm gal}=0.4$--$0.7$ from \cite{Qu:2024} is adopted for the study.  The redshift range ensures the coverage of the \ion{O}{VI} $\lambda \lambda 1031, 1037$ and \ion{C}{IV} $\lambda \lambda 1548, 1550$ in the COS and STIS spectra, respectively, enabling an accurate characterization of the ionization structure and chemical pattern of the circumgalactic gas. Galaxies may be found in overdense environments, which are identified using the CUBS galaxy survey catalog (see S.D. Johnson et al. 2026, to be submitted). 
The steps for identifying likely neighbors for each galaxy in the CUBS-midz sample are summarized below. An initial galaxy sample is identified based on $d_\mathrm{proj}$ 
being within three times the virial radius ($R_\mathrm{vir}$) of the galaxy. Galaxies in the sample are then grouped into systems using an iterative approach, where two systems are merged if any of their members are within $500\,\kms$ from each other in line-of-sight velocity. The grouping is repeated until the number of unique systems converges. Following this exercise, each galaxy in the sample is assigned a unique system ID.

\begin{figure*}
    \centering \includegraphics[width=\textwidth]{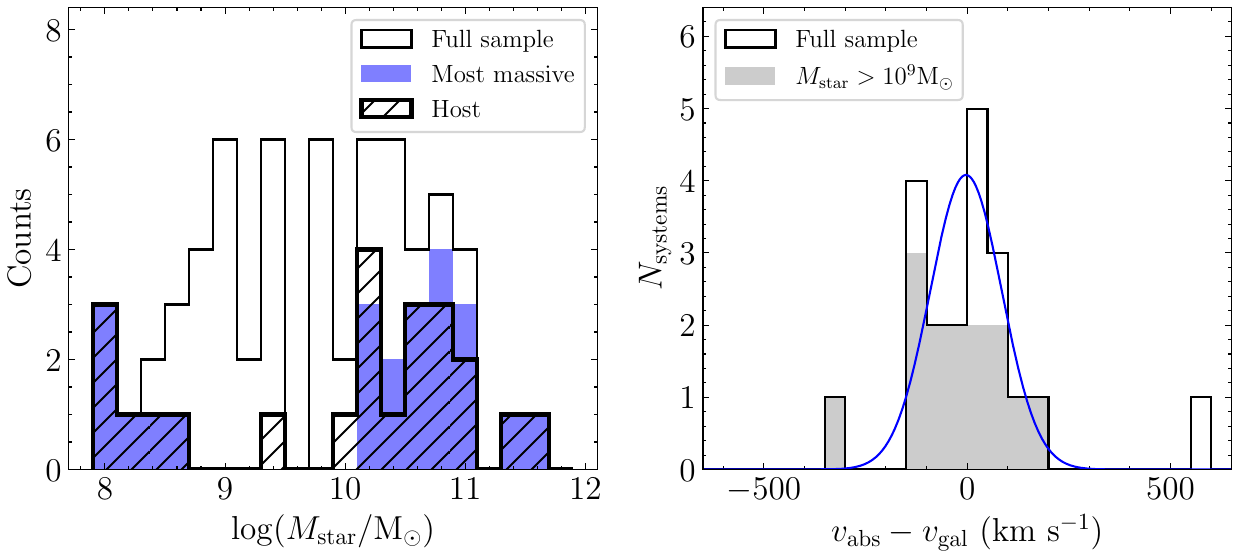}
\caption{(\textit{Left}): Stellar mass distributions of the CUBS-midz galaxies. The full sample is shown as open histograms, the most massive galaxy in each group as blue shaded histograms, and the galaxies with the smallest halo-radius-normalized projected distance, $d_\mathrm{proj}/R_\mathrm{vir}$, as hatched histograms. (\textit{Right}): Velocity distribution of the absorbing gas, characterized by the strongest \ion{H}{I} absorption component associated with the CUBS-midz host galaxies. The distribution is well described by a Gaussian with a centroid of $\mu\approx-3~\kms$ and a dispersion of $\sigma_v\approx88~\kms$. The distribution for massive hosts with $\mstar>10^9\,\msun$ is shown as the gray shaded histogram. See the main text in \S\,\ref{sec:gal_survey} for details.}
\label{fig:sample}
\end{figure*}

CUBS-midz galaxy systems are listed in Table \ref{tab:gal_props}. For each system, the galaxy host is designated as the member with the smallest $R_{\rm vir}$-normalized projected distance from the QSO sightline ($d_\mathrm{proj}/R_\mathrm{vir}$). To ascertain galaxy-absorber connection, only systems where the host satisfies $d_\mathrm{proj} < R_\mathrm{vir}$ are chosen for this study. The system IDs in Table \ref{tab:gal_props} follow an increasing order in $d_\mathrm{proj}$ for the galaxy host. Because the physical properties of the absorbing gas will likely only be affected by immediate satellites of the host, only system members that are within the virial radius of the host (or encompass the host within their virial radius) are specified in Table \ref{tab:gal_props}. In 11 out of 23 CUBS-midz systems, at least one satellite galaxy can be identified following the above mentioned criteria for neighbor identification. The stellar mass distribution of the entire galaxy sample (including satellites) is shown in the left panel of Figure \ref{fig:sample}. Although the stellar mass range for the full sample is
$M_\mathrm{star} \approx 10^8 \msun$--$3 \times 10^{11} \msun$, lower-mass halos are typically identified as satellites of more massive systems. Host galaxies can also be classified as star-forming (SF) or quiescent (Q) based on the presence of prominent emission features in the galaxy spectrum (e.g., [\ion{O}{II}] $\lambda \lambda 3727, 3729$, H$\beta$, see Appendix for host galaxy spectra). The spectral type for galaxy hosts has also been indicated in Table \ref{tab:gal_props}.

To identify CGM absorption associated with these galaxies, QSO spectra are searched in a wide window of $\pm 1000 \ \kms$ centered at the host galaxy. Out of 23 CUBS-midz galaxies, corresponding absorption is identified for 20 systems. The velocity centroid of the strongest \ion{H}{I} absorption component relative to the host galaxy redshift is stated in the last column of Table \ref{tab:gal_props}, and the corresponding velocity distribution is presented in the right panel of Figure \ref{fig:sample}. All but one of the absorbers are found to be within $500 \ \kms$ of the galaxy. In fact, excluding the two outliers at $|v_\mathrm{abs}-v_\mathrm{gal}| > 200 \ \kms$, the absorber velocity distribution for the remaining 18 galaxies is well-described by a Gaussian of dispersion $\sigma_v \approx 88 \ \kms$ (blue curve in Figure \ref{fig:sample}). For a dark matter halo of $5\times 10^{11}\,\msun$, roughly corresponding to the median halo mass of the CUBS-midz sample, the anticipated line-of-sight velocity dispersion due to virial motions is $\sigma_\mathrm{vir} \approx 69 \ \mathrm{km \ s}^{-1}$. At the same time, considering only massive galaxies with $\log\,\mstar/\msun>9$ (14 out of 20), the median halo mass is $\approx 10^{12}\,\msun$ and the anticipated $\sigma_\mathrm{vir} \approx 92 \ \kms$, while the observed velocity dispersion is $\sigma_v \approx 96 \ \kms$. Therefore, gas motions in star-forming CUBS-midz galaxies are comparable to expectations from virial motions \citep[see also][]{Huang:2021}, in contrast to sub-virial motions noted in massive quiescent halos \citep[see e.g.,][]{Huang:2016}.

In examining the contribution of correlated halos to observed metal-line absorption through line-of-sight projection, \citet{Ho:2020} report a false-positive rate of 15–30\% for \ion{Mg}{II} within $|v_c-v_\mathrm{gal}|<500 \ \kms$ at $d_\mathrm{proj} \approx 180 \ \mathrm{kpc}$ around galaxies with $10^{10} \msun \lesssim \mstar \lesssim 10^{11} \msun$, with greater contamination predicted for higher-ionization species such as \ion{O}{VI} \citep[][]{Ho:2021}. While some contamination by correlated halos cannot be ruled out, two observational signatures suggest that its impact on the low-ionization absorbers that are the primary focus of this work is limited. First, although the predicted contamination remains similar for velocity windows as narrow as $\pm300$ \kms, the observed absorber–galaxy velocity distribution is more concentrated, with a dispersion of only $\sigma_v \approx 90$ \kms\ \citep[Figure \ref{fig:sample}; see also][]{Zahedy:2019}, a velocity regime not explicitly tested in these simulations. Second, metal-line absorption declines steeply with increasing $d_{\rm proj}$ \citep[e.g.,][]{Chen:2010, Liang:2014, Qu:2023}, whereas contamination by correlated halos would tend to dilute such a radial trend. Taken together, these observations suggest that correlated-halo contamination is unlikely to significantly affect the presented results.

\section{Methods} \label{sec:methods}

Estimates of ionizing conditions from the observed ionic ratios in the diffuse CGM rely on accurate measurements of column densities from the absorption line profiles.  Additionally, the measured line broadening for different metals can be used to constrain thermal and non-thermal motions, providing insights into pressure support in the gas. In this section, the steps for the Voigt profile fitting, photoionization modeling, and line width analysis are presented.

\subsection{Voigt profile fitting} \label{sec:vp_fit}

The line profiles of ionic transitions are described using the minimum number of kinematic components necessary to reproduce the observed spectrum. Each component is modeled using a Voigt function with an associated velocity centroid $v_c-v_\mathrm{gal}$ (relative to the host galaxy redshift), ionic column density $N_c$, and Doppler parameter $b_c$ (or absorption line width). The profile fitting is performed using custom software described in \cite{Kumar:2024}. Kinematically matched ions are fit simultaneously to obtain an accurate description of the component structure. COS transitions may require wavelength-dependent velocity offsets of $\delta v \lesssim 7.5 \ \kms$ on average,  \citep[see e.g.,][]{Johnson:2013,Liang:2014,Wakker:2015,Plesha:2018}, but the wavelength calibration can be particularly uncertain at the segment edges, requiring offsets of $10$--$15\ \kms$. To estimate uncertainties on the best-fit Voigt profile parameters, a Markov Chain Monte Carlo (MCMC) sampling is performed using the \citep[\texttt{EMCEE} package in \texttt{Python},][]{F-M:2013}. Each MCMC run consists of 200 walkers seeded around the maximum likelihood solution, with 2500 steps assigned per walker. The median value and uncertainties corresponding to the 68\% (1\,$\sigma$) confidence intervals are tabulated for the ionic column density and line width for absorbing components in Tables \ref{tab:s1}--\ref{tab:s23}, and the best-fit absorption profiles are shown in Figures \ref{fig:s1}--\ref{fig:s23}. For saturated transitions, the line width is tied to an appropriate unsaturated transition during the fitting, and conservative 99.7\% (3\,$\sigma$) lower limits are reported on the column density. Non-absorbing transitions are not included in the fitting, but a 3\,$\sigma$ upper limit is placed on the ionic column density by comparing the error spectrum with models adopting a fixed $b$ value. For non-detections of low- and intermediate ions, the Doppler parameter is chosen to reflect the instrument line-spread function ($b = 12 \ \kms$ for COS, $6 \ \kms$ for STIS, and $4 \ \kms$ for MIKE). For non-detections of high-ions (e.g., \ion{O}{VI}), a fixed value of $b = 30 \ \kms$ is adopted. 

\subsection{Photoionization modeling and line width analysis} \label{sec:ion_model}

The ionizing conditions of individual gas clumps are reflected in the relative ionic column density ratios. Photoionization equilibrium (PIE) models assuming ionization balance with the metagalactic background radiation have been shown to reproduce ionic columns measured in circumgalactic absorbers \citep[see e.g.,][]{Bergeron:1986,Chen:2000,Werk:2014,Zahedy:2019,Qu:2022,Sameer:2024}. The photoionized origin of these clumps is also corroborated by the observed line widths, which suggest the presence of cool $T \sim 10^4 \ \mathrm{K}$ gas \citep[see e.g.,][]{Rauch:1996CIV,Rudie:2019,Chen:2023}, consistent with the expected thermal balance between photoheating from the extragalactic ultraviolet background (UVB) and radiative cooling \citep[see e.g.,][]{Oppenheimer:2013a,Gnat:2017}. Note that ionizing radiation escaping from star-forming regions can be important out to $\approx 50 \ \mathrm{kpc}$ \citep[see e.g.,][]{Bland-Hawthorn:2019}. Consequently, UVB-only models may underestimate the gas density for such components (e.g., system s1 in the CUBS-midz sample).

\begin{figure*}
    \centering    \includegraphics[width=\textwidth]{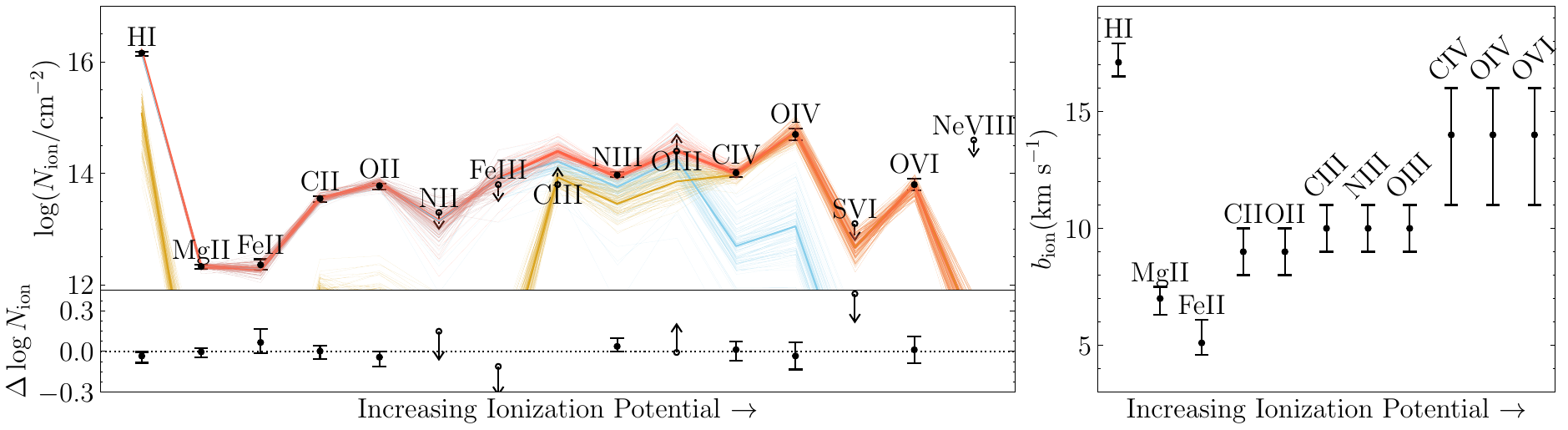}
  \caption{Column density measurements and detection limits in component c3 (at $v_c -v_\mathrm{gal} = -70 \ \kms$) in system s10 (Figure \ref{fig:s10}) are shown in the \textit{top-left} panel. A two-phase photoionization model is necessary to reproduce both low- and high-ions detected in this absorption component. The high-density phase ($n_\mathrm{c3h} \approx 0.003 \ \mathrm{cm^{-3}}$) is shown in blue, the low-density phase ($n_\mathrm{c3l} \approx 10^{-4} \ \mathrm{cm^{-3}}$) is shown in yellow, and the combined contribution from both phases is shown in red. Residuals 
  between the measurements and the best-fit ionization model are depicted in 
  the \textit{bottom-left} panel. Line width measurements for various ionic species in the component are included in the 
  \textit{right} panel. The distinct physical origins of low- and high-ionization species are apparent from their significantly different line widths.} 
  \label{fig:s10_PIE}
\end{figure*}

Photoionization modeling and line width analysis for CUBS-midz absorption components is performed using a custom software described in \cite{Kumar:2024}. Briefly, a grid of photoionization models is compiled using \texttt{CLOUDY} \citep[version 22.01][]{Ferland:2017, Chatzikos:2023}, assuming plane-parallel geometry and irradiation by the extragalactic ultraviolet background (UVB) from \citet[hereafter FG20]{F-G:2020}. For reference, a clump of density $n_\mathrm{H} = 0.001 \ \mathrm{cm}^{-3}$ at a redshift of $z=0.55$ corresponds to an ionization parameter of $\log\,U = -2.48$. The model grid is defined in terms of the \ion{H}{I} column density $N(\mathrm{H\ I})$, gas density $n_\mathrm{H}$, and gas metallicity $\mathrm{[O/H]}$. A likelihood function comparing the predicted ionic column densities with the measured values and detection limits resulting from Voigt profile analysis is used to derive the best-fit ionization model parameters. Note that a solar abundance pattern \citep[][]{Asplund:2009} is assumed during photoionization calculations because the ion fractions depend only weakly on the chemical pattern of the gas \cite[see][]{Kumar:2025}. However, the likelihood function incorporates departures from the solar pattern when fitting column density measurements from carbon, nitrogen, and iron ions (see S. Kumar et al. 2026b, to be submitted, for further discussion). 

Following the ionization analysis, the thermal and non-thermal motions in individual kinematic components can be separated. Line width measurements from different metal ions can be modeled using a single gas temperature $T$ and non-thermal broadening $b_\mathrm{NT}$ \citep[see e.g.,][]{Rauch:1996CIV,Rudie:2019,Qu:2022,Kumar:2024}, and a likelihood function comparing the predicted values with measurements from the Voigt profile analysis yields constraints on $T$ and $b_\mathrm{NT}$. Note that \ion{H}{I} lines from COS FUV data are analyzed jointly with low-ionization metal transitions (e.g., \ion{Mg}{II}, \ion{C}{II}) observed at high spectral resolution in optical and NUV echelle spectra, combining the large mass contrast with high resolving power needed to separate the thermal and non-thermal line-broadening \citep[see also][]{Qu:2022}.

As an illustrative example, the photoionization and line width analysis of component c3 at $v_c-v_\mathrm{gal} = -70 \ \mathrm{km \ s}^{-1}$ in system s10 is discussed here. This absorption system is identified in the vicinity of an isolated star-forming galaxy with stellar mass $M_\mathrm{star} \approx 6 \times 10^{10} \msun$ at a projected separation of $d_\mathrm{proj} \approx 66 \ \mathrm{kpc}$. Component c3 exhibits kinematically aligned absorption in low-, intermediate-, and high-ions, with the corresponding column density measurements presented in Figure \ref{fig:s10_PIE}. 
The low- and intermediate-ionization species can be reproduced by a high-density phase, denoted by the suffix h, with a hydrogen volume density of $n_\mathrm{H, c3h} \approx 0.003 \ \mathrm{cm}^{-3}$ (blue curves in Figure \ref{fig:s10_PIE}). This phase is constrained using stringent 3\,$\sigma$ limits on the \ion{O}{III}/\ion{O}{II}, \ion{N}{III}/\ion{N}{II}, and \ion{Fe}{III}/\ion{Fe}{II} ratios. Meanwhile, the measured line widths for this component indicate a gas temperature of $T_\mathrm{c3h} \approx 2 \times 10^4 \ \mathrm{K}$ and a modest non-thermal broadening of $b_\mathrm{NT,c3h} \approx 6 \ \kms$ for the high-density phase. The temperature is constrained by the measured line width for \ion{H}{I}, while the non-thermal broadening is sensitive to metal ions (e.g., \ion{Mg}{II}, \ion{C}{II}).

While the absorption from low- and intermediate-species is well-explained by the high-density phase, it underpredicts the amount of associated high-ionization species observed in this component. An additional low-density phase, denoted by the suffix l, with $n_\mathrm{H, c3l} \approx 10^{-4} \ \mathrm{cm}^{-3}$ (yellow curves in Figure \ref{fig:s10_PIE}) is necessary to explain the full range of ionic absorption observed for this component, particularly the observed \ion{O}{VI}/\ion{O}{IV} ratio. The associated \ion{H}{I} content for the low-density phase is uncertain because most of the measured \ion{H}{I} is expected to arise in the high-density phase. Among absorption components that are well-described by a single-phase solution, those with $3 \times 10^{15} \ \mathrm{cm}^{-2}<N(\mbox{H\ I})<3\times 10^{18} \mathrm{cm}^{-3}$ exhibit a steady increase in the inferred density from $n_\mathrm{H} \approx 0.001 \ \mathrm{cm}^{-3}$--$0.03 \ \mathrm{cm}^{-3}$. Meanwhile, components with $N(\mbox{H\ I})<3 \times 10^{15} \ \mathrm{cm}^{-3}$ have densities $n_\mathrm{H}<0.001 \ \mathrm{cm}^{-3}$. Accordingly, the observed neutral hydrogen column density of $N(\mbox{H I})\approx10^{16}~\mathrm{cm^{-2}}$ for component c3 is likely associated entirely with the high-density phase, which implies a characteristic line-of-sight thickness of $l_{\mathrm{c3h}}\approx40~\mathrm{pc}$. Plausible 3\,$\sigma$ limits of $N($\ion{H}{I}$)=4\times10^{13}$--$5 \times 10^{15} \ \mathrm{cm}^{-2}$ can be placed for the low-density phase by requiring the associated cloud size to be more extended than the high-density phase, without violating the error budget provided by the measurement uncertainty of $0.05 \ \mathrm{dex}$ on $N(\mbox{H I})$ from the Voigt profile analysis \citep[see e.g.,][]{Kumar:2024}. The two-phase structure is also reflected in the measured absorption line widths, where low-ionization species exhibit comparitively narrower profiles than the high-ions \citep[right panel of Figure \ref{fig:s10_PIE}, see also,][]{Cooper:2021}, although the thermal and non-thermal motions cannot be separated in the low-density phase due to similar masses of \ion{C}{IV} and \ion{O}{IV}. 

Ionization modeling and line width analysis results for the remaining CUBS-midz components are presented in Table \ref{tab:ion_model}. Ionization modeling constraints are available for 31 CUBS-midz components identified near 12 unique CUBS-midz galaxies. The median values and 1\,$\sigma$ uncertainties are presented for the \ion{H}{I} column density, inferred total hydrogen column density, gas density, gas temperature, and non-thermal broadening. In the absence of constraints, appropriate 3\,$\sigma$ limits are quoted. Constraints on multi-element abundances will be reported in S. Kumar et al. (2026b).

\section{Physical conditions of the CGM at $z\approx 0.4$--0.7} \label{sec:results}

Simultaneous wavelength coverage of multiple ionic species in QSO absorption systems provides powerful diagnostics of the ionization conditions in the absorbing gas. Comparing observed ionic ratios in individual absorption components with predictions from photoionization models yields constraints on the gas ionization parameter $U$, or equivalently, the volume density $n_\mathrm{H}$. Additionally, comparisons between the absorption-line widths of species with sufficiently different atomic masses allow the thermal and non-thermal contributions to the line broadening, parameterized by $b_T$ and $b_{\rm NT}$, to be separated. Physical conditions of the CGM in CUBS-midz galaxies are derived following the procedure described in \S\,\ref{sec:ion_model} and are summarized in Table~\ref{tab:ion_model}.

\startlongtable
\begin{deluxetable*}{lrrrrrrr}
\tablecaption{Summary of absorption properties for CUBS-midz galaxy sample\tablenotemark{a}\label{tab:ion_model}}
\tablehead{
\multicolumn{1}{l}{} & \multicolumn{1}{c}{$v_c-v_\mathrm{gal}$} & \multicolumn{1}{r}{} & \multicolumn{1}{r}{} & \multicolumn{1}{c}{$b_\mathrm{NT}$} & \multicolumn{1}{r}{} & \multicolumn{1}{r}{} & \multicolumn{1}{r}{} \\
\multicolumn{1}{l}{Component} & \multicolumn{1}{c}{(\kms)} & \multicolumn{1}{r}{$\log(N_\mathrm{HI}/\mathrm{cm}^{-2})$} & \multicolumn{1}{r}{$\log(T/\mathrm{K})$} & \multicolumn{1}{c}{(\kms)} & \multicolumn{1}{r}{$\log(P_\mathrm{NT}/P_\mathrm{T})$} & \multicolumn{1}{r}{$\log(N_\mathrm{H}/\mathrm{cm}^{-2})$} & \multicolumn{1}{r}{$\log(n_\mathrm{H}/\mathrm{cm}^{-3})$} 
}
\startdata
s1c1 & $-95$ & $16.12\pm0.02$ & $3.82_{-0.03}^{+0.02}$ & $6.3\pm0.1$ & $-0.66_{-0.02}^{+0.03}$ & $18.05\pm0.08$ & $-2.20\pm0.07$ \\
s1c3 & $+25$ & $16.72_{-0.02}^{+0.03}$ & $4.18\pm0.02$ & $3.6\pm0.1$ & $-1.51\pm0.03$ & $18.8\pm0.4$ & $-2.3\pm0.4$ \\
s1c4 & $+49$ & $17.52_{-0.01}^{+0.02}$ & $4.03_{-0.08}^{+0.07}$ & $6.2\pm0.1$ & $-0.9\pm0.1$ & $19.5\pm0.1$ & $-2.1\pm0.1$ \\
s1c5 & $+73$ & $16.5\pm0.2$ & $4.4_{-0.2}^{+0.1}$ & $2\pm1$ & $-2.2\pm0.5$ & $17.9_{-0.2}^{+0.3}$ & $-1.7_{-0.3}^{+0.2}$ \\
s1c6 & $+91$ & $16.0\pm0.3$ & $4.20\pm0.05$ & $3.8\pm0.2$ & $-1.5\pm0.1$ & $18.26_{-0.09}^{+0.07}$ & $-2.42\pm0.09$ \\
s2c1 & $+5$ & $14.47\pm0.03$ & ... & ... & ... & ... & ... \\
s3 & $-76$ & $20.1\pm0.1$ & ... & ... & ... & ... & ... \\
s4c2 & $+592$ & $13.99\pm0.03$ & ... & ... & ... & ... & ... \\
s5c2h & $-22$ & $16.04_{-0.03}^{+0.02}$ & $4.47_{-0.06}^{+0.05}$ & $12\pm1$ & $-0.8\pm0.1$ & $18.44_{-0.05}^{+0.04}$ & $-2.58\pm0.04$ \\
s5c2l & ... & $[12.6, 15.3]$ & $<6.0$ & $<34$ & ... & $[17.0, 20.1]$ & $-4.58_{-0.05}^{+0.06}$ \\
s5c3h & $+32$ & $14.90_{-0.05}^{+0.07}$ & $4.3\pm0.1$ & $16\pm2$ & $-0.3\pm0.1$ & $17.27\pm0.05$ & $-2.79\pm0.07$ \\
s5c3l & ... & $[13.2, 14.7]$ & $<5.5$ & $<20$ & ... & $[17.3, 19.1]$ & $-4.32_{-0.04}^{+0.03}$ \\
s5c4h & $+166$ & $15.09_{-0.02}^{+0.03}$ & $4.24_{-0.09}^{+0.08}$ & $3_{-2}^{+1}$ & $-1.7_{-0.6}^{+0.3}$ & $17.21\pm0.04$ & $-2.53\pm0.06$ \\
s5c4l & ... & $[13.1, 14.5]$ & $<5.6$ & $<23$ & ... & $[17.3, 18.9]$ & $-4.42\pm0.02$ \\
s6c1 & $-10$ & $14.84\pm0.04$ & $4.9\pm0.1$ & $<13$ & $<-1.1$ & $18.8\pm0.2$ & $-3.9\pm0.1$ \\
s6c2 & $+48$ & $15.23\pm0.02$ & $3.9_{-0.4}^{+0.3}$ & $26_{-2}^{+1}$ & $+0.5_{-0.3}^{+0.4}$ & $18.8_{-0.2}^{+0.1}$ & $-3.6\pm0.1$ \\
s7c1 & $-129$ & $15.18\pm0.04$ & ... & ... & ... & ... & ... \\
s8 & ... & $<13.6$ & ... & ... & ... & ... & ... \\
s9c1 & $+10$ & $14.29\pm0.03$ & ... & ... & ... & ... & ... \\
s10c1 & $-141$ & $16.10_{-0.03}^{+0.02}$ & $4.17_{-0.08}^{+0.06}$ & $15\pm1$ & $-0.3\pm0.1$ & $18.28\pm0.04$ & $-2.41\pm0.03$ \\
s10c2h & $-102$ & $17.10\pm0.04$ & $4.17_{-0.07}^{+0.06}$ & $6.9\pm0.7$ & $-0.9\pm0.1$ & $19.8_{-0.2}^{+0.1}$ & $-2.8\pm0.1$ \\
s10c2l & ... & $[13.5, 16.2]$ & $<5.9$ & $<27$ & ... & $[17.9, 21.0]$ & $-4.6\pm0.1$ \\
s10c3h & $-69$ & $16.16_{-0.05}^{+0.03}$ & $4.20\pm0.04$ & $5.8\pm0.5$ & $-1.1\pm0.1$ & $18.5\pm0.1$ & $-2.6\pm0.1$ \\
s10c3l & ... & $[13.6, 15.7]$ & $<5.5$ & $<19$ & ... & $[17.4, 20.1]$ & $-4.1\pm0.1$ \\
s10c4h & $-40$ & $14.35_{-0.09}^{+0.07}$ & $4.6\pm0.2$ & $14\pm2$ & $-0.8\pm0.2$ & $17.1\pm0.3$ & $-3.2_{-0.2}^{+0.3}$ \\
s10c4l & ... & $[12.5, 14.3]$ & $<6.0$ & $<30$ & ... & $[17.0, 19.0]$ & $-4.58_{-0.08}^{+0.07}$ \\
s10c6h & $+154$ & $13.95_{-0.03}^{+0.04}$ & $4.1_{-0.3}^{+0.2}$ & $16\pm2$ & $-0.1_{-0.2}^{+0.3}$ & $16.7\pm0.2$ & $-3.2\pm0.2$ \\
s10c6l & ... & $[12.1, 13.5]$ & $<5.9$ & $<27$ & ... & $[16.5, 18.0]$ & $-4.48\pm0.07$ \\
s11c2 & $+54$ & $14.32\pm0.03$ & $4.8_{-0.2}^{+0.1}$ & $17\pm4$ & $-0.8_{-0.2}^{+0.3}$ & $18.1\pm0.2$ & $-4.0_{-0.2}^{+0.1}$ \\
s12c1h & $-112$ & $14.86\pm0.02$ & $4.4_{-0.2}^{+0.1}$ & $17_{-4}^{+3}$ & $-0.4\pm0.2$ & $17.7_{-0.4}^{+0.3}$ & $-2.9_{-0.2}^{+0.3}$ \\
s12c1l & ... & $[12.5, 14.1]$ & $<6.0$ & $<45$ & ... & $[16.8, 18.7]$ & $-4.33_{-0.07}^{+0.06}$ \\
s12c2 & $-3$ & $14.42\pm0.05$ & $5.0_{-0.2}^{+0.1}$ & $<13$ & $<-1.2$ & $19.0\pm0.1$ & $-4.41\pm0.08$ \\
s13c1 & $-73$ & $13.67\pm0.03$ & ... & ... & ... & ... & ... \\
s14c1 & $-142$ & $14.10\pm0.04$ & ... & ... & ... & ... & ... \\
s15c1 & $+26$ & $14.95\pm0.03$ & $4.7_{-0.2}^{+0.1}$ & $25_{-4}^{+3}$ & $-0.3\pm0.2$ & $19.42\pm0.06$ & $-4.40\pm0.05$ \\
s15c2h & $+78$ & $14.76\pm0.07$ & $4.0_{-0.4}^{+0.3}$ & $15\pm2$ & $-0.1_{-0.3}^{+0.4}$ & $17.6\pm0.2$ & $-3.3\pm0.1$ \\
s15c2l & ... & $[13.0, 14.6]$ & $<5.5$ & $<20$ & ... & $[17.0, 19.1]$ & $-4.34_{-0.08}^{+0.07}$ \\
s15c3 & $+118$ & $14.94_{-0.06}^{+0.07}$ & $4.8_{-0.3}^{+0.2}$ & $32\pm5$ & $-0.2_{-0.2}^{+0.3}$ & $18.50\pm0.08$ & $-3.70\pm0.06$ \\
s15c4 & $+197$ & $14.71\pm0.03$ & $3.6\pm0.3$ & $12\pm2$ & $+0.1\pm0.3$ & $18.0\pm0.4$ & $-3.4_{-0.3}^{+0.2}$ \\
s16c1 & $-49$ & $15.85\pm0.02$ & $4.0_{-0.4}^{+0.2}$ & $19_{-4}^{+3}$ & $+0.1_{-0.3}^{+0.4}$ & $18.5_{-0.2}^{+0.3}$ & $-2.7\pm0.2$ \\
s17c2 & $+8$ & $15.40_{-0.03}^{+0.02}$ & $4.2\pm0.1$ & $20\pm1$ & $0.0\pm0.1$ & $20.1_{-0.2}^{+0.1}$ & $-4.4\pm0.1$ \\
s17c3 & $+124$ & $18.19\pm0.06$ & $4.09_{-0.06}^{+0.05}$ & $8.8\pm0.3$ & $-0.6\pm0.1$ & $19.8_{-0.2}^{+0.1}$ & $-2.1_{-0.1}^{+0.2}$ \\
s17c4 & $+151$ & $16.2_{-0.1}^{+0.3}$ & $4.5_{-0.2}^{+0.1}$ & $5\pm1$ & $-1.5_{-0.2}^{+0.3}$ & $18.2_{-0.4}^{+0.3}$ & $-2.1\pm0.3$ \\
s18 & ... & $<14.0$ & ... & ... & ... & ... & ... \\
s19c2h & $+21$ & $15.81\pm0.02$ & $4.26_{-0.06}^{+0.05}$ & $5.5\pm0.8$ & $-1.2\pm0.1$ & $18.12\pm0.06$ & $-2.63\pm0.05$ \\
s19c2l & ... & $[13.3, 15.0]$ & $<5.6$ & $<21$ & ... & $[17.2, 19.5]$ & $-4.27\pm0.07$ \\
s19c3 & $+45$ & $14.53_{-0.07}^{+0.06}$ & $4.95_{-0.09}^{+0.08}$ & $8_{-5}^{+4}$ & $-1.6_{-0.6}^{+0.4}$ & $18.0\pm0.1$ & $-3.7\pm0.1$ \\
s20 & ... & $<13.7$ & ... & ... & ... & ... & ... \\
s21c2h & $-309$ & $15.82_{-0.02}^{+0.03}$ & $4.66_{-0.04}^{+0.03}$ & $<14$ & $<-0.8$ & $18.1_{-0.2}^{+0.1}$ & $-2.5\pm0.1$ \\
s21c2l & ... & $[12.9, 15.1]$ & $<6.0$ & $<37$ & ... & $[16.8, 19.8]$ & $-3.8\pm0.4$ \\
s22c1h & $+167$ & $16.38\pm0.03$ & $3.99\pm0.03$ & $5.5_{-0.6}^{+0.5}$ & $-0.9\pm0.1$ & $18.42_{-0.09}^{+0.08}$ & $-2.22_{-0.06}^{+0.07}$ \\
s22c1l & ... & $[14.0, 15.8]$ & $<4.9$ & $<10$ & ... & $[17.9, 20.4]$ & $-4.0\pm0.2$ \\
s22c2 & $+187$ & $13.9_{-0.5}^{+0.3}$ & $4.3_{-0.3}^{+0.2}$ & $4_{-3}^{+2}$ & $-1.5_{-0.7}^{+0.5}$ & $17.0_{-0.4}^{+0.3}$ & $-3.4_{-0.3}^{+0.4}$ \\
s23c1 & $+99$ & $14.79\pm0.02$ & ... & ... & ... & ... & ... \\
\enddata
\tablenotetext{a}{For galaxies with no absorption detected, a 3-$\sigma$ upper limit to the neutral hydrogen column density is presented. Similarly, table entries with ... indicate that no measurements are possible due to a lack of metal absorption.}
\end{deluxetable*} 

Derived constraints on the thermodynamic state and internal motions within cool clumps provide a unique opportunity to examine the physical processes shaping the CGM at intermediate redshifts. Combining absorption-line measurements and ionization modeling results for individual components with the available galaxy data, the physical origin of the absorbing gas and its connection to the host galaxies is now investigated. 

\subsection{A joint CUBS-midz and literature galaxy sample} \label{sec:galaxies}

CUBS-midz galaxies are typically star-forming, with typical halo masses between $M_h \approx 10^{11}$--$10^{12} \msun$.
To extend the investigation of the CGM to more massive, quiescent halos, physical properties are derived for absorption components associated with nine luminous red galaxies with halo masses $M_h \approx 10^{12} \msun$--$10^{13.5} \msun$ at $z=0.2$--$0.6$ from \citet[hereafter COS-LRGs]{Chen:2018} and are reported in Table \ref{tab:ion_model_cos_lrg}. \cite{Zahedy:2019} assumed an updated version of the UVB from \citet[hereafter HM05]{Haardt:2001} for the photoionization analysis of COS-LRG absorbers. Since the CUBS-midz sample is analyzed using the FG20 UVB (\S\,\ref{sec:ion_model}), the photoionization analysis for COS-LRG components is re-performed using FG20 to enable comparison with CUBS-midz components. On average, inferred gas densities assuming the FG20 UVB are found to be lower by $0.2 \ \mathrm{dex}$ compared to the HM05 estimates. 

Another important feature of the CUBS-midz sample is the spectroscopic identification of satellite galaxies (\S\,\ref{sec:gal_survey}). For COS-LRGs, companion galaxies are cross-matched within the Sloan Digital Sky Survey \citep[SDSS,][]{York:2000}, Data Release 1 of the Dark Energy Spectroscopic Instrument \citep[DESI DR1,][]{DESI:2025}, and the COS-Halos spectroscopic survey \citep[][]{Werk:2012}. Due to the limited survey depth in several of these fields, galaxies with photometric redshifts $z_\mathrm{phot}$ that are consistent with the LRG are also accepted as its satellites if the photometric redshift uncertainty satisfies $\sigma (z_\mathrm{phot}) < 0.1$ and the galaxy colors are similar to the LRG.

\begin{figure*}              
    \centering \includegraphics[width=\textwidth]{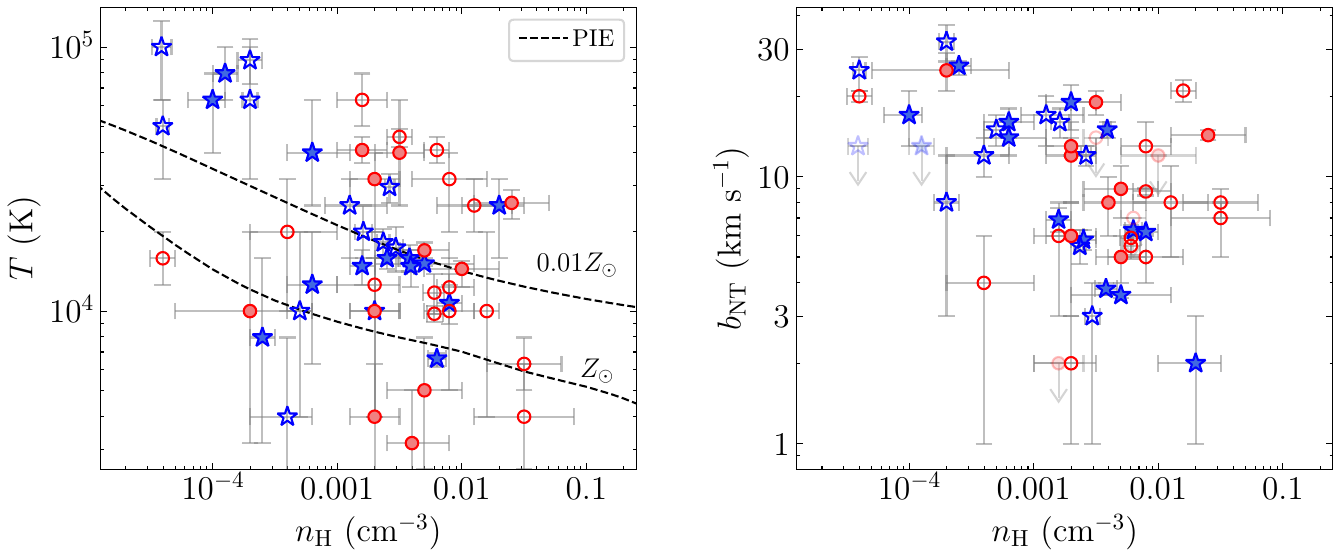}
  \caption{The density-temperature distribution of absorption components in CUBS-midz galaxies ($z = 0.4$--$0.7$) and COS-LRGs \citep[$z=0.2$--$0.6$,][]{Chen:2019,Zahedy:2019} are presented in the \textit{left} panel. The panel includes 25 components identified near nine star-forming galaxies (blue stars) and 26 components near 12 quiescent galaxies (red circles). Clumps around galaxy hosts with known neighbors are indicated with open markers. The expected relation from photoionization equilibrium (PIE) adopting the FG20 UVB at $z=0.55$ is indicated as a dashed line for different gas metallicities. Constraints on the non-thermal broadening in cool clumps, derived jointly with the gas temperature from absorption line widths, are shown in the \textit{right} panel. For components without a constraint on the non-thermal broadening, corresponding 99.7\% (3\,$\sigma$) upper limits are shown.} \label{fig:phase}
\end{figure*}

\startlongtable
\begin{deluxetable*}{rrrlrrrrrr}
\tablecaption{Ionization modeling results for COS-LRG absorbers\label{tab:ion_model_cos_lrg}}
\tablewidth{\textwidth}
\tabletypesize{\footnotesize}
\tablehead{
\multicolumn{1}{r}{} & \multicolumn{1}{c}{$d_\mathrm{proj}$} & \multicolumn{1}{c}{$v_c-v_\mathrm{gal}$} & \multicolumn{1}{l}{} & \multicolumn{1}{r}{} & \multicolumn{1}{r}{} & \multicolumn{1}{c}{$b_\mathrm{NT}$} & \multicolumn{1}{r}{} & \multicolumn{1}{r}{} & \multicolumn{1}{r}{} \\
\multicolumn{1}{r}{$\log(M_\mathrm{star}/M_\odot)$} & \multicolumn{1}{c}{$(\mathrm{kpc})$} & \multicolumn{1}{c}{$(\mathrm{km \ s^{-1}})$} & \multicolumn{1}{l}{Comp.} & \multicolumn{1}{r}{$\log(N_\mathrm{HI}/\mathrm{cm}^{-2})$} & \multicolumn{1}{r}{$\log(T/\mathrm{K})$} & \multicolumn{1}{c}{$(\mathrm{km \ s^{-1}})$} & \multicolumn{1}{r}{$\log(P_\mathrm{NT}/P_\mathrm{T})$} & \multicolumn{1}{r}{$\log(N_\mathrm{H}/\mathrm{cm}^{-2})$} & \multicolumn{1}{r}{$\log(n_\mathrm{H}/\mathrm{cm}^{-3})$}
}
\startdata
11.2 & 42 & $-104$ & s1c3 & $16.0_{-0.5}^{+0.3}$ & $3.5_{-0.3}^{+0.2}$ & $8\pm2$ & $-0.1_{-0.3}^{+0.4}$ & $17.9\pm0.2$ & $-2.4_{-0.2}^{+0.3}$ \\
 &  & $-65$ & s1c4 & $15.9\pm0.1$ & $4.5_{-0.2}^{+0.1}$ & $12\pm4$ & $-0.8\pm0.3$ & $18.1_{-0.2}^{+0.3}$ & $-2.7\pm0.2$ \\
11.1\tablenotemark{a} & 47 & $-7$ & s2c3 & $17.0_{-0.4}^{+0.5}$ & $4.4\pm0.1$ & $8\pm3$ & $-1.0\pm0.3$ & $18.5\pm0.2$ & $-1.9\pm0.3$ \\
 &  & $+15$ & s2c4 & $18.0_{-0.1}^{+0.2}$ & $3.6\pm0.2$ & $7_{-2}^{+1}$ & $-0.3_{-0.3}^{+0.2}$ & $18.9\pm0.3$ & $-1.5\pm0.4$ \\
11.4\tablenotemark{a} & 77 & $+97$ & s3c2 & $16.32_{-0.08}^{+0.06}$ & $4.0_{-0.4}^{+0.3}$ & $21\pm2$ & $+0.2_{-0.3}^{+0.4}$ & $17.8\pm0.1$ & $-1.8\pm0.1$ \\
 &  & $+127$ & s3c3 & $17.80\pm0.01$ & $3.8\pm0.1$ & $8\pm1$ & $-0.4\pm0.1$ & $18.8_{-0.2}^{+0.3}$ & $-1.5\pm0.3$ \\
11.0 & 94 & $-156$ & s6c1 & $15.4\pm0.1$ & $4.0_{-0.5}^{+0.3}$ & $25\pm4$ & $+0.4_{-0.3}^{+0.5}$ & $18.7\pm0.6$ & $-3.7_{-0.6}^{+0.5}$ \\
 &  & $-91$ & s6c2 & $16.9_{-0.7}^{+0.4}$ & $4.0_{-0.4}^{+0.3}$ & $13\pm1$ & $-0.2_{-0.3}^{+0.4}$ & $19.0\pm0.2$ & $-2.7\pm0.2$ \\
 &  & $-59$ & s6c3 & $18.5\pm0.1$ & $4.41_{-0.06}^{+0.05}$ & $14.3\pm0.6$ & $-0.54_{-0.06}^{+0.07}$ & $19.0_{-0.1}^{+0.2}$ & $-1.6\pm0.3$ \\
11.2\tablenotemark{a} & 99 & $+99$ & s7c1 & $15.42_{-0.06}^{+0.07}$ & $4.61\pm0.05$ & $<7$ & $<-1.4$ & $17.1_{-0.2}^{+0.1}$ & $-2.2\pm0.1$ \\
 &  & $+171$ & s7c2 & $15.6\pm0.1$ & $4.0_{-0.3}^{+0.2}$ & $13_{-4}^{+3}$ & $-0.2_{-0.3}^{+0.4}$ & $17.4\pm0.2$ & $-2.1\pm0.2$ \\
11.2 & 107 & $-25$ & s9c1 & $15.5_{-0.2}^{+0.1}$ & $4.61\pm0.05$ & $<2$ & $<-2.5$ & $17.7\pm0.2$ & $-2.8\pm0.2$ \\
 &  & $+10$ & s9c2 & $16.12_{-0.09}^{+0.08}$ & $4.6\pm0.2$ & $19\pm2$ & $-0.5\pm0.2$ & $18.2\pm0.2$ & $-2.5\pm0.2$ \\
 &  & $+59$ & s9c3 & $16.37_{-0.08}^{+0.07}$ & $3.7_{-0.3}^{+0.2}$ & $9\pm2$ & $-0.2_{-0.3}^{+0.4}$ & $18.3\pm0.3$ & $-2.3\pm0.3$ \\
11.4 & 109 & $+30$ & s10c2 & $16.49_{-0.05}^{+0.06}$ & $4.23\pm0.03$ & $5\pm1$ & $-1.3\pm0.2$ & $18.5\pm0.2$ & $-2.3\pm0.2$ \\
 &  & $+259$ & s10c4 & $17.09_{-0.02}^{+0.01}$ & $4.16_{-0.07}^{+0.03}$ & $<12$ & $<-0.4$ & $18.8\pm0.3$ & $-2.0\pm0.3$ \\
 &  & $+278$ & s10c5 & $15.6_{-0.3}^{+0.2}$ & $3.6_{-0.3}^{+0.2}$ & $6_{-3}^{+2}$ & $-0.5_{-0.5}^{+0.4}$ & $18.1\pm0.2$ & $-2.7\pm0.2$ \\
11.2\tablenotemark{a} & 140 & $-171$ & s12c1 & $16.6_{-0.2}^{+0.4}$ & $4.07_{-0.08}^{+0.07}$ & $5.9_{-0.8}^{+0.7}$ & $-1.0\pm0.1$ & $18.24\pm0.09$ & $-2.22_{-0.09}^{+0.08}$ \\
11.0\tablenotemark{a} & 156 & $+359$ & s14c1 & $15.0_{-0.2}^{+0.1}$ & $4.8\pm0.1$ & $6_{-3}^{+2}$ & $-1.7_{-0.5}^{+0.3}$ & $17.4\pm0.2$ & $-2.8\pm0.2$ \\
 &  & $+371$ & s14c2 & $15.57\pm0.07$ & $4.1\pm0.1$ & $2\pm1$ & $-1.9\pm0.5$ & $18.0_{-0.2}^{+0.3}$ & $-2.7_{-0.3}^{+0.2}$ \\
\enddata
\tablenotetext{a}{Satellite galaxies identified within the LRG halo (see \S\,\ref{sec:galaxies})}
\end{deluxetable*}

\subsection{Thermodynamic properties of the clumpy CGM} \label{sec:gas_props}

The photoionization modeling outlined in \S\,\ref{sec:methods} results in constraints on the clump density $n_\mathrm{H}$, while the line width analysis of individual resolved components leads to constraints on the gas temperature $T$ and non-thermal motions independent of the ionization models. The density-temperature distribution for CGM components in the joint galaxy sample described in \S\,\ref{sec:galaxies} is shown in the left panel of Figure \ref{fig:phase}. Components are grouped according to whether their galaxy host is star-forming or quiescent based on the optical spectrum as described in \S\,\ref{sec:gal_survey}. As a reference, the expected relation between the gas density and temperature assuming photoionization equilibrium (PIE) evaluated using \texttt{CLOUDY} is also shown. The PIE temperature is defined by the balance between UVB heating and radiative cooling in the gas. Due to the enhancement of cooling efficiency with metal content in the gas \citep[see e.g.,][]{Gnat:2007,Oppenheimer:2013a}, metal-rich clumps are expected to be cooler at a fixed gas density. Curves for the PIE density-temperature relation are shown in Figure \ref{fig:phase} for gas metallicities of $0.01Z_\odot$ and $Z_\odot$, which span the metallicity range for COS-LRG and CUBS-midz absorbers \citep[][S. Kumar et al. 2026b, to be submitted]{Zahedy:2019}. The measured gas temperatures for most components are broadly consistent with the PIE relation within the measurement uncertainties, although the observed scatter at a fixed density is larger compared to the expectation given variations in clump metallicity. This may reflect fluctuations in the ionizing background radiation, both in terms of its amplitude and spectral slope. In particular, significant radiation flux from local sources in the EUV to soft X-ray regime can substantially reduce gas cooling rates by altering the ionization states of the dominant coolants, while also providing additional heating \citep[see, e.g.,][]{Cantalupo:2010,Werk:2016,Qu:2023,Cadiou:2025}. These effects could explain why some systems exhibit temperatures higher than expected under PIE with the UVB alone, suggesting that local radiation sources may need to be included to fully model their thermal and ionization states. A detailed treatment of these effects is deferred to future studies.

\begin{figure*}              
    \centering \includegraphics[width=\textwidth]{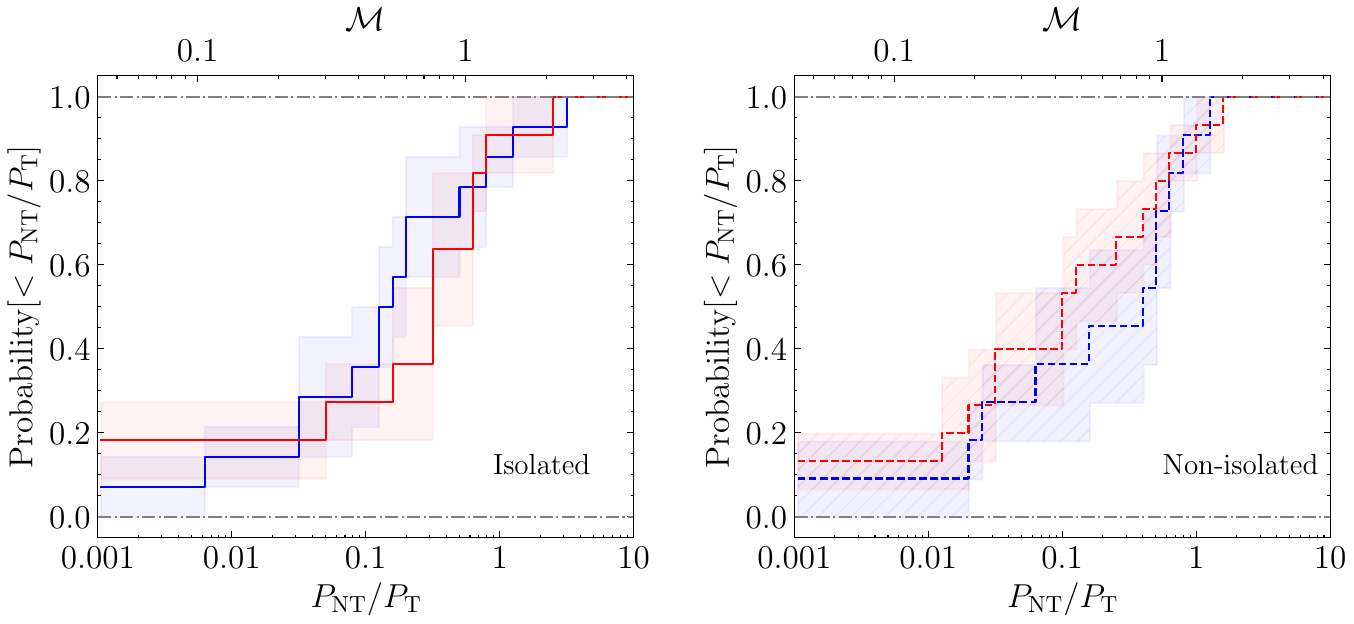}
  \caption{Cumulative distribution functions (CDFs) for the non-thermal to thermal pressure ratio in cool clumps identified near star-forming and quiescent galaxies are shown in these panels. Measurements in the \textit{left} panel correspond to  14 absorption components in five isolated star-forming galaxies (blue) and 11 components in four isolated quiescent galaxies (red). CDFs in the \textit{right} panel are constructed with 11 clumps in the vicinity of four non-isolated star-forming galaxies and 15 clumps associated with eight non-isolated quiescent galaxies. The 1\,$\sigma$ uncertainties shown in shaded bands are evaluated using a bootstrap method. The corresponding Mach numbers for a given pressure ratio following Equation \ref{eqn:logP_ratio} are indicated on the top axis.  
  } \label{fig:logP_ratio}
\end{figure*}

The observed line broadening in metal ions cannot be explained with thermal motions alone, indicating the presence of non-thermal motions within the gas. Constraints for the non-thermal motions, parametrized by $b_\mathrm{NT}$, for CGM components is shown in the right panel of Figure \ref{fig:phase}, with the observed range between $2$--30 \kms. An interesting feature of Figure \ref{fig:phase} that hints towards the turbulent nature of the gas is the apparent lack of points with low $b_\mathrm{NT}$ at low density. While two stringent 3\,$\sigma$ upper limits at $b_\mathrm{NT} \lesssim 10 \ \kms$ can be noted at $n_\mathrm{H} \approx 10^{-4} \ \mathrm{cm^{-3}}$, the remaining six clumps in this low-density regime exhibit $b_\mathrm{NT} \gtrsim 20 \ \kms$. Typical cloud size for low-density gas traced by high-ions is $l \gtrsim 1 \ \mathrm{kpc}$ within the CUBS-midz sample \citep[see e.g.,][]{Kumar:2024}, compared to the characteristic $l \sim 100 \ \mathrm{pc}$ for the denser clumps \citep[see e.g.,][]{Zahedy:2021}. Therefore, the observed $n_\mathrm{H}-b_\mathrm{NT}$ distribution may reflect a velocity-size relation due to turbulent cascade \citep[see e.g.,][]{Chen:2023,Qu:2026}, although contributions from bulk velocity gradients may be important for larger clouds \citep[see e.g.,][]{Rauch:1997}.

To investigate the role of turbulent pressure support (and equivalently heating) in the CGM, contributions from thermal and non-thermal motions can be evaluated within individual clumps as follows --
\begin{equation} \label{eqn:P_T_NT}
    P_\mathrm{T} = \left(\frac{\rho}{\mu m_\mathrm{H}}\right) k_\mathrm{B} T, \ P_\mathrm{NT} = \rho \sigma_\mathrm{NT}^2,
\end{equation}
where $\rho = 1.34 m_\mathrm{H} n_\mathrm{H}$, $\mu = 0.6$ for an ionized plasma, and $\sigma_\mathrm{NT} = b_\mathrm{NT}/\sqrt{2}$ is the velocity dispersion due to turbulent motions along the line-of-sight. The ratio of the non-thermal to thermal pressure can be expressed using the Mach number $\mathcal{M} \equiv \sqrt{3}\,\sigma_\mathrm{NT}/c_s$ as follows --
\begin{equation} \label{eqn:logP_ratio}
    P_\mathrm{NT}/P_\mathrm{T} = \frac{m_\mathrm{H} \sigma_\mathrm{NT}^2}{k_\mathrm{B} T / \mu} = \frac{\gamma \mathcal{M}^2}{3}
\end{equation} 
where $c_s = \sqrt{\gamma k_\mathrm{B}T/\mu m_\mathrm{H}}$ is the sound speed for a monatomic gas with an adiabatic index of $\gamma = 5/3$. 

The relative contributions of thermal and non-thermal pressure are now compared. To account for upper limits in $P_{\rm NT}$, the method described in \cite{Chen:2009} is adopted to construct the cumulative fraction of cool clumps with $P_{\rm NT}/P_{\rm T}$ not to exceed the designated value, which is akin to the Kaplan-Meier estimator described in \citet{Feigelson:1985}. Separate CDFs are presented for isolated and non-isolated galaxies to investigate the role of galaxy environment on the thermodynamic state of the CGM. To gauge the underlying uncertainty in the CDFs, a bootstrap method is used, resampling each distribution with replacement. The resulting 1\,$\sigma$ uncertainties on the CDF are included in Figure \ref{fig:logP_ratio}. A key conclusion from Figure \ref{fig:logP_ratio} is that most clumps exhibit subsonic turbulence across different galaxy environments, with $\approx20$--$35$\% of the clumps corresponding to Mach numbers $\mathcal{M}>1$. This result is consistent with similar analyses conducted using the thermal-to-total energy ratio in kinematically resolved absorption components at $z \approx 1$ by \cite{Qu:2022} and $z \approx 2-3$ by \cite{Rudie:2019}. Isolated quiescent halos also appear to host more turbulent clumps than corresponding star-forming systems. This may be due to the presence of a hot halo in massive ellipticals  \citep[][see also \S\,\ref{sec:pressure}]{Qu:2022}. Nevertheless, the median pressure ratio in both isolated star-forming (corresponding to $\mathcal{M}_\mathrm{iso,SF} = 0.5 \pm0.1$) and quiescent systems ($\mathcal{M}_\mathrm{iso,Q} = 0.8 \pm0.3$) is consistent with sub-sonic motions within the clumps and no significant distinction can be made between different types of halos with the current sample size. 

\begin{figure*}              
\includegraphics[width=\textwidth]{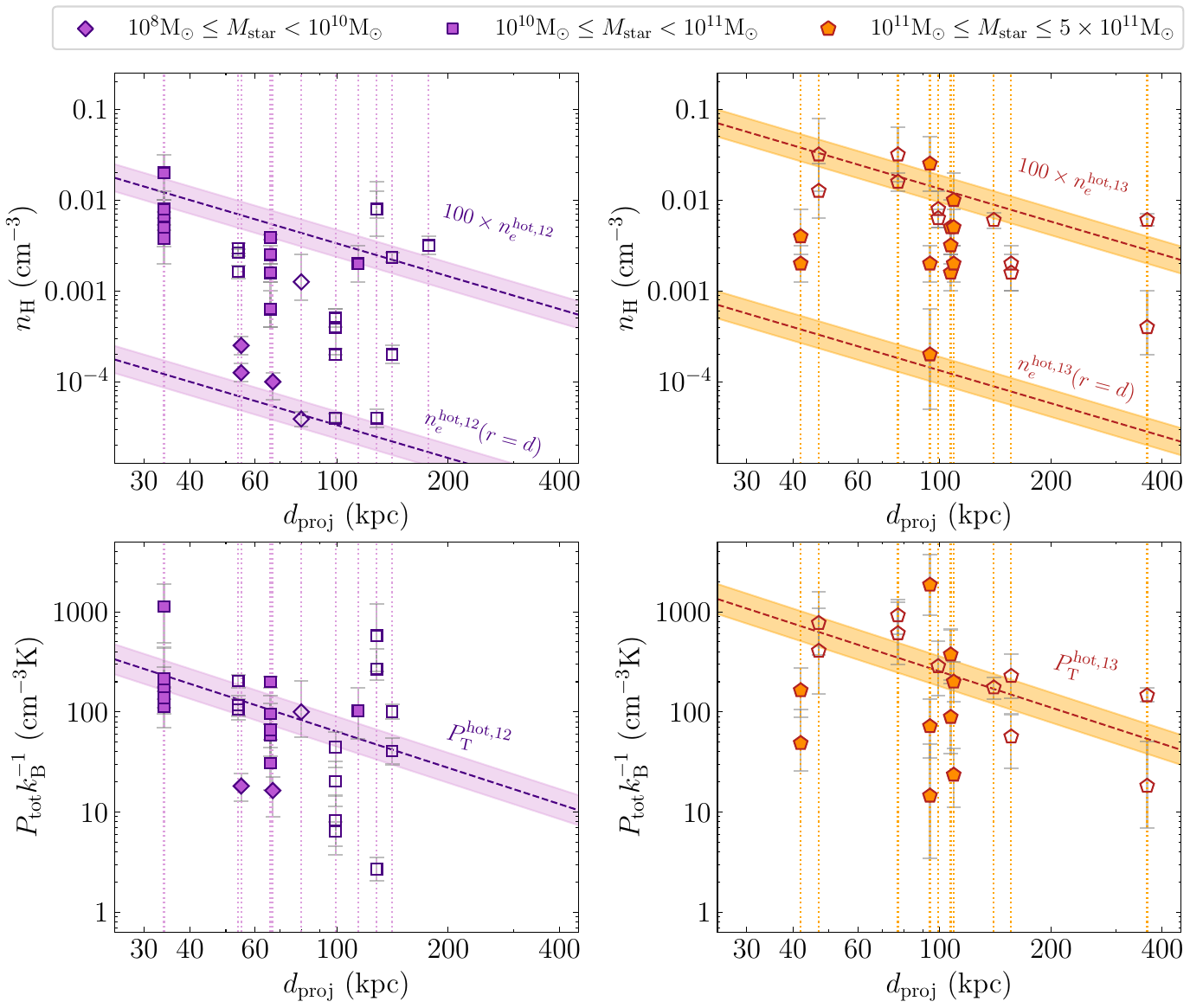}
  \caption{The cool gas density profiles around CUBS-midz galaxies and COS-LRGs are shown in the \textit{top} panels. Different symbols indicate distinct stellar mass ranges for the galaxy host. Multiple cool clumps resolved in individual halos are connected by a vertical dotted line. Non-isolated systems are indicated using open markers. The hot gas density profile from \cite{Singh:2018} for typical halo masses in both samples is also shown. A scaled-up density profile from the hot halo, $100 \times n_e^\mathrm{hot}$, is also displayed to examine pressure equilibrium between cool clumps with $T_\mathrm{cool} \approx 10^4 \ \mathrm{K}$ and the ambient hot medium at $T_\mathrm{hot} \approx 10^6 \ \mathrm{K}$. The total pressure within these cool clumps, adding the thermal and non-thermal contributions following Equation \ref{eqn:P_T_NT}, is shown in the \textit{bottom} panels. The indicated hot gas pressure profiles adopt the density profiles shown in the top panels with a fiducial gas temperature of $T_\mathrm{hot} = 10^6 \ \mathrm{K}$.} \label{fig:profiles}
\end{figure*}

\subsection{Cool gas density and pressure profiles} \label{sec:pressure}

Spatial variations in cool gas density derived in \S\,\ref{sec:ion_model} and total gas pressure derived in \S\,\ref{sec:gas_props} are now investigated. The cool gas density profiles for the combined galaxy sample are presented in the top panels of Figure \ref{fig:profiles}. For non-isolated systems, the projected distance depicted is for the galaxy host at minimum $d_\mathrm{proj}/R_\mathrm{vir}$ from the sightline. Recall that CUBS-midz galaxies typically have dark matter halo masses of $M_h \lesssim 10^{12} \ \msun$, while COS-LRGs are more massive. Note that low-density phases in multiphase absorbers (components with suffix ``l'' in Table \ref{tab:ion_model}) are omitted here. Large density variations of up to $2 \ \mathrm{dex}$ are observed within individual halos. Despite the large scatter, there is a striking dependence of the cool gas density on the halo mass at fix projected distances. This amplitude is distinctly higher for COS-LRGs than for CUBS-midz galaxies. Even within the CUBS-midz sample, massive star-forming galaxies (squares) exhibit denser clumps than lower-mass systems (diamonds). This trend is consistent with the expectations that these cool CGM clouds are pressure confined in a hot ambient medium which reduces in density and pressure support with decreasing halo mass \citep[see e.g.,][]{Zahedy:2019,Qu:2023}.

To further investigate this scenario, density profiles for hot gas are included in Figure \ref{fig:profiles}.
Key observational probes of the hot CGM are the X-ray luminosity and thermal Sunyaev-Zel'dovich (tSZ) signal. Direct measurements for these probes have only been made in a small number of external massive galaxies, although stacking efforts can provide insights for lower-mass halos. \cite{Singh:2018} jointly model stacked X-ray and tSZ observations from galaxies at $z \approx 0.1$ with dark matter halo masses ranging between $M_h \approx 10^{12}$--$10^{13}\msun$. Adopting a power law model for the electron density profile $n_e^\mathrm{hot}(r) \propto r^{-3\beta}$, where $r$ is the galactocentric radius and $\beta=0.4$, the density amplitude then scales with the X-ray luminosity. The corresponding density profile for both halo mass regimes are depicted in Figure \ref{fig:profiles}. Note that the hot gas profile is evaluated at $r=d_\mathrm{proj}$. There is a temperature contrast between the absorption components at a typical temperature of $T_\mathrm{cool} \sim 10^4 \ \mathrm{K}$ (Figure \ref{fig:phase}) and the hot medium, expected to be at $T_\mathrm{hot} \sim 10^6 \ \mathrm{K}$ \citep[see e.g.,][]{Maller:2004}. To evalute pressure balance between the cool and hot phases, an elevated hot halo profile corresponding to $100 \times n_e^\mathrm{hot}$ is also presented in the top panels of Figure \ref{fig:profiles}. There is reasonable agreement between the elevated hot halo profile and the densest clumps along massive CUBS-midz galaxies (square symbols) and COS-LRGs. This would suggest that the densest clumps along the sightline are situated at a physical distance of $r \approx d_\mathrm{proj}$ and are in pressure equilibrium with the hot halo \citep[see e.g.,][]{Zahedy:2019,Qu:2023}. Note that cool clumps in lower-mass halos (diamonds in Figure \ref{fig:profiles}) exhibit systematically lower densities compared to more massive systems \citep[see e.g.,][]{Piacitelli:2025,Polzin:2026}. The inferred density of $n_\mathrm{H} \sim 10^{-4} \ \mathrm{cm}^{-3}$ for such clumps is due to the presence of intermediate ions like \ion{O}{III} and \ion{O}{IV} \citep[see e.g.,][N. Mishra et al. 2026, to be submitted]{Johnson:2026}. This may reflect reduced pressure support in the hot CGM of low-mass galaxies, although the hot phase is not expected to be stable in such halos \citep[see e.g.,][]{Stern:2020,Kakoly:2025} and observational constraints are challing to obtain due to low X-ray emissivity \citep[see e.g.,][]{Ott:2005,He:2025}.

If the densest clumps along individual sightlines are indeed situated at $r \approx d_\mathrm{proj}$, more diffuse clumps will have to be located at $r>d_\mathrm{proj}$ to maintain pressure equilibrium with the hot halo. Using the hot gas density profile as a guide, clumps with a density of $n_\mathrm{H} \lesssim 0.001 \ \mathrm{cm}^{-3}$ should extend out to $r \gtrsim 300 \ \mathrm{kpc}$, likely beyond the halo radius. It is possible that the diffuse clumps may be tracing lower-mass satellites, akin to the measurements for low-mass halos depicted in Figure \ref{fig:profiles}. However, the presence of low-density components is noted in isolated and non-isolated systems alike. To enable a more precise comparison between cool and hot gas pressures, the total pressure constraints derived in \S\,\ref{sec:pressure} for individual cool clumps are presented in the bottom panels of Figure \ref{fig:profiles}. Clumps without a constraint on the non-thermal broadening $b_\mathrm{NT}$ are excluded from this exercise. The corresponding hot gas pressure $P_\mathrm{T}^\mathrm{hot} \equiv \rho_\mathrm{hot} k_\mathrm{B}T_\mathrm{hot}/\mu$, with $\rho_\mathrm{hot} = 1.34 m_\mathrm{H} n_\mathrm{H}^\mathrm{hot}$, $\mu = 0.6$, and a fiducial hot gas temperature of $T_\mathrm{hot} = 10^6 \ \mathrm{K}$ is also presented in Figure \ref{fig:profiles}.

Reasonable agreement is found between cool clump pressure and the expected hot gas profile for both CUBS-midz and LRG samples, supporting the need of accounting for warmer gas temperatures and turbulence when considering pressure support in diffuse components (see Figure \ref{fig:phase}). Nevertheless, a large scatter around the expected hot gas pressure profile is observed. It is worth acknowledging that the adopted hot gas temperatures bear significant uncertainties.  Direct measurements of hot gas temperature are restricted to nearby galaxies, typically within the inner $20\,\mathrm{kpc}$ \citep[see e.g.,][]{Hodges-Kluck:2018}, with limited constraints for the outer halo. In the Milky Way, \ion{O}{VII} and \ion{O}{VIII} emission profiles also suggest the presence of temperature gradients \citep[see e.g.,][]{Miller:2015}. Finally, contaminating signals from satellite galaxies and unresolved sources may further skew the hot gas density and temperature measurements \citep[see e.g.,][]{Zhang:2024, He:2026}. Accounting for these systematic effects in the hot gas density may be able to accommodate higher temperatures for the hot gas. 

Despite the detailed accounting of the pressure budget in individual absorption components, there are still clumps that exhibit a lower pressure support than the hot halo expectation at $r=d_\mathrm{proj}$ according to the pressure profiles in Figure \ref{fig:profiles}. Escaping ionizing flux from local stellar populations can be significant out to $r\lesssim 50 \ \mathrm{kpc}$ in star-forming galaxies \citep[see e.g.,][]{Chen:2017b,Bland-Hawthorn:2019}, causing UVB-only models to underestimate gas pressure (\S\,\ref{sec:ion_model}). However, low-density clumps with $n_\mathrm{H} < 0.001 \ \mathrm{cm}^{-3}$ in the CUBS-midz sample primarily arise at $d_\mathrm{proj}>50 \ \mathrm{kpc}$, and in fact exhibit a profile steeper than $d_\mathrm{proj}^{-2}$ expected for ionizing flux from the galaxy. Therefore, it is unlikely that local radiation can account for the observed pressure deficit across the sample, although its effects cannot be ruled out for any particular clump at small impact parameter. Al9ternatively, underpressured clumps at smaller $d_\mathrm{proj}$ may be situated at $r>d_\mathrm{proj}$ \citep[see e.g.,][]{Yang:2025}. Additional sources of non-thermal pressure like magnetic fields \citep[see e.g.,][]{vandeVoort:2021,Ramesh:2024} and cosmic rays (CRs) can also be important. The role of CR electrons has been invoked to explain the observed X-ray surface brightness profiles of Milky-Way like galaxies \citep[see e.g.,][]{Hopkins:2025}. Cosmic ray pressure can also promote isochoric thermal instabilities in the cool gas, leading to the prevalence of low-density clumps \citep[see e.g.,][]{Butsky:2020}. However, theoretical predictions regarding the effect of CRs on the CGM are sensitive to resolution effects \citep[see e.g.,][]{Weber:2025} and assumptions about the transport mechanism \citep[see e.g.,][]{Quataert:2022b,Quataert:2022a}.

\subsection{Baryon census in galaxy halos} \label{sec:baryon}

Independent of uncertainties in detailed 3D spatial profiles of the cool CGM, a robust measurement of the surface gas mass density can be made \citep[e.g.,][]{Werk:2014, Zahedy:2019}. Specifically, the best-fit photoionization models provide the necessary ionization fraction for different species, including \ion{H}{I}, which in turn enable estimates of the total hydrogen column density, $N_{\rm H, c}$, in individual clumps (see Tables \ref{tab:ion_model} and \ref{tab:ion_model_cos_lrg}).  The integrated total hydrogen column density $N_\mathrm{H}$ encountered along the sightline can be evaluated by adding contributions from all components \citep[see also][]{Zahedy:2019}. 

\begin{figure}              
    \centering \includegraphics[width=\columnwidth]{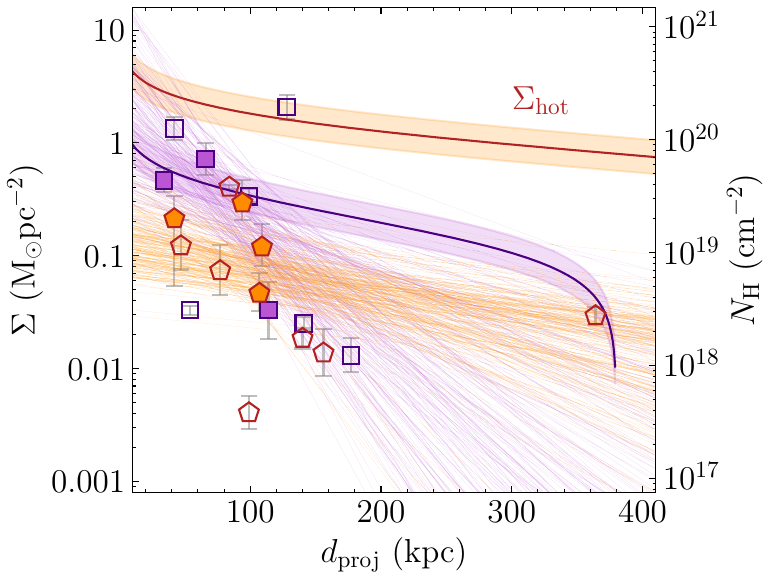}
  \caption{Surface mass density constraints for cool gas in $M_h \approx 10^{12} \msun$ halos (squares) and $M_h \approx 10^{13} \msun$ (pentagons) are shown in this panel, following symbols from Figure \ref{fig:profiles}. The corresponding total hydrogen column densities are indicated on the right. Randomly drawn exponential profile fits for $\Sigma_\mathrm{cool}$ resulting from  a likelihood analysis are shown for both galaxy samples using indigo and orange lines. The corresponding hot gas surface density profiles are presented as shaded bands.} \label{fig:surface}
\end{figure}

These constraints are presented for $M_h \approx 10^{12} \msun$ and $10^{13}\msun$ in Figure \ref{fig:surface}, following the symbol legend from Figure \ref{fig:profiles}. The hydrogen column density is converted to cool gas surface mass density following the relation $\Sigma_\mathrm{cool} = 1.34\,m_\mathrm{H}\,N_\mathrm{H}$ with a mass correction for the helium content in the gas. The steep decline observed in the cool gas surface density profiles are not well matched by a simple-power law, so exponential profiles of the form $\Sigma_\mathrm{cool} = \Sigma_0 \exp(-d_\mathrm{proj}/d_s)$ are fit for both samples using a likelihood analysis incorporating intrinsic scatter in the data \citep[c.f.,][]{Zahedy:2019}. Random draws from the posterior distribution for individual models are also depicted in Figure \ref{fig:surface}. The best-fit profile for a $M_h \approx 10^{12}\,\msun$ halo is characterized by $\Sigma_0 \approx 1.5 \,\msun\,\mathrm{pc}^{-2}$ and $d_s \approx 50 \,\mathrm{kpc}$, while for a $M_h \approx 10^{13}\,\msun$ halo the best-fit parameters are $\Sigma_0 \approx 0.1 \,\msun \,\mathrm{pc}^{-2}$ and $d_s \approx 200 \,\mathrm{kpc}$. The significantly flatter profile for LRG halos is largely driven by the detection for system s22 in the CUBS-midz sample at a projected distance of $d_\mathrm{proj} = 364 \,\mathrm{kpc}$. Excluding s22 from this exercise leads to a steeper profile with a $\Sigma_0 \approx 1 \,\msun \,\mathrm{pc}^{-2}, d_s \approx 40 \,\mathrm{kpc}$, more consistent with the profile of star-forming halos, while the enclosed gas mass within the halo radius remains the same. 

\begin{figure*}              
    \centering \includegraphics[width=\textwidth]{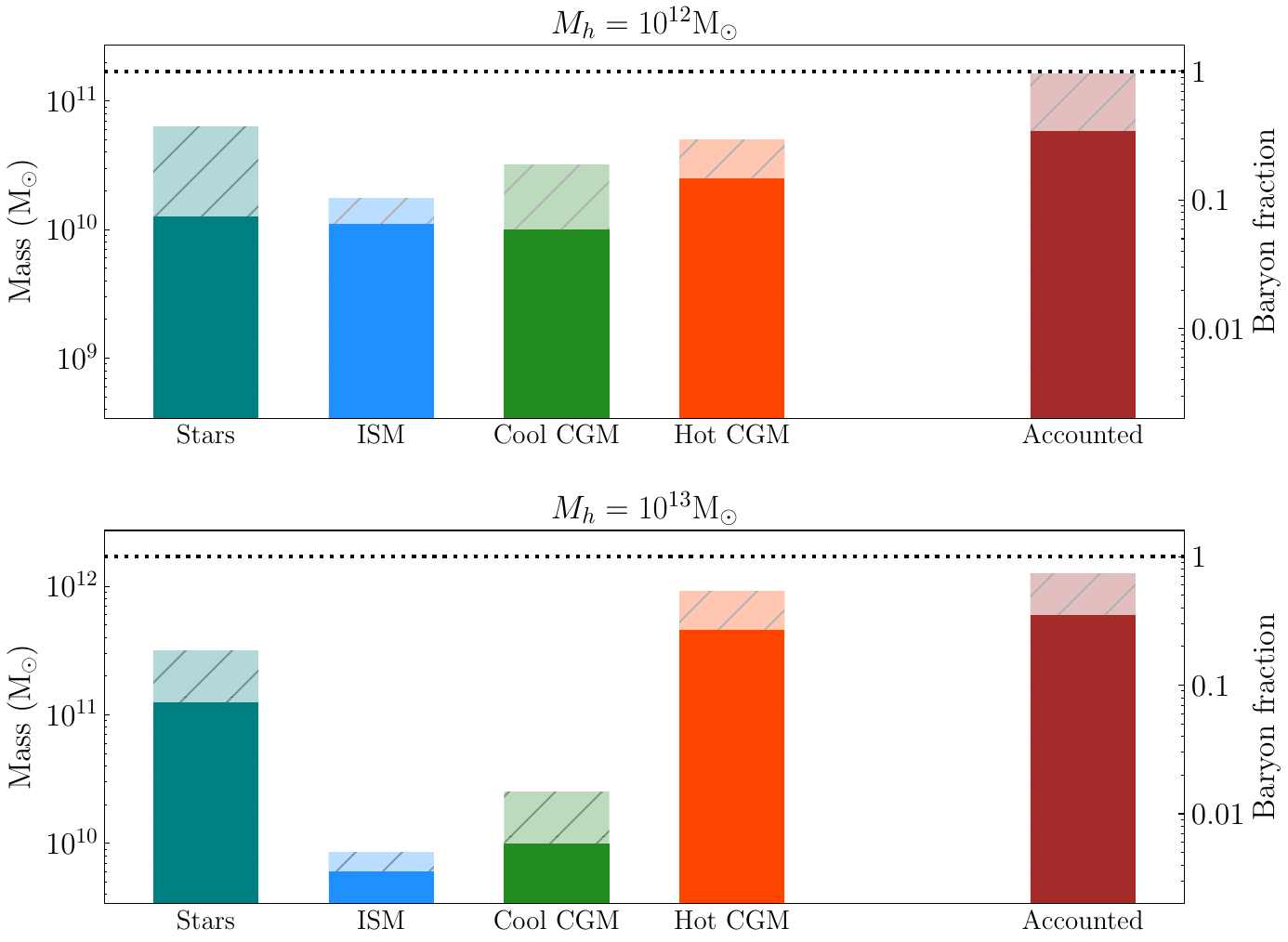}
  \caption{Census of baryonic mass in different reservoirs within $M_h \approx 10^{12} \msun$ halos (\textit{top} panel) and $10^{13} \msun$ halos (\textit{bottom} panel). The height of the solid bar in each category represents the 16\% confidence interval, while the hatched region extend to the 84\% interval. The expected baryonic mass in the halo is depicted using a dotted line \citep[e.g.,][]{Planck:2020}. The stellar mass range captures the observed variation in CUBS-midz galaxies and COS-LRGs. The ISM mass is evaluated using \ion{H}{I} surveys from \cite{DePalma:2025,Bianchetti:2025b}. The cool CGM mass is evaluated using the surface mass density profiles in Figure \ref{fig:surface}, integrated out to the virial radius. The hot CGM mass is evaluated using hot gas profiles from \cite{Singh:2018}. Note that contributions from warm-hot gas are not accounted for here \citep[e.g.,][]{Qu:2023, Chen:2026}. Consequently, the combined mass in stars, the ISM, the cool CGM, and the hot halo represents a lower limit to the total baryon content of galaxy halos.}\label{fig:baryon_budget}
\end{figure*}

In contrast, the total hot gas mass differs substantially between massive LRG halos and lower-mass star-forming halos. This difference is evident in their respective surface mass density profiles, constructed from the three-dimensional density distributions of the hot gas \citep[see e.g.,][]{Singh:2018}. Following previous studies, the expected hot gas surface mass density is computed by integrating along the line of sight out to $2\times R_{\rm vir}$ \citep[see also][]{Zhang:2024}\footnote{Restricting the line-of-sight integration to within $R_{\rm vir}$ lowers the overall  $\Sigma_{\rm hot}(d_{\rm proj})$ but the amplitude remains consistent to within the uncertainty.}. The strong halo-mass dependence of the hot gas profile, contrasted with the comparable cool gas surface mass density profiles, therefore implies a clear decline in the fraction of baryons residing in the cool phase from $M_h \approx 10^{12},\msun$ to $10^{13},\msun$ \citep[see also][]{Sultan:2025,Davies:2026}.

The baryon census between stars, the ISM, the cool CGM, and the hot halo is evaluated more precisely for the CUBS-midz and COS-LRG samples. Specifically, the exponential profile model for the surface mass density profiles of the cool gas, $\Sigma$, can be used to examine the gas mass in the cool and hot phases in both samples. For $M_h \approx 10^{12} \msun$, the 1\,$\sigma$ range on the cool gas mass out to the virial radius ($\approx 190 \ \mathrm{kpc}$) is $M_\mathrm{cool,12} \approx$\,(1.0--3.2)\,$\times 10^{10} \msun$, accounting for a covering fraction of $\approx 75\%$. Note that this estimate includes system s21 at $d_\mathrm{proj}=177 \ \mathrm{kpc}$, which shows absorption at $v_c-v_\mathrm{gal} = -309 \ \kms$ (Figure \ref{fig:sample}).  A comparable cool gas mass range, $M_\mathrm{cool,12} \approx$\,(1.1--3.2)\,$\times 10^{10} \msun$, is achieved even if this system is excluded.  The 1\,$\sigma$ range on the hot gas mass out to the virial radius is $M_\mathrm{hot,12} \approx$\,(2.5--5.0)\,$\times 10^{10} \msun$. For more massive halos with $M_h \approx 10^{13} \msun$ (with $R_\mathrm{vir} \approx 440 \ \mathrm{kpc}$), the cool gas mass is $M_\mathrm{cool,13} \approx$\,(1.0--2.5)\,$\times 10^{10} \msun$, incorporating a covering fraction of $\approx 80\%$, while the 1\,$\sigma$ range on the hot gas mass out to the virial radius is $M_\mathrm{hot,13} \approx$\,(4.6--9.1)\,$\times 10^{11} \msun$, comparable to the hot gas mass observed out to virial radii of X-ray selected galaxy groups \citep[e.g.,][]{Khalil:2026}.

These estimates are presented in Figure \ref{fig:baryon_budget}, alongside corresponding baryonic mass present in stars and ISM. For the stellar mass, reported values for CUBS-midz and COS LRG hosts are used to determine the 84\% upper bound, 
corresponding to the hatched region in Figure \ref{fig:baryon_budget}. For the ISM of $M_h \approx 10^{12} \msun$ halos, estimates from the \ion{H}{I} survey of CUBS galaxies are used \citep[see][]{DePalma:2025}. Although an \ion{H}{I} survey is not available for COS-LRGs, results from \cite{Bianchetti:2025b} constraining the \ion{H}{I} mass in quiescent $M_h \approx 10^{13} \msun$ halos is used because their companion survey for lower-mass star-forming halos  \citep{Bianchetti:2025a} agrees with the \ion{H}{I} mass in CUBS-midz galaxies.

The cool CGM plays an important role in completing the baryon census, with the anticipated total baryonic mass $M_b=0.17\,M_h$ from the cosmic mean \citep[e.g.,][]{Planck:2020}. The near constant cool gas mass observed across a broad mass range directly translates to a declining baryon mass fraction from $X_{\rm cool}\equiv M_{\rm cool}/M_{\rm b}\approx 6$--19\% in halos of $M_h \approx 10^{12}\, \msun$ to $X_{\rm cool}\approx 0.6$--1.5\% in halos of $M_h \approx 10^{13}\, \msun$. Note that due to the multiphase nature of the CGM, warm halo gas is expected to be an additional baryonic reservoir in galaxies \citep[see e.g.,][]{Qu:2023, Qu:2024, Chen:2026}. Therefore, the sum of all four components in stars, ISM, cool CGM, and the hot halo shown in Figure \ref{fig:baryon_budget} constitutes a lower limit to the total baryon content in galaxy halos.

\subsection{Kinematically aligned low- and high-ionization gas} \label{sec:multiphase}

\begin{figure}              
    \centering \includegraphics[width=\columnwidth]{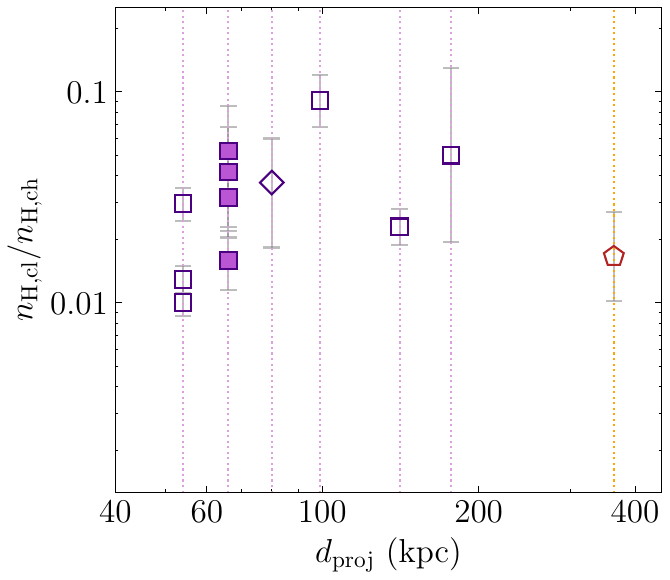}
  \caption{In this panel, the contrast in gas density $n_\mathrm{H}$ between the high-density phase (suffix h) and low-density phase (suffix l) in components exhibiting kinematically aligned low- and high-ions. The symbols represent different stellar mass ranges for the galaxy host following Figure \ref{fig:profiles}, and non-isolated systems are marked using open symbols.} \label{fig:multiphase}
\end{figure}

QSO absorption line studies frequently reveal the presence of highly ionized species (e.g., C$^{3+}$, O$^{5+}$) in gaseous halos. The physical origin of this phase is of great interest because it is expected to harbor a large fraction of baryons within galaxies \citep[see e.g.,][]{Chen:2026}. A collisionally ionized origin for \ion{O}{VI} absorbers has been proposed in previous studies \citep[see e.g.,][]{Tripp:2011,Meiring:2013,Hussain:2015,Pachat:2017,Nevalainen:2017,Rosenwasser:2018}. However, a significant population of high-ionization absorbers can also be decomposed into a complex of narrow components with $b \lesssim 20 \ \kms$ \citep[see e.g.,][]{Thom:2008,Savage:2014}. The observed high-ion column densities in such components can be reproduced with photoionization equilibrium models at densities $n_\mathrm{H} \lesssim 10^{-4} \ \mathrm{cm}^{-3}$ \citep[see e.g.,][]{Carswell:2002,Zahedy:2021,Kumar:2024}. However, it is also possible that narrow high-ionization components originate from rapidly cooling gas \citep[see e.g.,][]{Gnat:2007,Oppenheimer:2013a,Kumar:2025}. An empirical characterization of warm-hot absorbers around CUBS-midz galaxies can be found in \cite{Qu:2024}.

Meanwhile, photoionized gas of densities $n_\mathrm{H} \approx 0.001$--$0.01 \ \mathrm{cm}^{-3}$ accounting for absorption from low- and intermediate-ionization species
(\S\,\ref{sec:gas_props}) cannot reproduce the observed column densities of kinematically aligned high-ions \citep[see also][]{Zahedy:2021,Cooper:2021,Sameer:2021,Qu:2022}. In the CUBS-midz sample, 22 out of 31 components with available ionization models show kinematically aligned \ion{O}{VI} absorption. In eight of these components, the \ion{O}{VI} absorption is too broad ($b>30 \ \kms$) and the observed column densities cannot be produced using high-density photoionized gas. Such components may trace low-density photoionized gas or warmer collisionally ionized gas. In the remaining 14 components, the measured $b(\mbox{O VI})$ are narrower, and the corresponding $N(\mbox{O VI})$ can be reproduced using PIE models. Of these 14 components, 10 have a significant presence of low-ions, necessitating a two-phase model as shown in Figure \ref{fig:s10_PIE}. There are two additional non-\ion{O}{VI}-bearing components (s21c2 and s22c1) that require a two-phase solution. Note that most COS-LRG absorbers do not have the spectroscopic coverage for \ion{O}{IV}, which is required to constrain the low-density phase, but this does not rule out a photoionized origin for the observed \ion{O}{VI}.

Although some high-ionization components coincident in velocity space with the low-ionization gas, it is unclear whether the two gas phases are also co-spatial. It is possible that the two phases may only be superimposed in velocity space but reside at two distinct locations within the halo \citep[see e.g.,][]{Ticoras:2026}. Assuming pressure equilibrium with the hot halo (Figure \ref{fig:profiles}), the implied physical distance for the low-density phases ($n_\mathrm{H} \lesssim 10^{-4} \ \mathrm{cm}^{-3}$) would be $r \gtrsim 1 \ \mathrm{Mpc}$. However, QSO absorption surveys report covering fractions of $\lesssim10$\% for metal ions beyond the halo radius \citep[see e.g.,][]{Liang:2014,Johnson:2015a,Dutta:2025}. Additionally, in an effort to derive physical conditions of circumgalactic gas around CUBS galaxies at $z\approx1$, \cite{Qu:2023} report a steep drop in the incidence rate of components bearing both low- and high-ions from $\approx 80\%$ at $d_\mathrm{proj}<100 \ \mathrm{kpc}$ to $<15\%$ at $d_\mathrm{proj} = 100$--$200 \ \mathrm{kpc}$. Although projection effects cannot be ruled out, the above arguments, in addition to the presence of low- and high-ionization species in multiple components within individual absorption systems, favor such components tracing multiphase gas in the halo.

Figure \ref{fig:multiphase} depicts the density contrast between the low- and high-density phases in such components as a function of projected distance. The observed range of density fluctuations across systems is between one and two orders of magnitude. The density contrast has a direct consequence on the pressure balance between the two phases in such components. Although the thermal- and non-thermal motions in the low-density phase cannot be distinguished (see \S\,\ref{sec:ion_model}), the available 3\,$\sigma$ limits on the temperature and turbulent broadening can be used to assess pressure balance. Comparing with the total pressure for the high-density phase derived in \S\,\ref{sec:pressure} with limiting pressure for the low-density phase, pressure balance is ruled out for 8 out of 12 components in Figure \ref{fig:multiphase}. Therefore, the two phases are likely not in dynamical equilibrium if they are co-spatial \citep[see also][]{Haislmaier:2021}. 

The observed density fluctuations in CUBS-midz multiphase components \citep[see also,][]{Qu:2022, Sameer:2024} have been noted in simulations. In an analysis of CGM virialization, \cite{Stern:2021} predict that gas in the inner 50 kpc of simulated halos can show over two orders-of-magnitude in local density fluctuations \citep[see also][]{Kakoly:2025}. Disk settling is also achieved as a result of virialization in these simulations. The galaxy at $d_\mathrm{proj} \approx 66 \ \mathrm{kpc}$ in Figure \ref{fig:multiphase} with an extended rotating disk (system s10, $M_\mathrm{star} \approx 6 \times 10^{10} \msun$) is in good agreement with this scenario. Among the remaining six systems, satellite galaxies are identified for each absorber host within the galaxy survey. Satellite winds have also been proposed for generating density fluctuations in simulated gaseous halos \citep[see e.g.,][]{F-G:2016,Hafen:2019,Esmerian:2021}.

\section{Conclusions} \label{sec:conclusions}

In this paper, thermal and turbulent properties of circumgalactic absorbers are presented for 23 CUBS-midz galaxies at $z=0.4$--$0.7$ with background QSO sightlines intersecting within the halo radius. High-quality QSO spectra obtained with \textit{HST} COS and STIS enable accurate decomposition of absorption profiles across multiple ionic transitions. Ionizing conditions within individual clumps are derived using photoionization models, while thermal and non-thermal motions are separated using absorption-line width measurements. Deep galaxy surveys carried out in CUBS QSO fields using VLT MUSE alongside LDSS3C and IMACS instruments on the Magellan telescopes reveal the galaxy environment of absorber hosts. Combining constraints from the CUBS-midz sample with CGM measurements around intermediate redshift galaxies in the literature, the following key results arise --

\begin{enumerate}
    \item Cool clumps around star-forming and quiescent galaxies are consistent with a photoionization equilibrium (PIE) scenario, as evident from the agreement between measured gas temperatures and the PIE expectation at the inferred gas density (Figure \ref{fig:phase}). The presence of outliers, however, may point to variations in the intensity and spectral shape of the ionizing radiation field due to local sources.  

    \item The non-thermal to thermal pressure ratio in circumgalactic clumps is connected to the turbulent Mach number. The median pressure ratio suggests sub-sonic turbulence in cool clumps, with $\mathcal{M}\approx 0.5$ for star-forming galaxies and $\mathcal{M}\approx 0.8$ for quiescent galaxies (Figure \ref{fig:logP_ratio}).

    \item Cool gas densities exhibit a large variations of up to $2 \ \mathrm{dex}$ in individual halos, but the density profile distinctly depends on the halo mass $M_h$, with lower-mass halos exhibiting more diffuse gas (Figure \ref{fig:profiles}). The cool gas pressure profile (including thermal and turbulent contributions) around halos with $M_h \approx 10^{12} \msun$ and $10^{13} \msun$ is in approximate agreement with the anticipated hot gas profile from stacking analyses of X-ray and tSZ measurements of galaxy halos.

    \item A near constant total cool halo gas mass is found across a broad halo mass range, translating to a declining baryon mass fraction in the cool phase from $\approx 6$--19\% in halos of $M_h \approx 10^{12}\, \msun$ to only $\approx 0.6$--1.5\% in halos of $M_h \approx 10^{13}\, \msun$ (Figure \ref{fig:baryon_budget}). 
    
    \item Kinematically aligned absorption in low- and high-ionization species is identified in 12 out 31 CUBS-midz components, occurring in seven unique galaxies. Narrow line widths for high-ions (e.g., C$^{3+}$) are consistent with a photoionization scenario.
    The implied density contrast between the high- and low-density phases in such components range between one to two orders of magnitude, possibly driven by local pressure fluctuations within the halo (Figure \ref{fig:multiphase}).
\end{enumerate}

\begin{acknowledgments}
The authors thank an anonymous referee for a swift and constructive review process. The authors thank M.\ Voit for illuminating discussions regarding the contribution of the X-ray-detected hot phase to the halo baryon budget.  S.K.\ and H.W.C.\ acknowledge partial support from HST-GO-17517.001A and from NASA ADAP 80NSSC23K0479. G.C.R.\ acknowledges the support from Grant 63667 from the John Templeton Foundation. Any opinions, findings, and conclusions or recommendations expressed in this material are those of the authors and do not necessarily reflect the views of the John Templeton Foundation. S.C.\ gratefully acknowledges support from the European Research Council (ERC) under the European Union's Horizon 2020 Research and Innovation program grant agreement number 864361. M.C.C.\ is supported by the Brinson Foundation through the Brinson Prize Fellowship Program. D.D.\ acknowledges the support of the John A. Lyons Fellowship from MIT’s Office of Graduate Education. CAFG was supported by NSF through grants AST-2108230 and AST-2307327; by NASA through grants 80NSSC22k0809, 80NSSC22K1124 and 80NSSC24K1224; by STScI through grant JWST-AR-03252.001-A; and by BSF through grant \#2024262. S.L.\ acknowledges support by FONDECYT grant 1231187. This work is based on observations made with ESO Telescopes at the Paranal Observatory under program ID 0104.A-0147(A), observations made with the 6.5m Magellan Telescopes located at Las Campanas Observatory, and spectroscopic and imaging data gathered under the HST-GO-15163.01A, HST-GO-17517.001A, and HST-GO-17815-001A programs using the NASA/ESA Hubble Space Telescope operated by the Space Telescope Science Institute and the Association of Universities for Research in Astronomy, Inc., under NASA contract NAS 5-26555. This research has made use of the services of the ESO Science Archive Facility and the Astrophysics Data Service (ADS)\footnote{\url{https://ui.adsabs.harvard.edu}}. The analysis in this work was greatly facilitated by the following \texttt{python} packages:  \texttt{Numpy} \citep{Numpy:2020}, \texttt{Scipy} \citep{Scipy}, \texttt{Astropy} \citep{astropy:2013,astropy:2018}, \texttt{Matplotlib} \citep{Matplotlib}, and \texttt{emcee} \citep{Foreman-Mackey:2013aa}. 
\end{acknowledgments}

%
\facilities{\textit{HST} (COS/STIS), Magellan telescopes, VLT MUSE}

\software{ 
          Cloudy \citep{Ferland:2017}
          }


\bibliography{CUBSX}{}
\bibliographystyle{aasjournalv7}

\appendix \label{sec:appendix}

\counterwithin{figure}{section}
\counterwithin{table}{section}

\section{Notes on individual systems}

\begin{figure*}
    \centering
    \includegraphics[width=.8\textwidth]{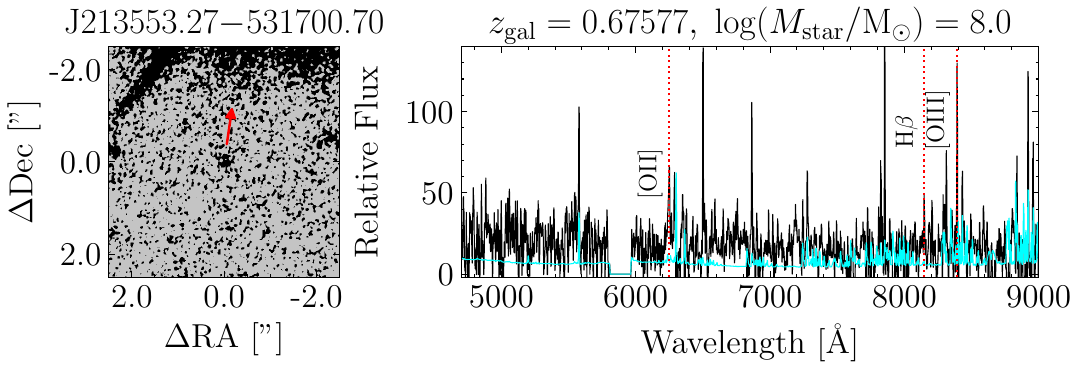}
    \includegraphics[width=.8\textwidth]{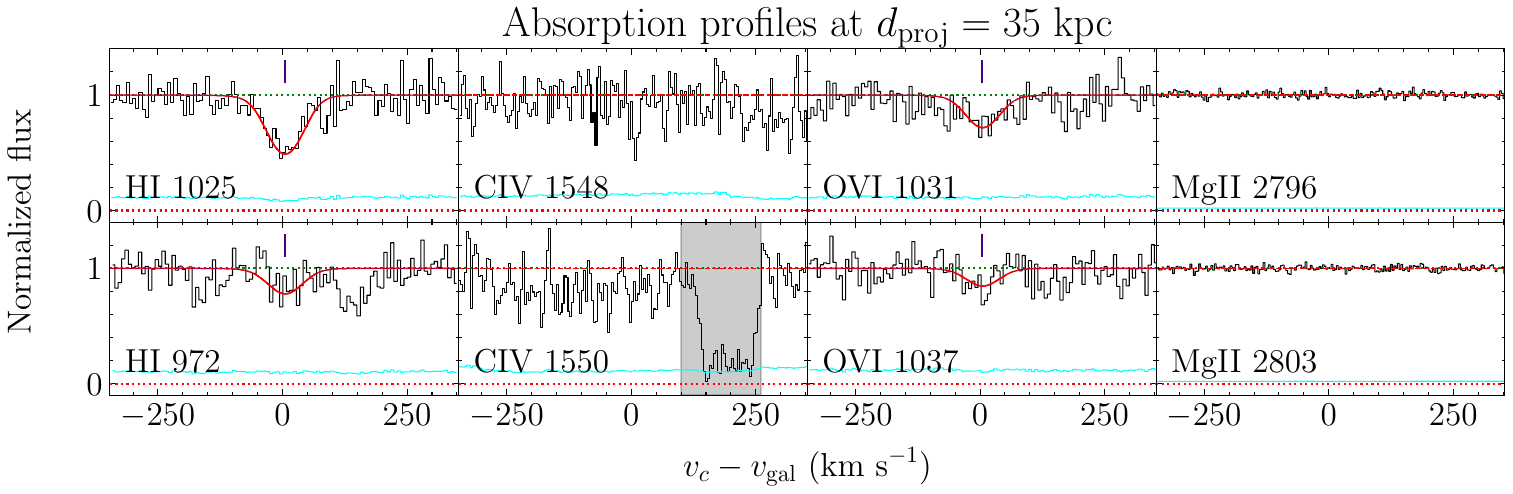}
  \caption{Galaxy image, optical spectrum, and QSO absorption spectra for system s2.} \label{fig:s2}
\end{figure*}

\begin{figure*}
    \centering
    \includegraphics[width=.8\textwidth]{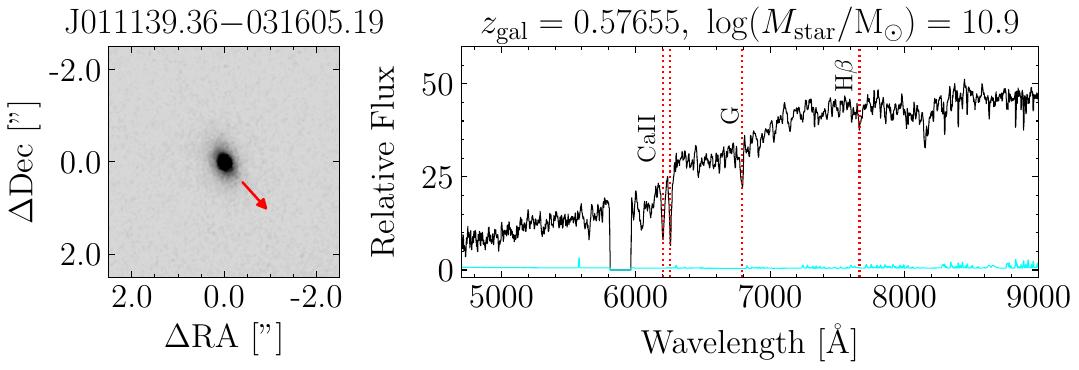}
    \includegraphics[width=.8\textwidth]{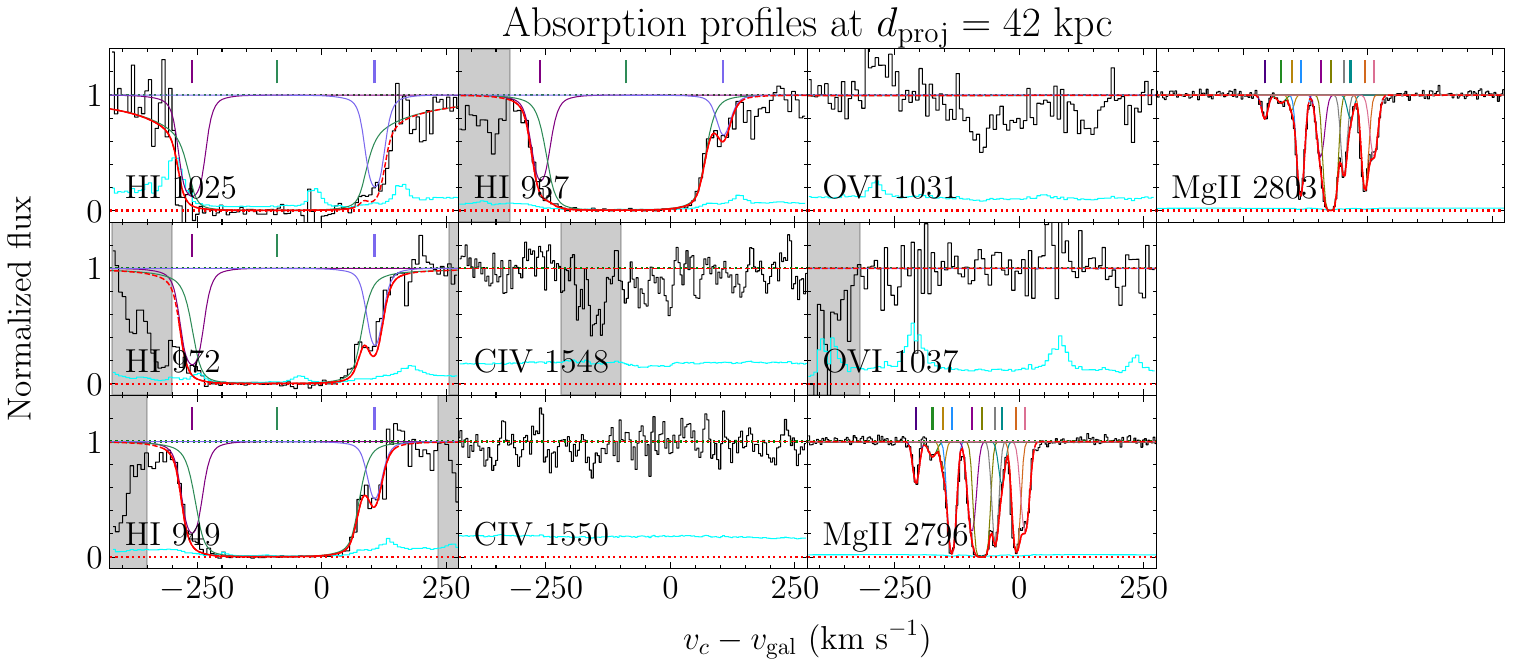}
  \caption{Galaxy image, optical spectrum, and QSO absorption spectra for system s3.} \label{fig:s3}
\end{figure*}

\begin{figure*}[t!]
    \centering
    \includegraphics[width=.8\textwidth]{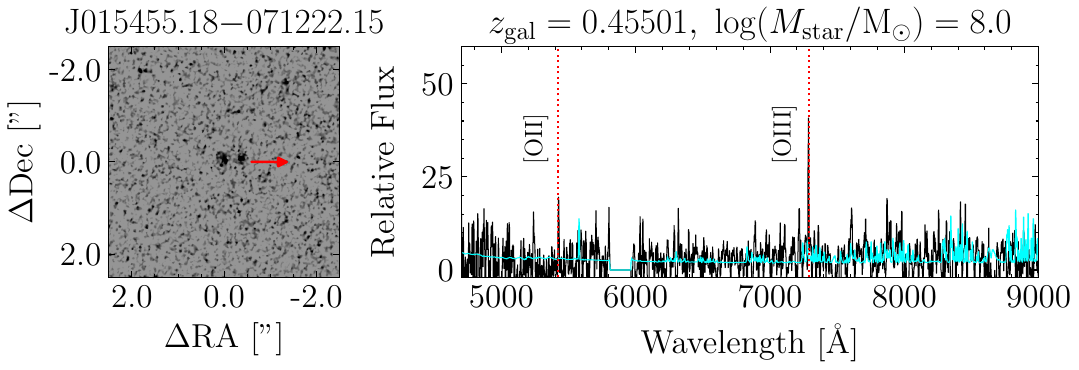}
    \includegraphics[width=.8\textwidth]{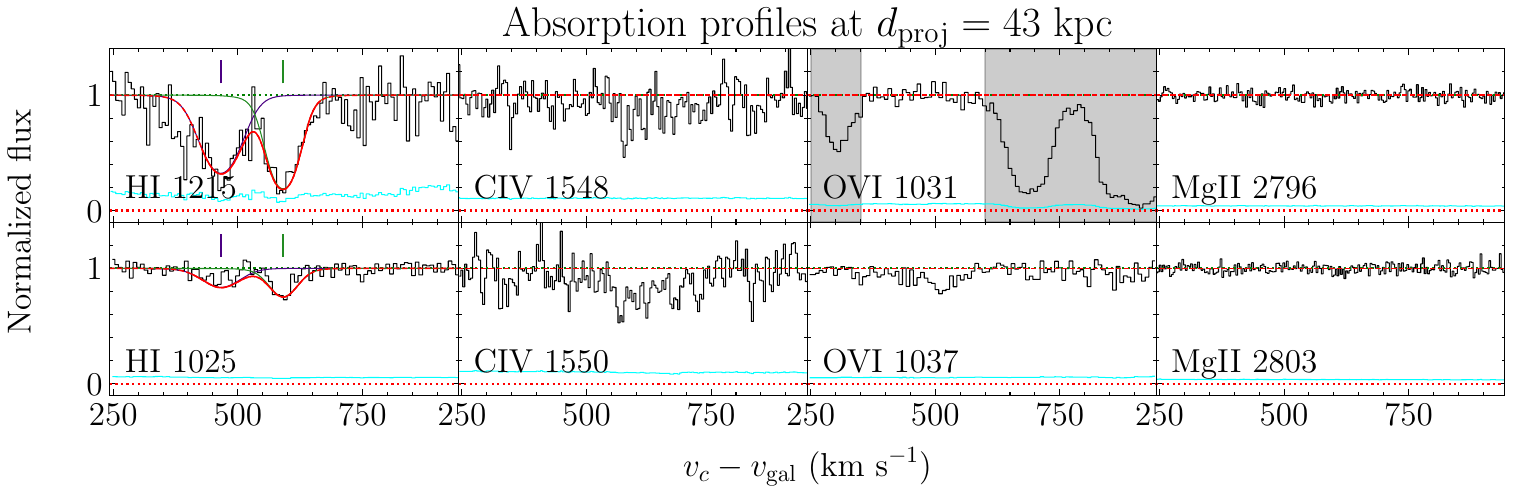}
  \caption{Galaxy image, optical spectrum, and QSO absorption spectra for system s4.} \label{fig:s4}
\end{figure*}

\begin{figure*}
    \centering
    \includegraphics[width=.8\textwidth]{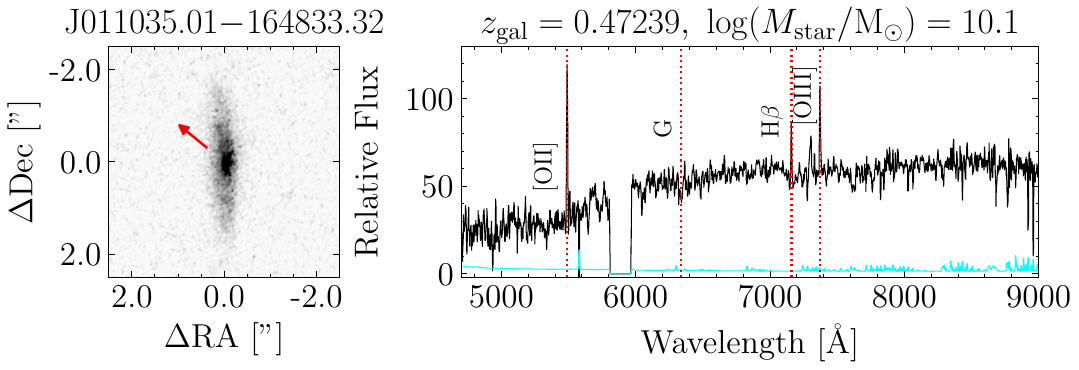}
    \includegraphics[width=.8\textwidth]{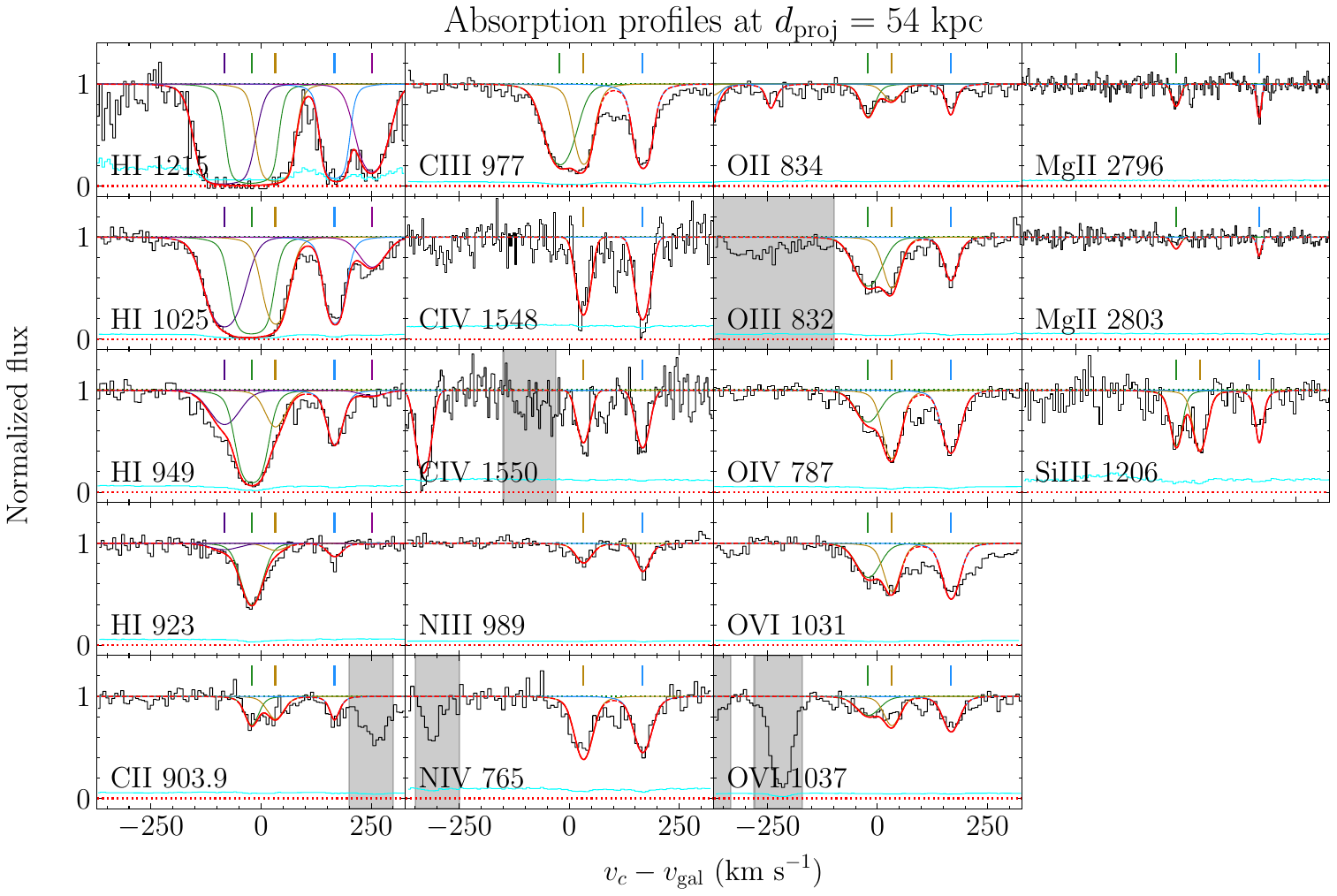}
  \caption{Galaxy image, optical spectrum, and QSO absorption spectra for system s5.} \label{fig:s5}
\end{figure*}

\begin{figure*}
    \centering
    \includegraphics[width=.8\textwidth]{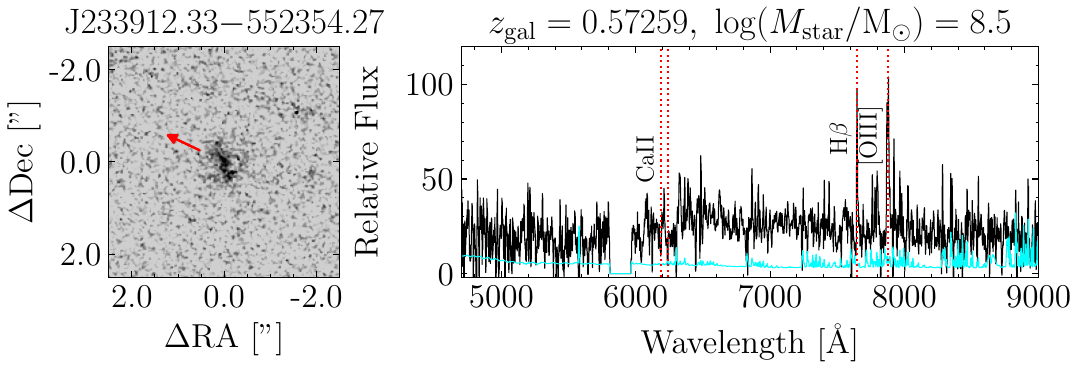}
    \includegraphics[width=.8\textwidth]{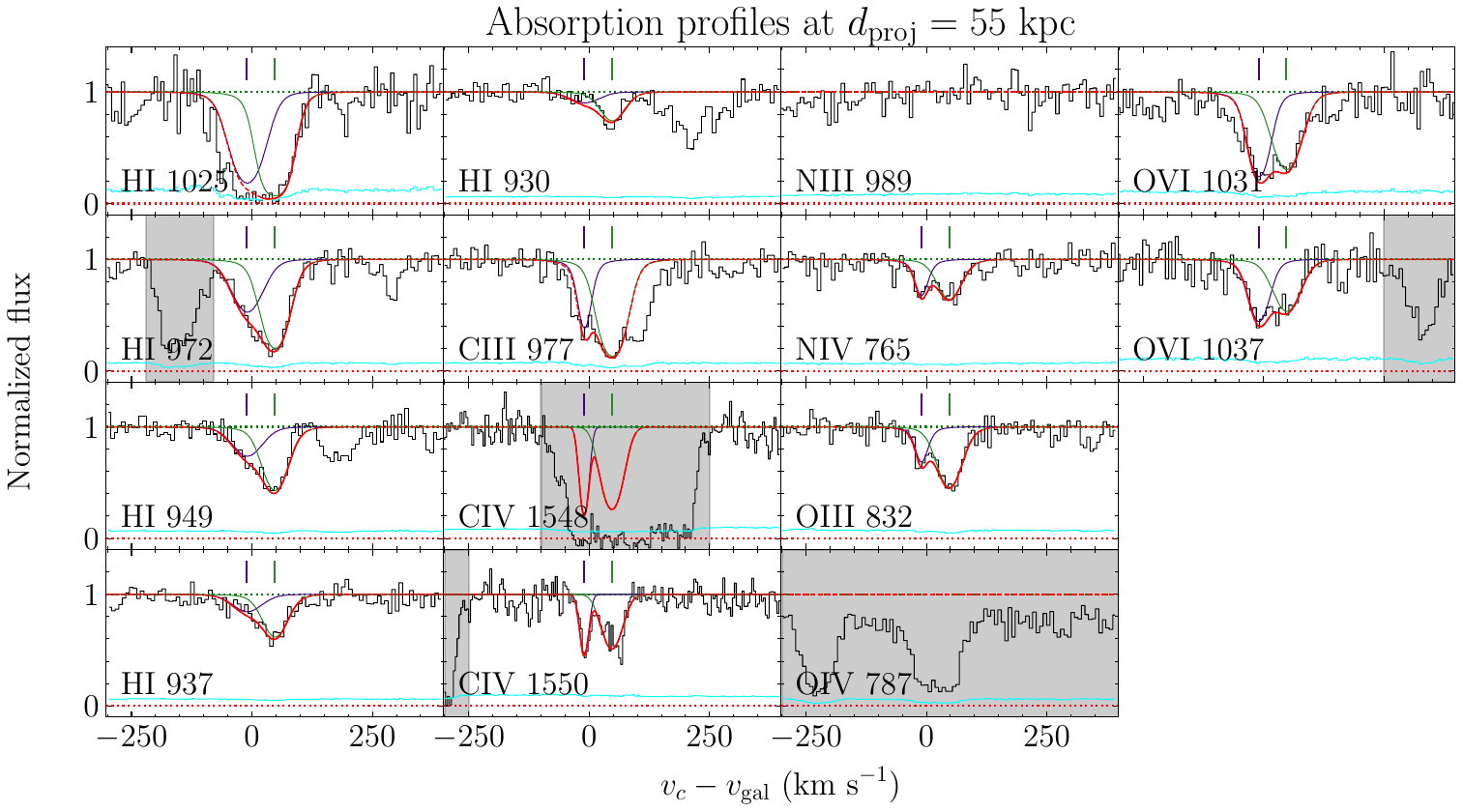}
  \caption{Galaxy image, optical spectrum, and QSO absorption spectra for system s6.} \label{fig:s6}
\end{figure*}

\begin{figure*}
    \centering
    \includegraphics[width=.8\textwidth]{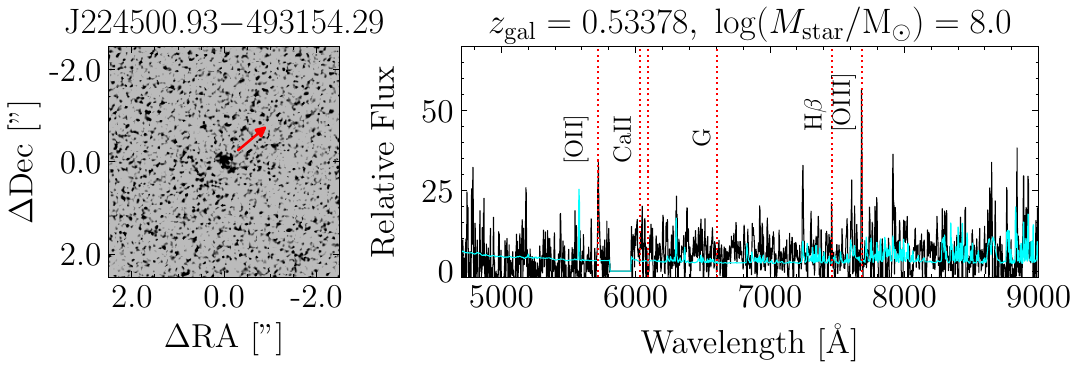}
    \includegraphics[width=.8\textwidth]{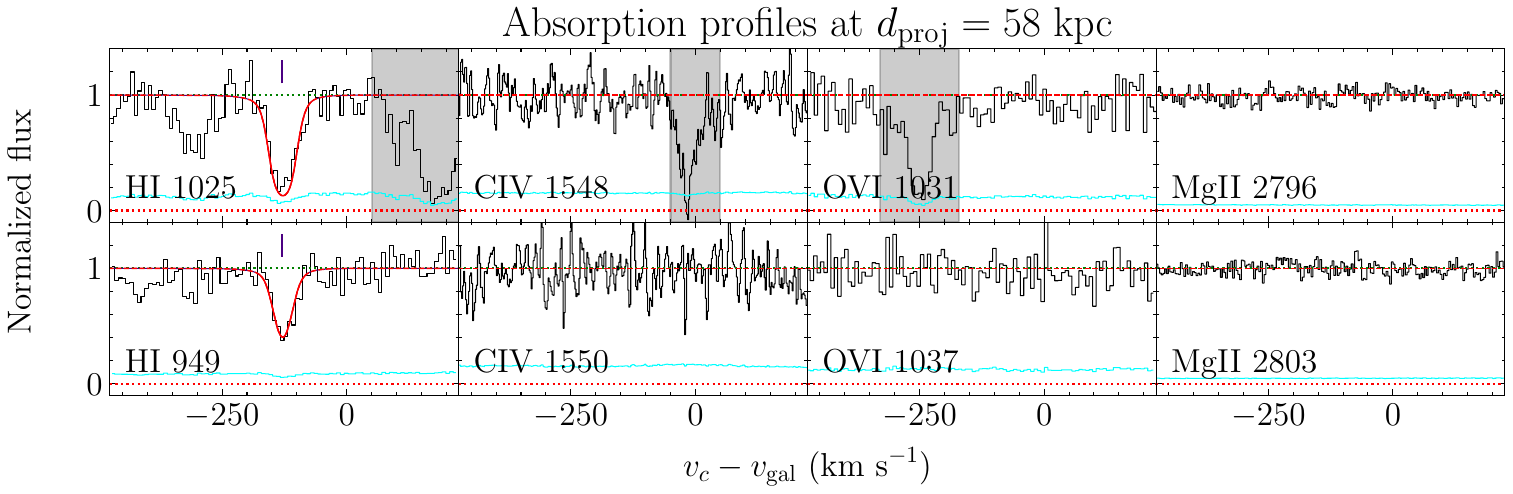}
  \caption{Galaxy image, optical spectrum, and QSO absorption spectra for system s7.} \label{fig:s7}
\end{figure*}

\begin{figure*}
    \centering
    \includegraphics[width=.8\textwidth]{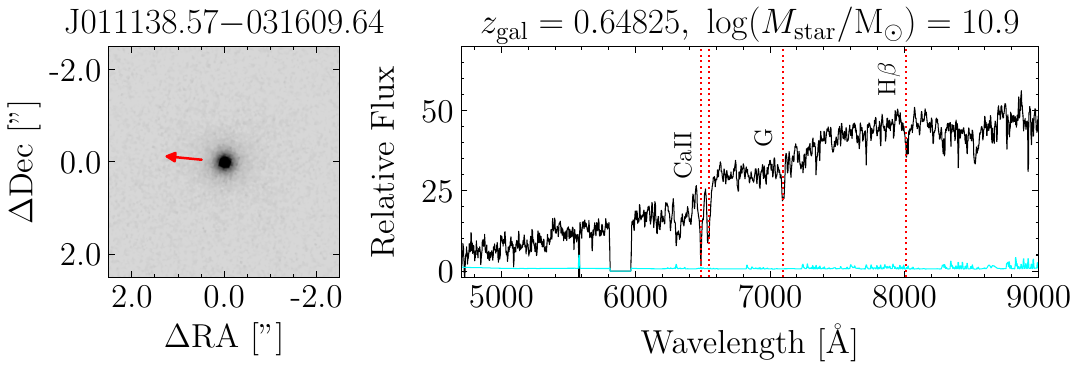}
    \includegraphics[width=.8\textwidth]{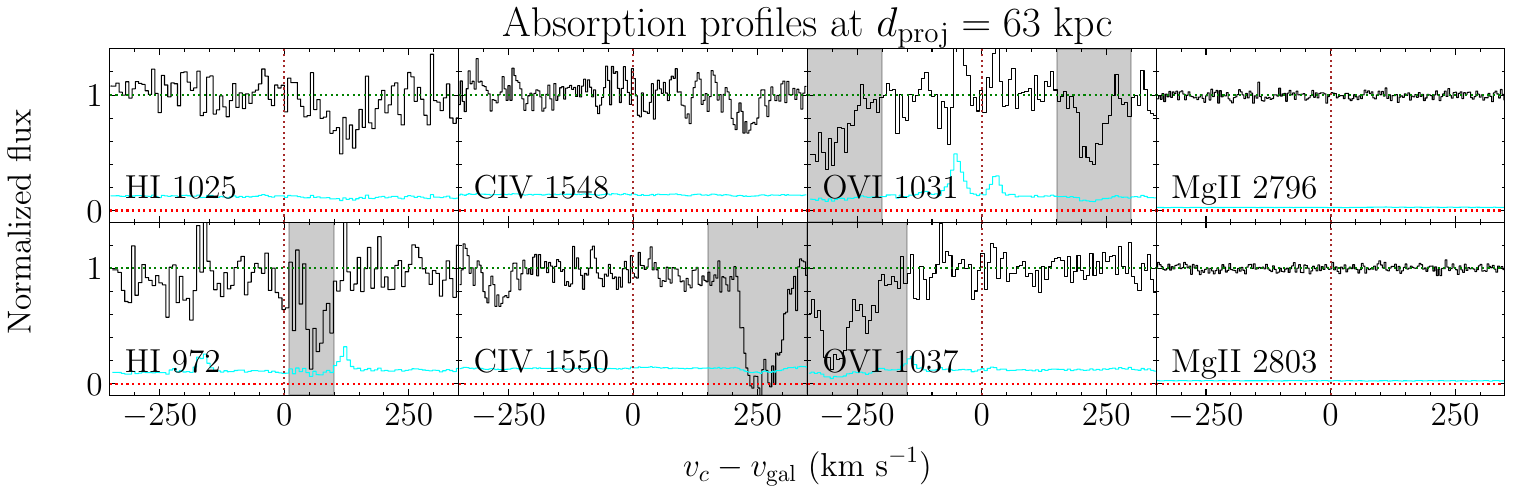}
  \caption{Galaxy image, optical spectrum, and QSO absorption spectra for system s8.} \label{fig:s8}
\end{figure*}

\begin{figure*}
    \centering
    \includegraphics[width=.8\textwidth]{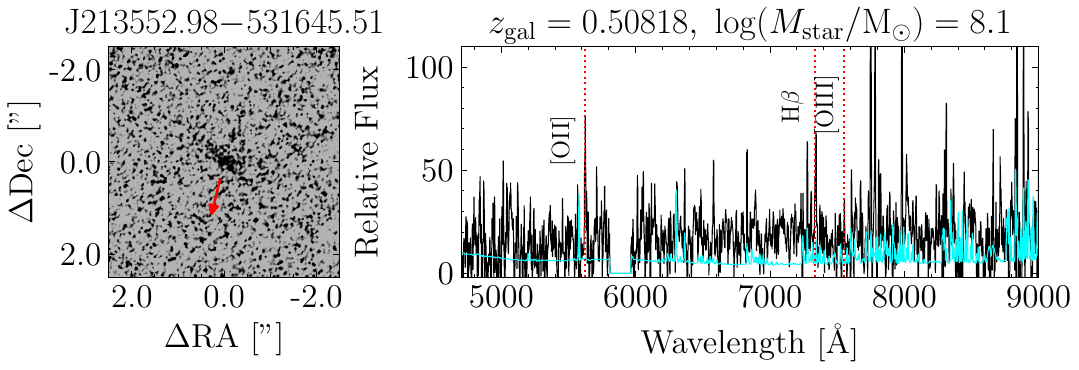}
    \includegraphics[width=.8\textwidth]{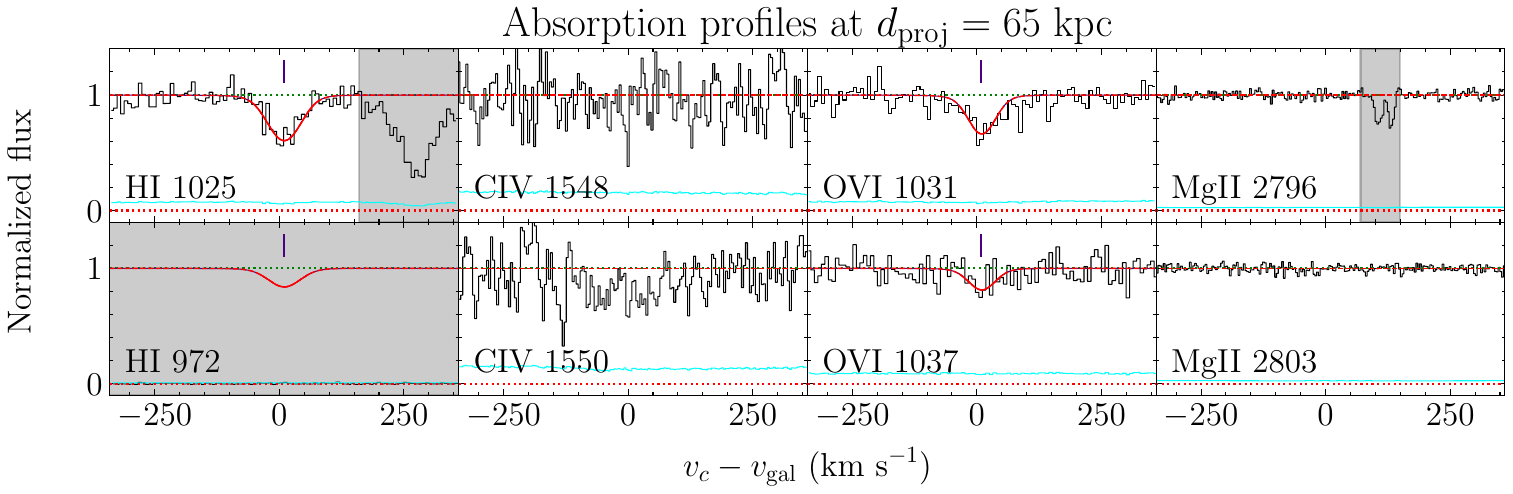}
  \caption{Galaxy image, optical spectrum, and QSO absorption spectra for system s9.} \label{fig:s9}
\end{figure*}

\begin{figure*}
    \centering
    \includegraphics[width=.8\textwidth]{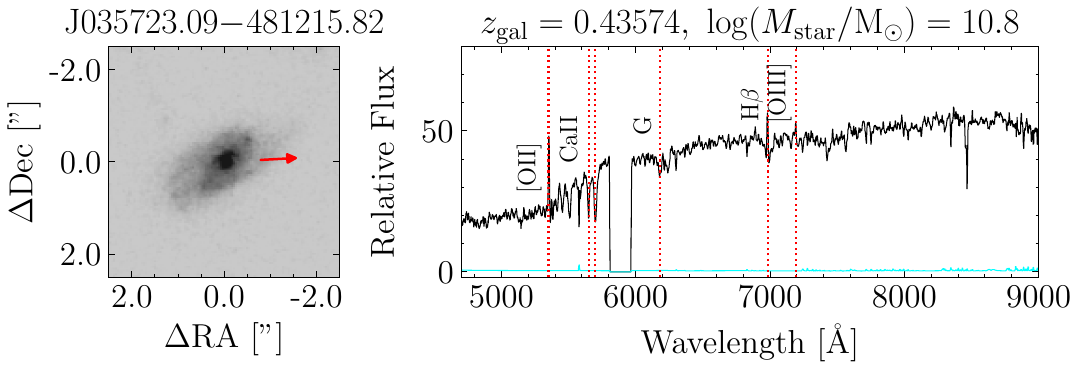}
    \includegraphics[width=.8\textwidth]{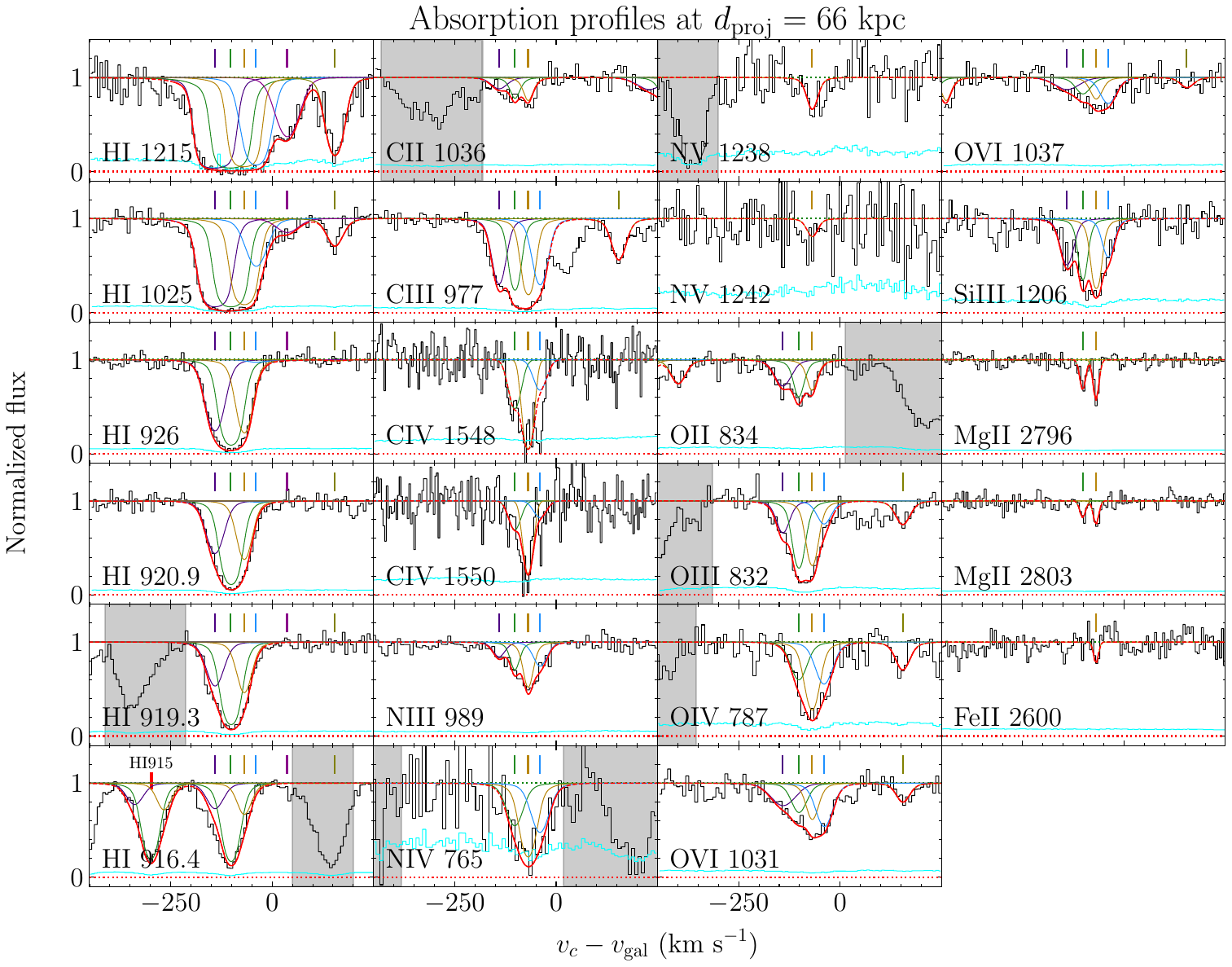}
  \caption{Galaxy image, optical spectrum, and QSO absorption spectra for system s10.} \label{fig:s10}
\end{figure*}

\begin{figure*}
    \centering
    \includegraphics[width=.8\textwidth]{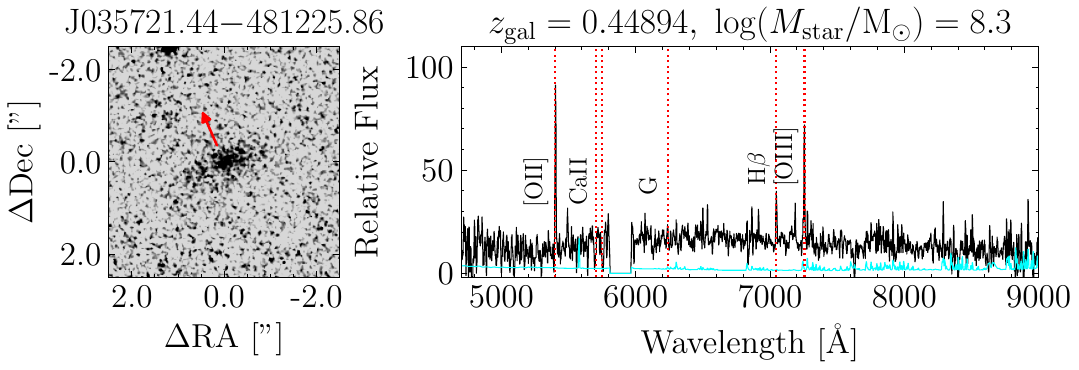}
    \includegraphics[width=.8\textwidth]{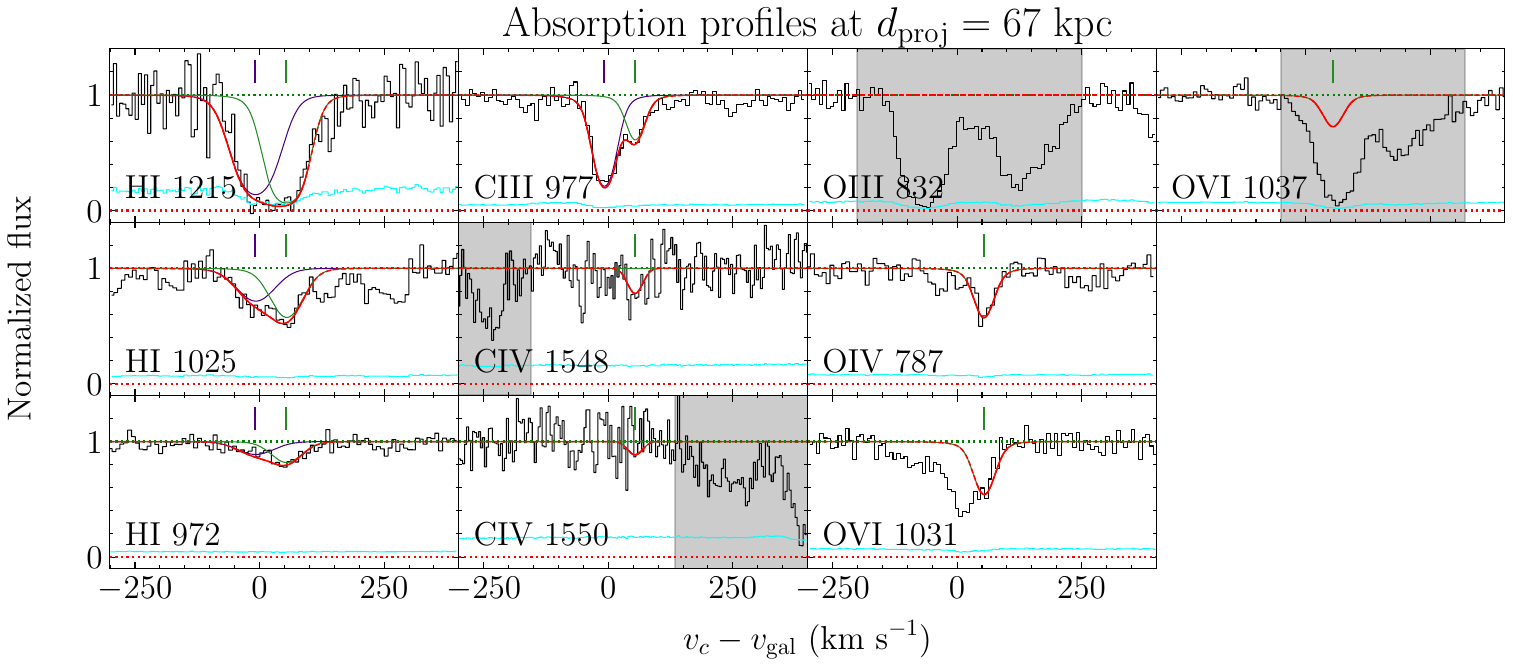}
  \caption{Galaxy image, optical spectrum, and QSO absorption spectra for system s11.} \label{fig:s11}
\end{figure*}

\begin{figure*}
    \centering
    \includegraphics[width=.8\textwidth]{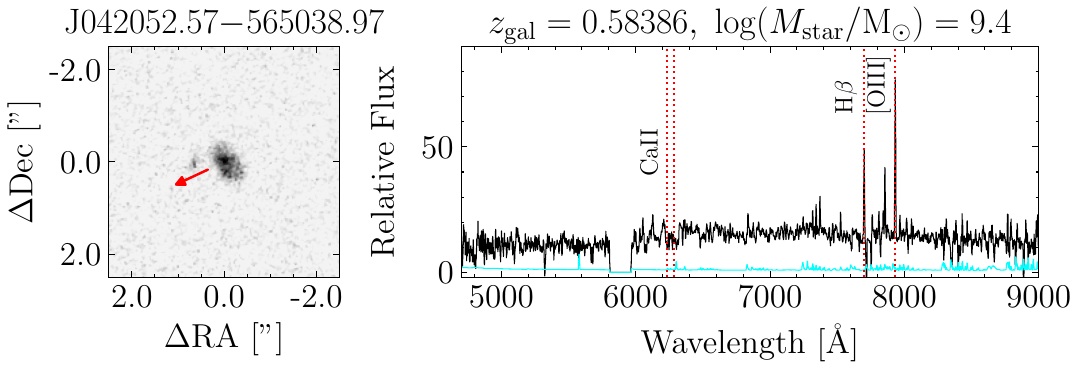}
    \includegraphics[width=.8\textwidth]{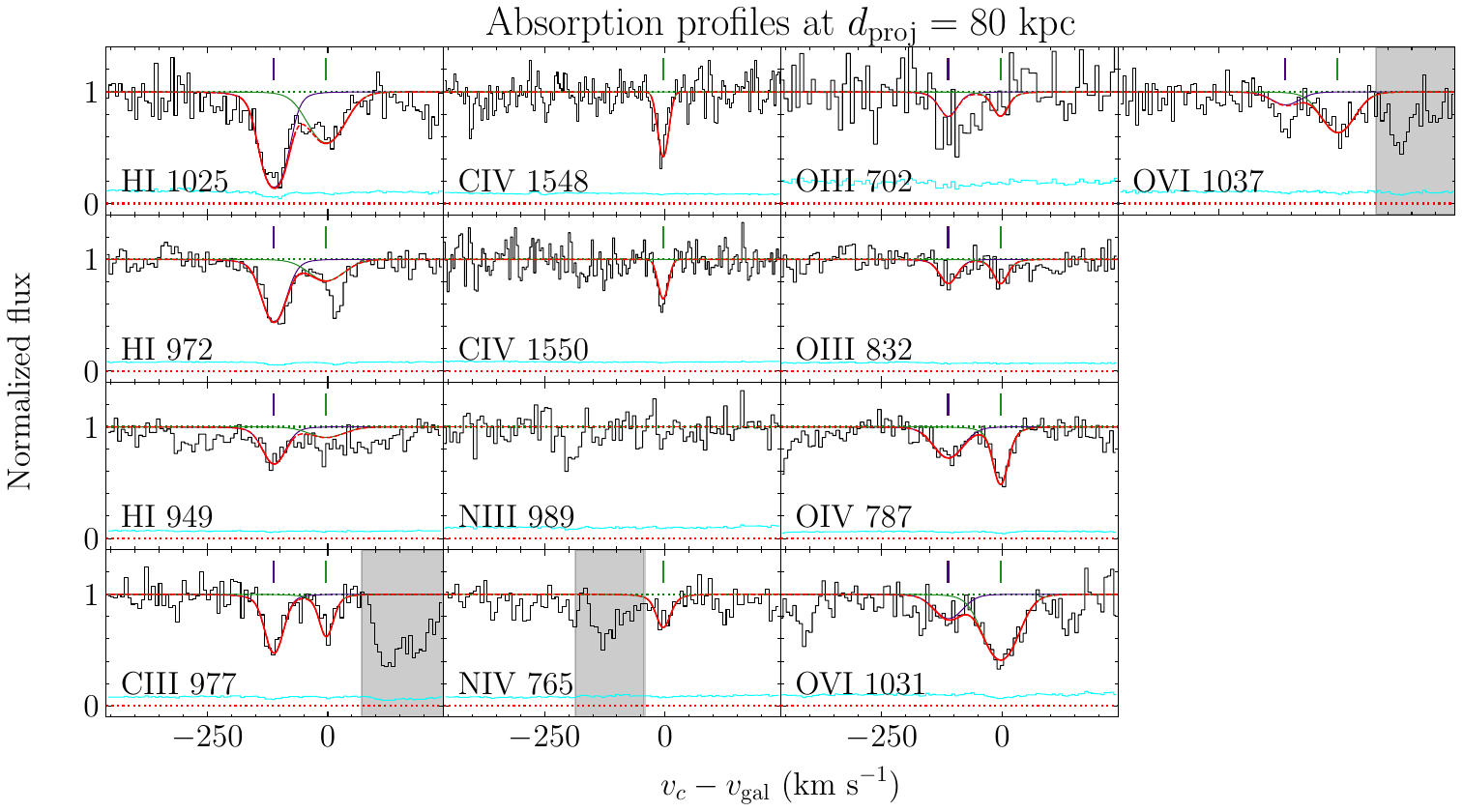}
  \caption{Galaxy image, optical spectrum, and QSO absorption spectra for system s12.} \label{fig:s12}
\end{figure*}

\begin{figure*}
    \centering
    \includegraphics[width=.8\textwidth]{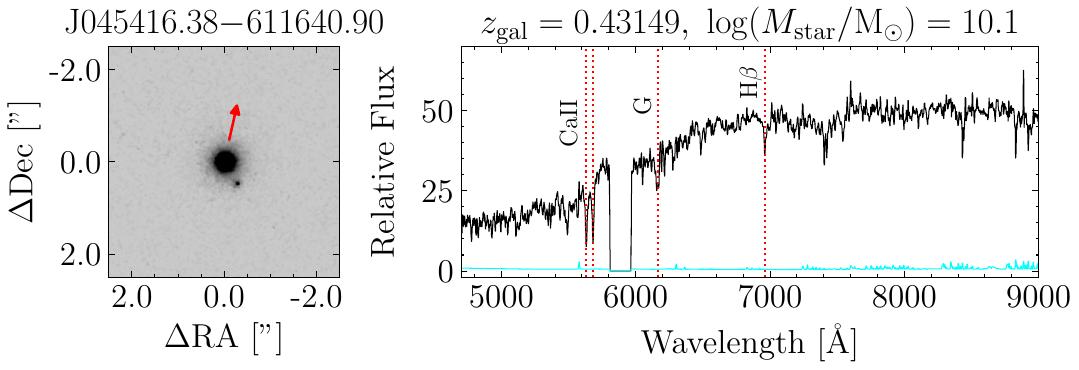}
    \includegraphics[width=.8\textwidth]{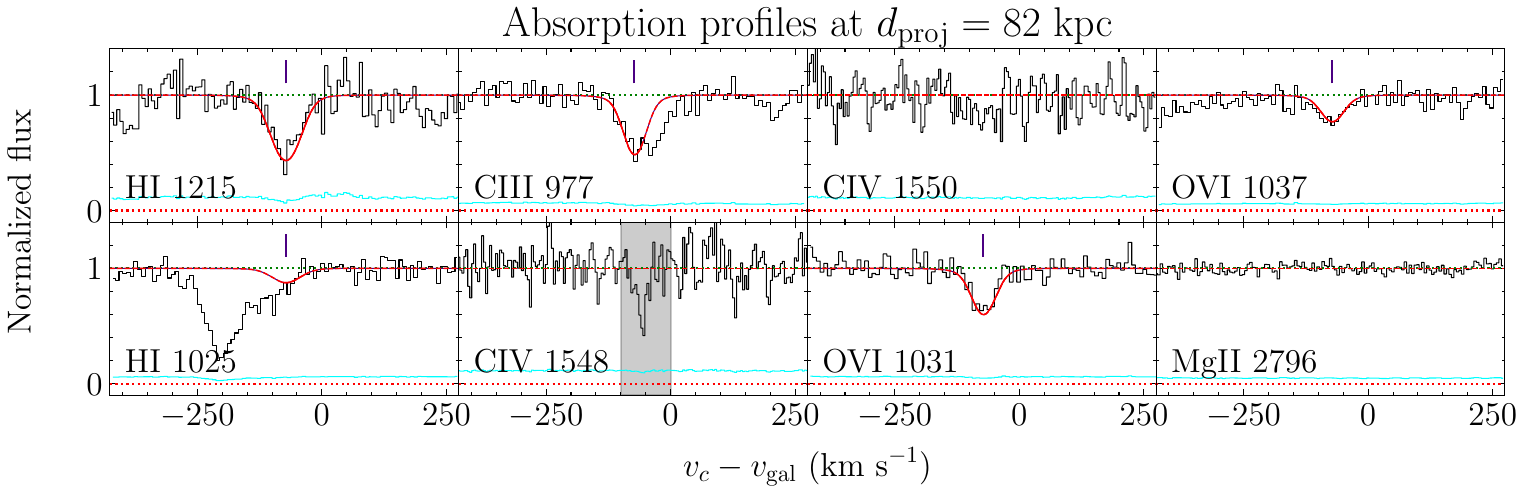}
  \caption{Galaxy image, optical spectrum, and QSO absorption spectra for system s13.} \label{fig:s13}
\end{figure*}

\begin{figure*}
    \centering
    \includegraphics[width=.8\textwidth]{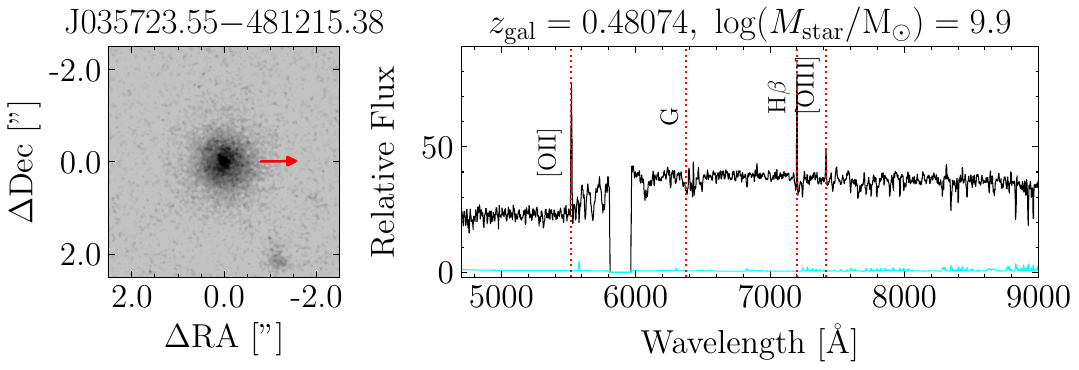}
    \includegraphics[width=.8\textwidth]{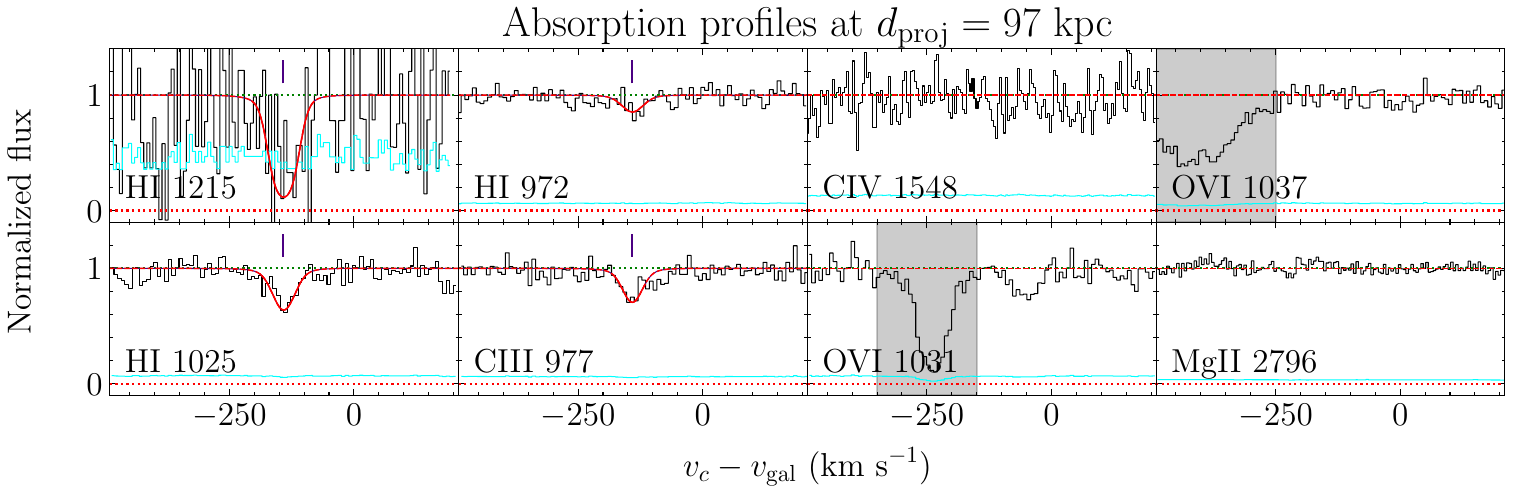}
  \caption{Galaxy image, optical spectrum, and QSO absorption spectra for system s14.} \label{fig:s14}
\end{figure*}

\begin{figure*}
    \centering
    \includegraphics[width=.8\textwidth]{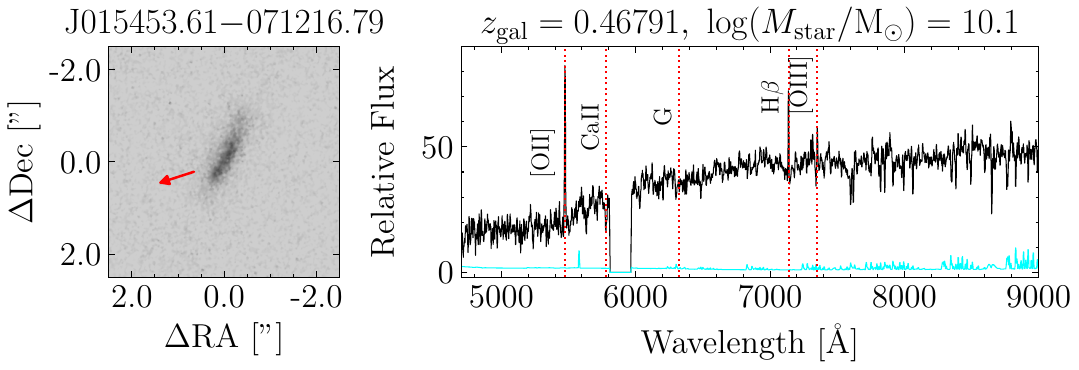}
    \includegraphics[width=.8\textwidth]{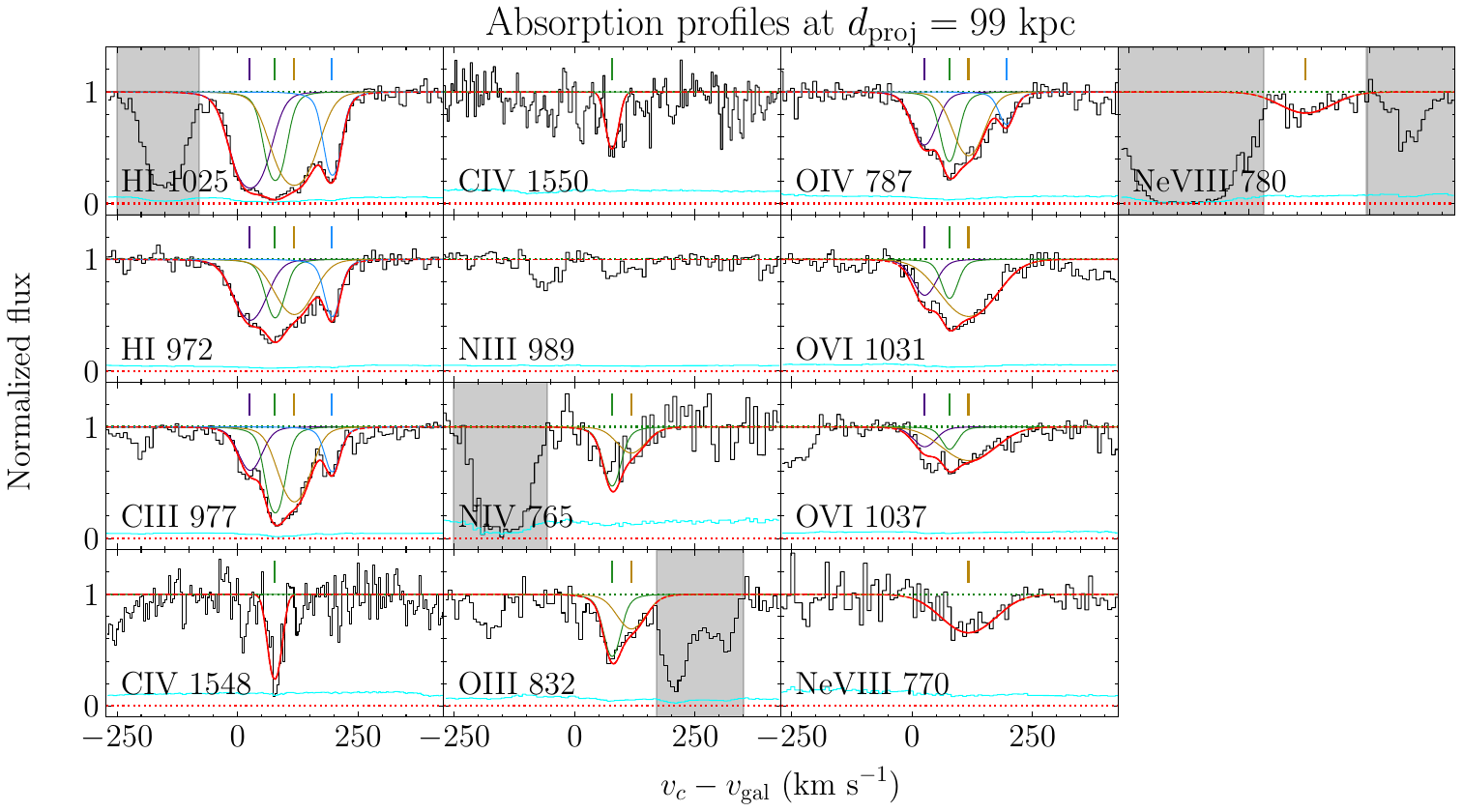}
  \caption{Galaxy image, optical spectrum, and QSO absorption spectra for system s15.} \label{fig:s15}
\end{figure*}

\begin{figure*}
    \centering
    \includegraphics[width=.8\textwidth]{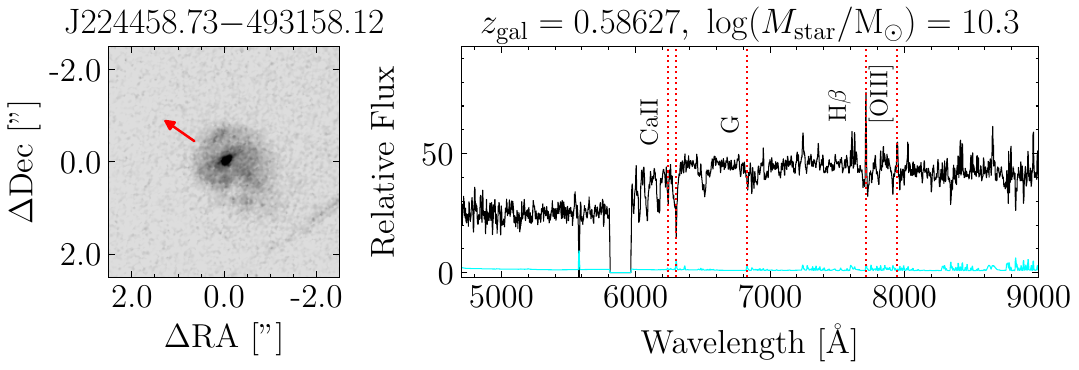}
    \includegraphics[width=.8\textwidth]{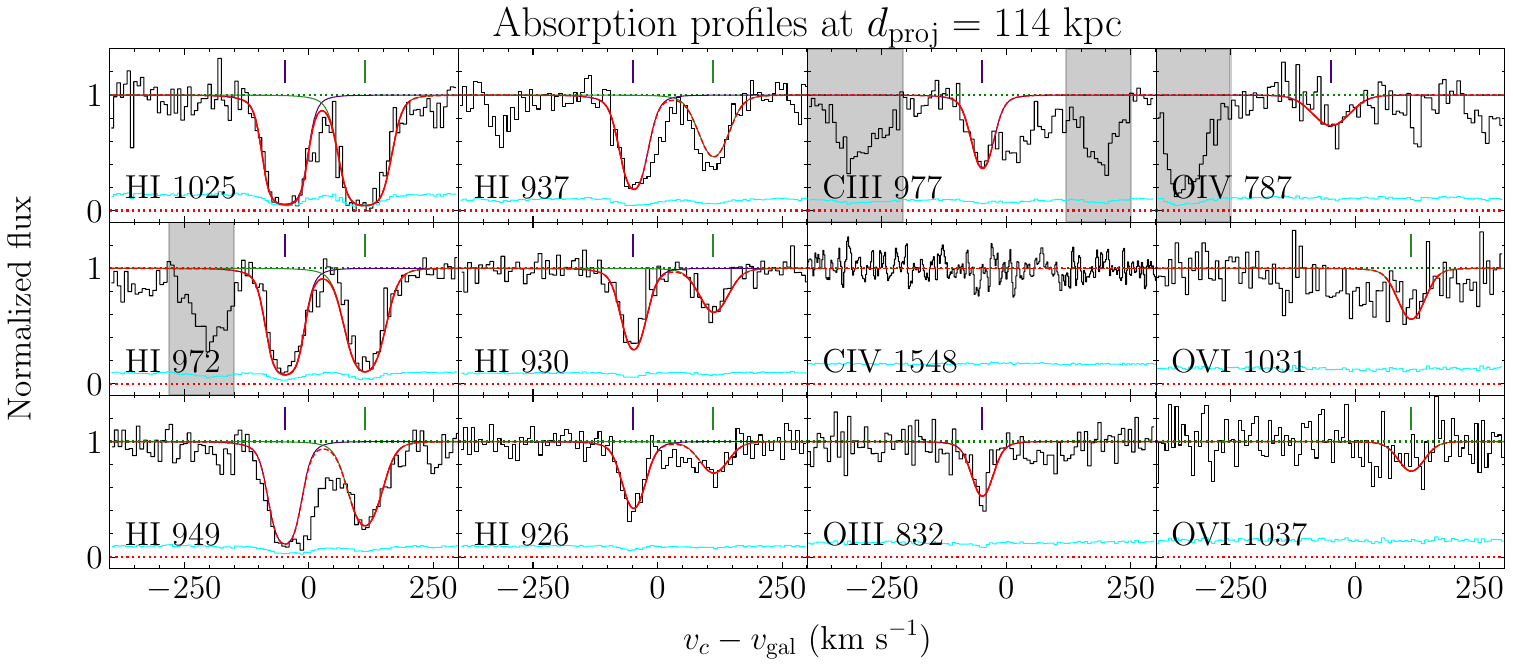}
  \caption{Galaxy image, optical spectrum, and QSO absorption spectra for system s16.} \label{fig:s16}
\end{figure*}

\begin{figure*}
    \centering
    \includegraphics[width=.8\textwidth]{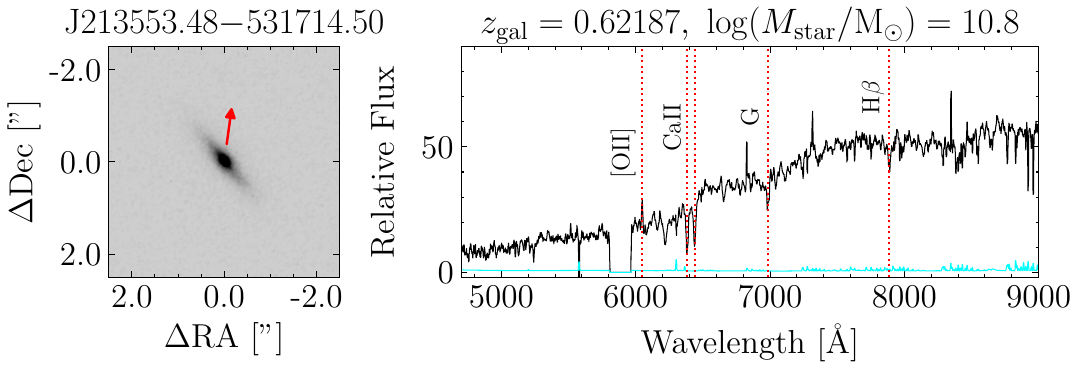}
    \includegraphics[width=.8\textwidth]{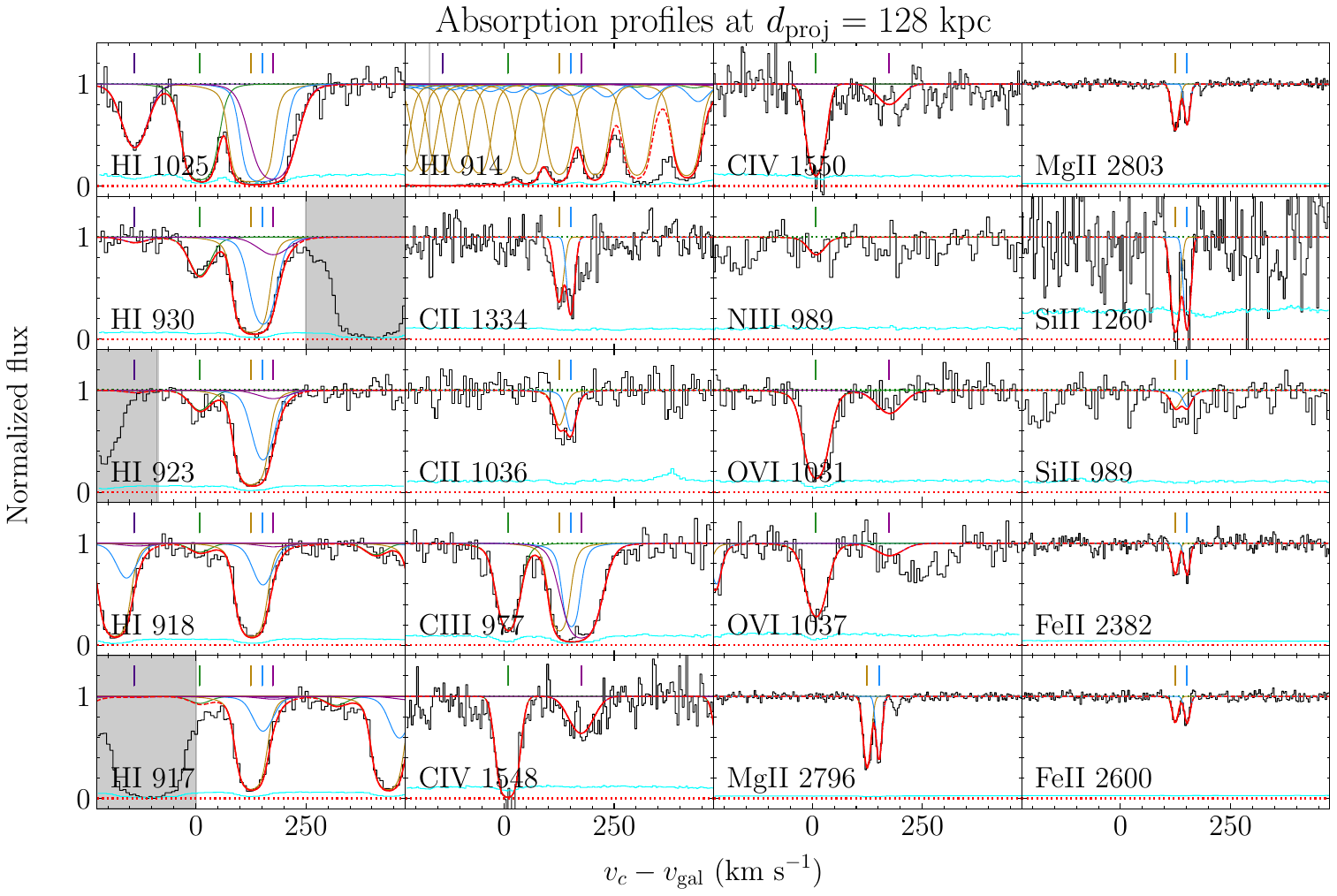}
  \caption{Galaxy image, optical spectrum, and QSO absorption spectra for system s17.} \label{fig:s17}
\end{figure*}

\begin{figure*}
    \centering
    \includegraphics[width=.8\textwidth]{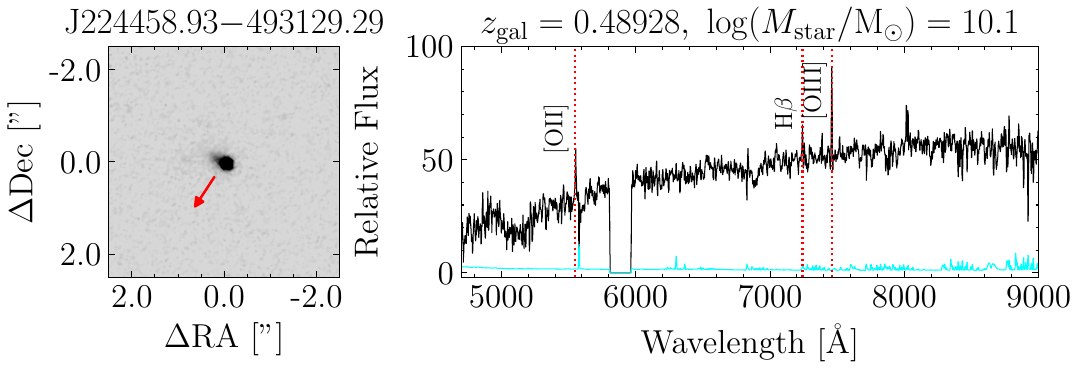}
    \includegraphics[width=.8\textwidth]{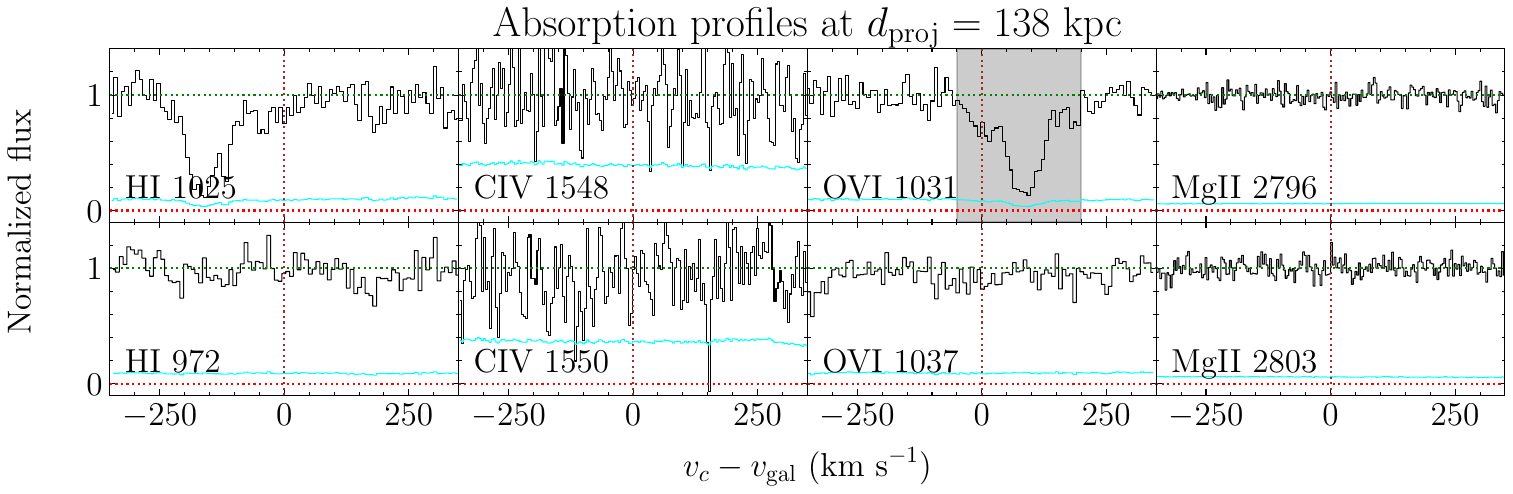}
  \caption{Galaxy image, optical spectrum, and QSO absorption spectra for system s18.} \label{fig:s18}
\end{figure*}

\begin{figure*}
    \centering
    \includegraphics[width=.8\textwidth]{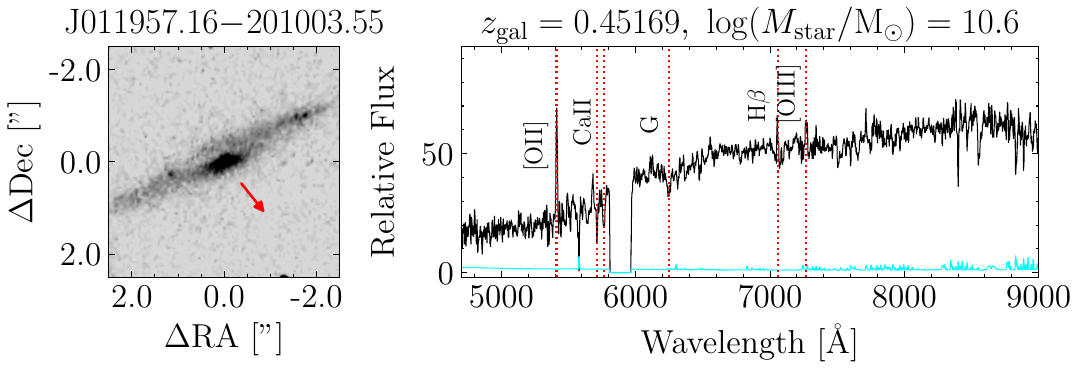}
    \includegraphics[width=.8\textwidth]{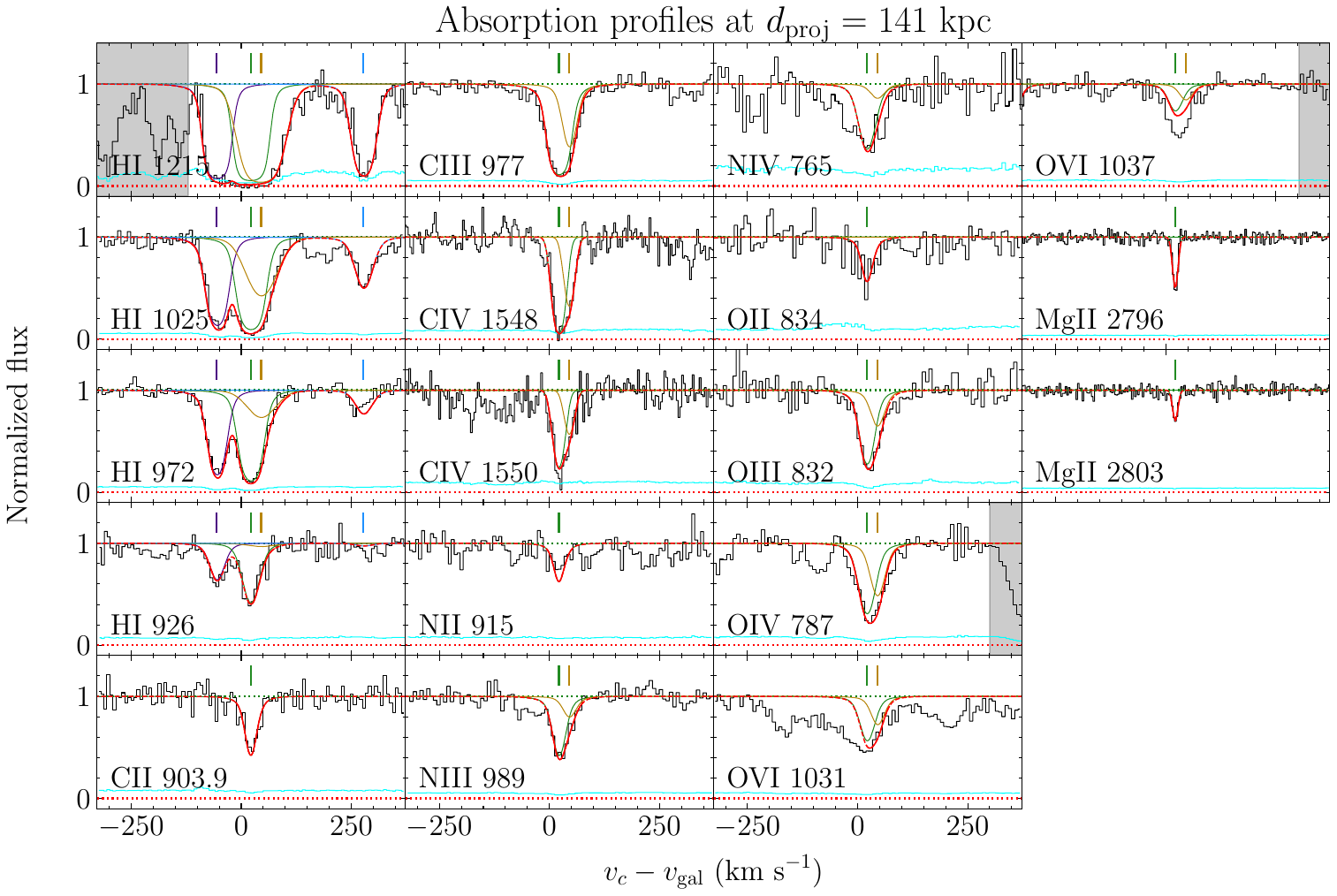}
  \caption{Galaxy image, optical spectrum, and QSO absorption spectra for system s19.} \label{fig:s19}
\end{figure*}

\begin{figure*}
    \centering
    \includegraphics[width=.8\textwidth]{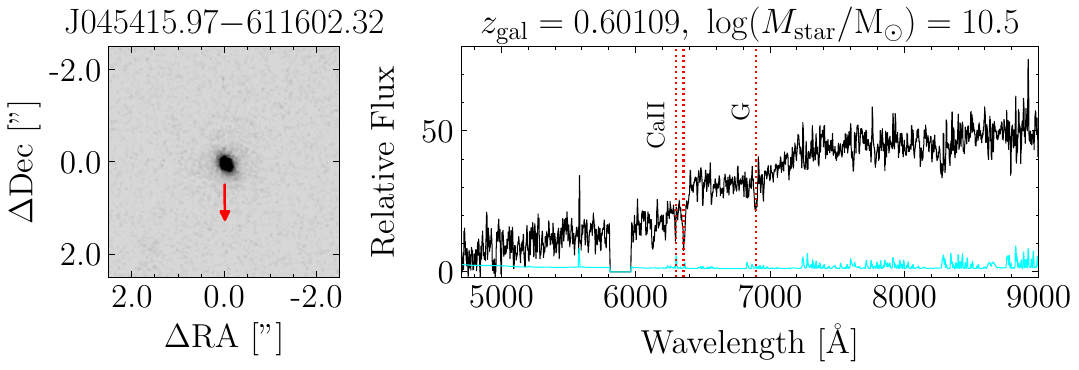}
    \includegraphics[width=.8\textwidth]{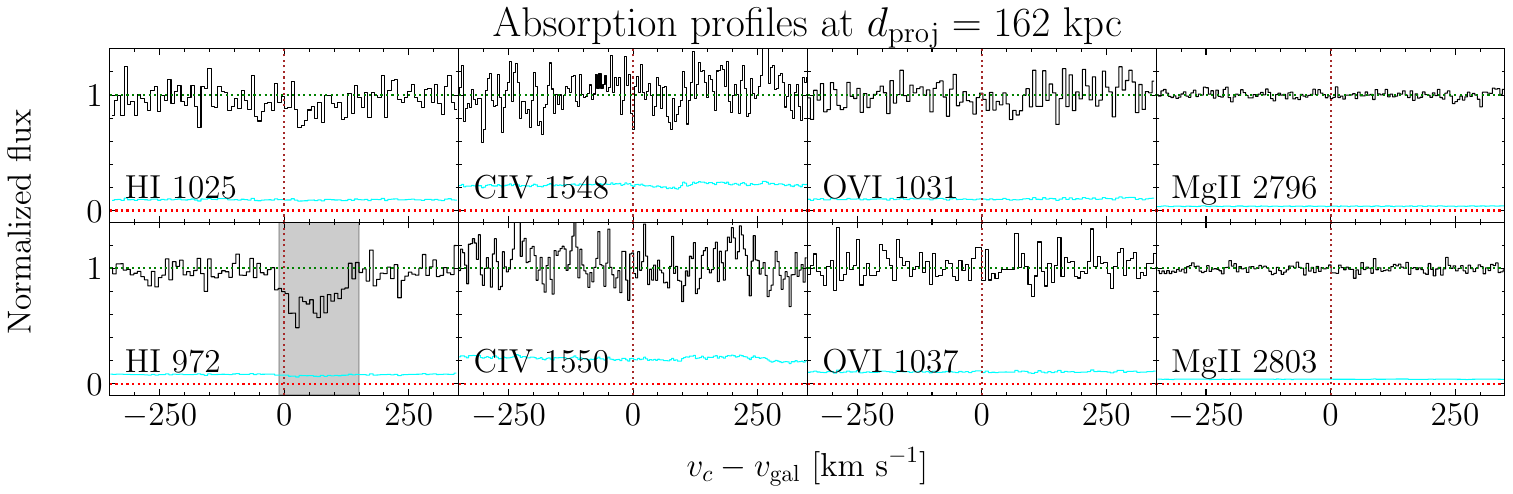}
  \caption{Galaxy image, optical spectrum, and QSO absorption spectra for system s20.} \label{fig:s20}
\end{figure*}

\begin{figure*}
    \centering
    \includegraphics[width=.8\textwidth]{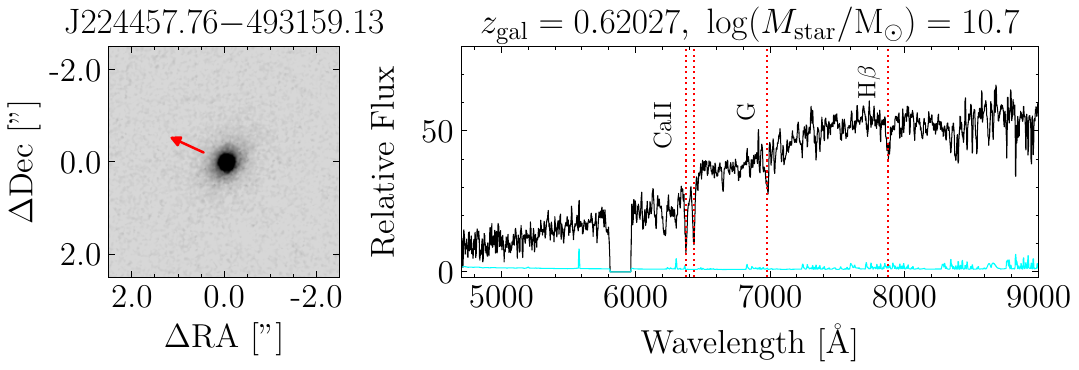}
    \includegraphics[width=.8\textwidth]{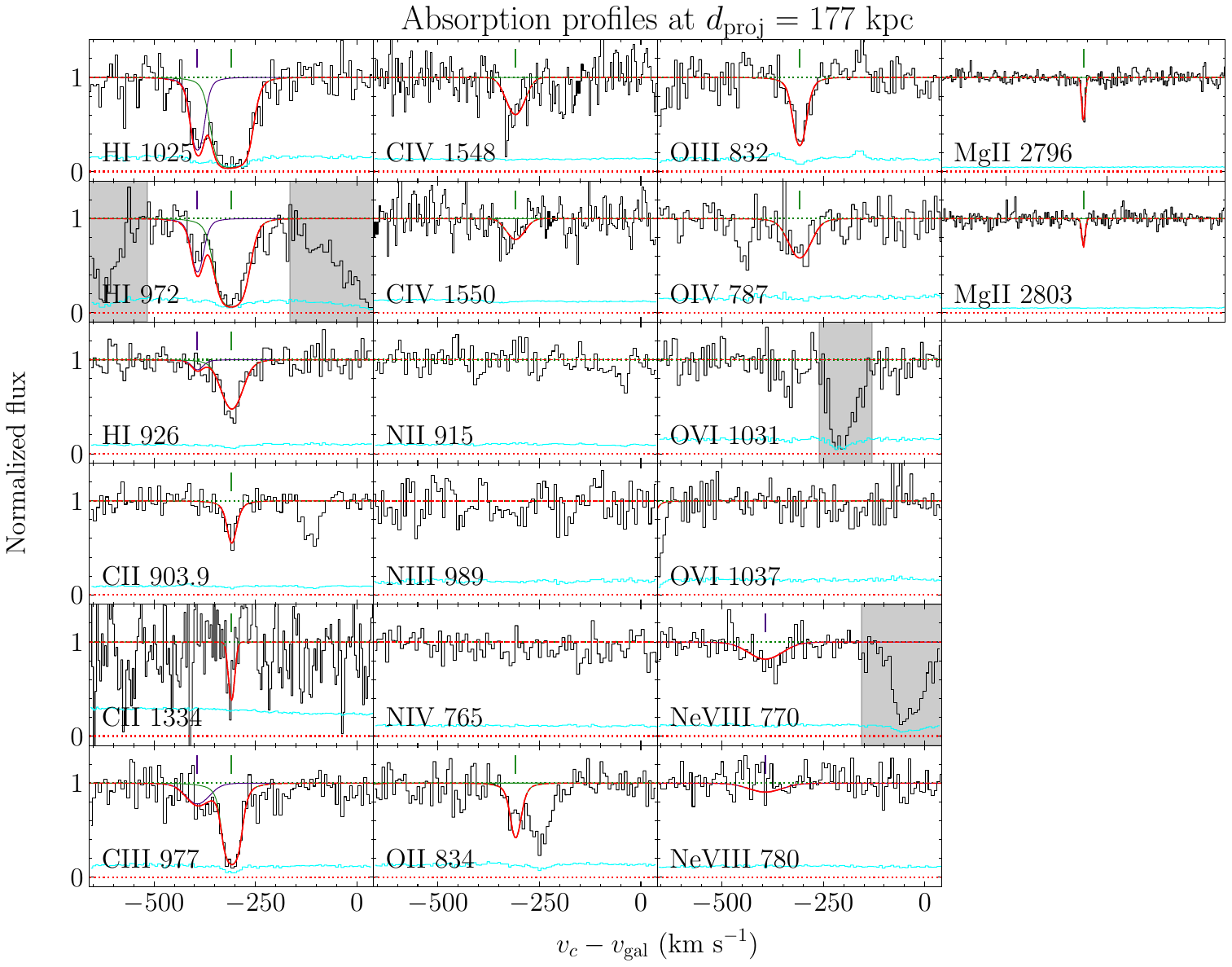}
  \caption{Galaxy image, optical spectrum, and QSO absorption spectra for system s21.} \label{fig:s21}
\end{figure*}

\begin{figure*}
    \centering
    \includegraphics[width=.8\textwidth]{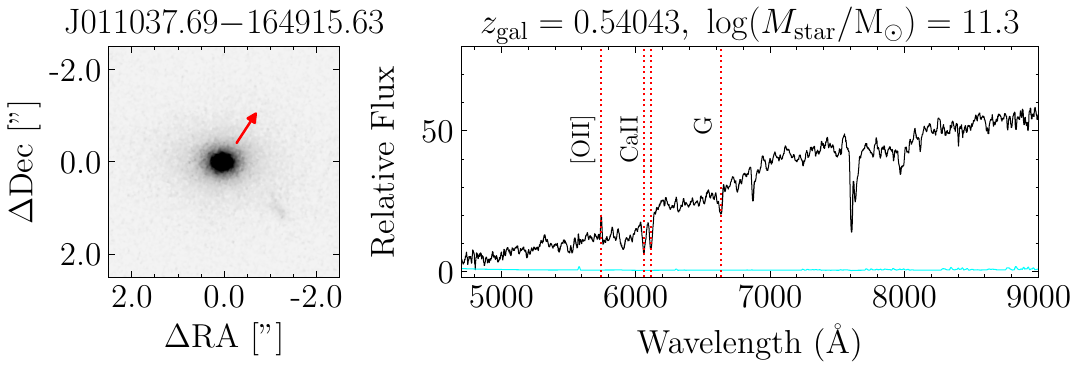}
    \includegraphics[width=.8\textwidth]{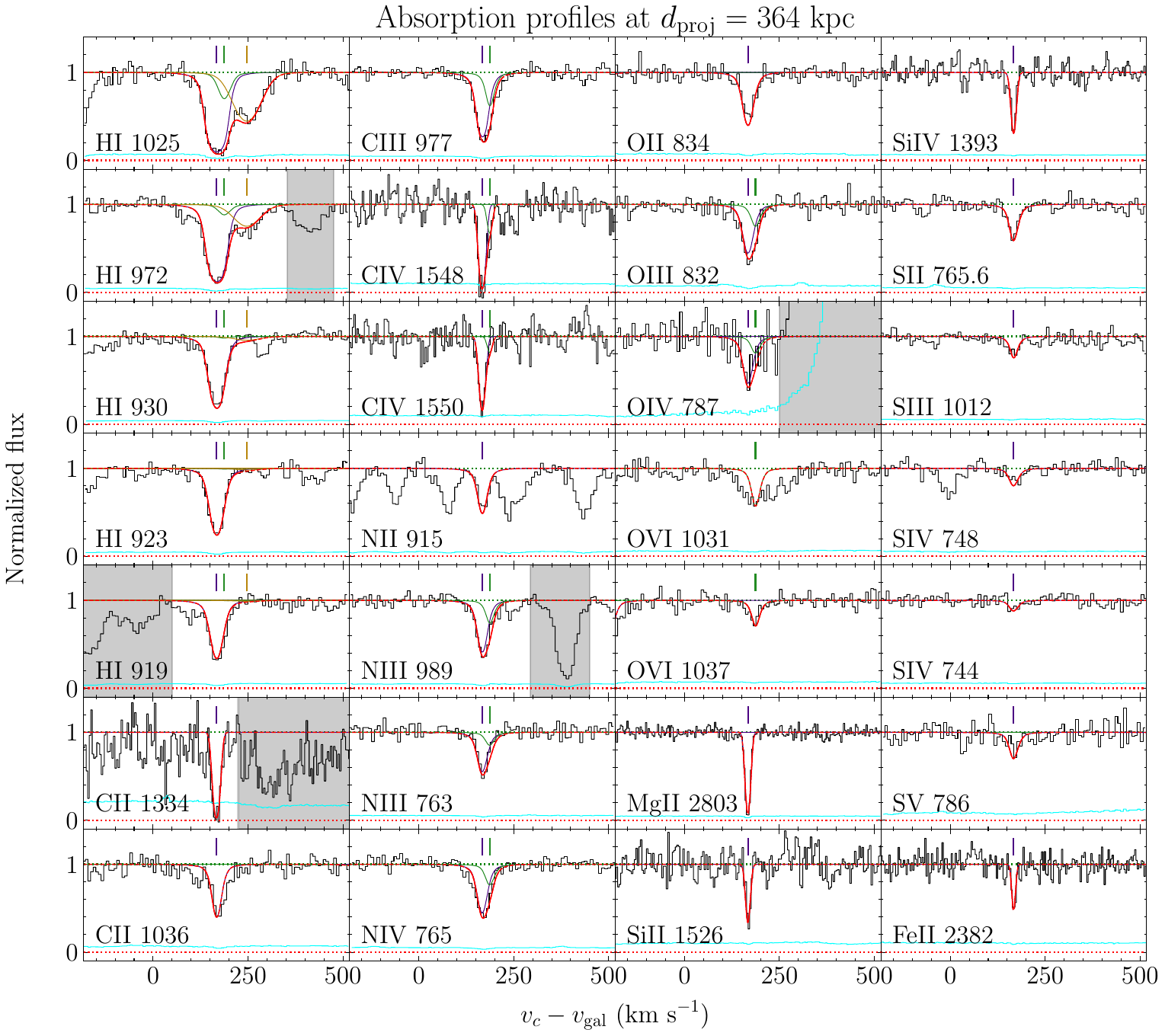}
  \caption{Galaxy image, optical spectrum, and QSO absorption spectra for system s22.} \label{fig:s22}
\end{figure*}

\begin{figure*}
    \centering
    \includegraphics[width=.8\textwidth]{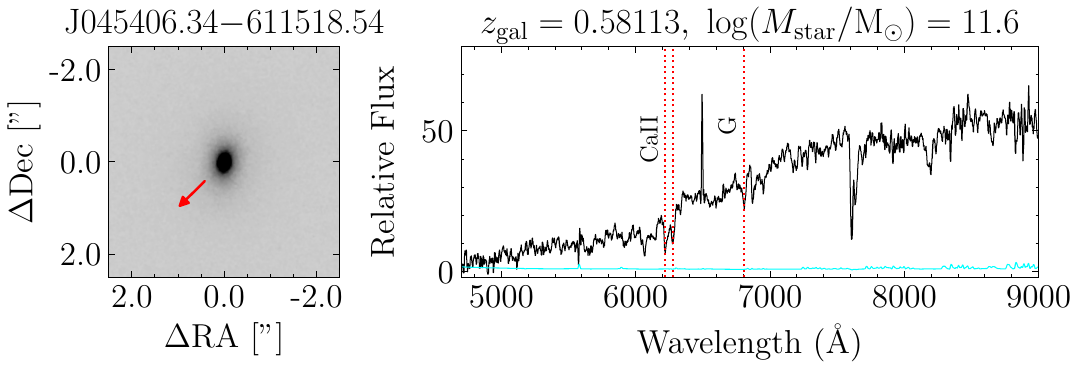}
    \includegraphics[width=.8\textwidth]{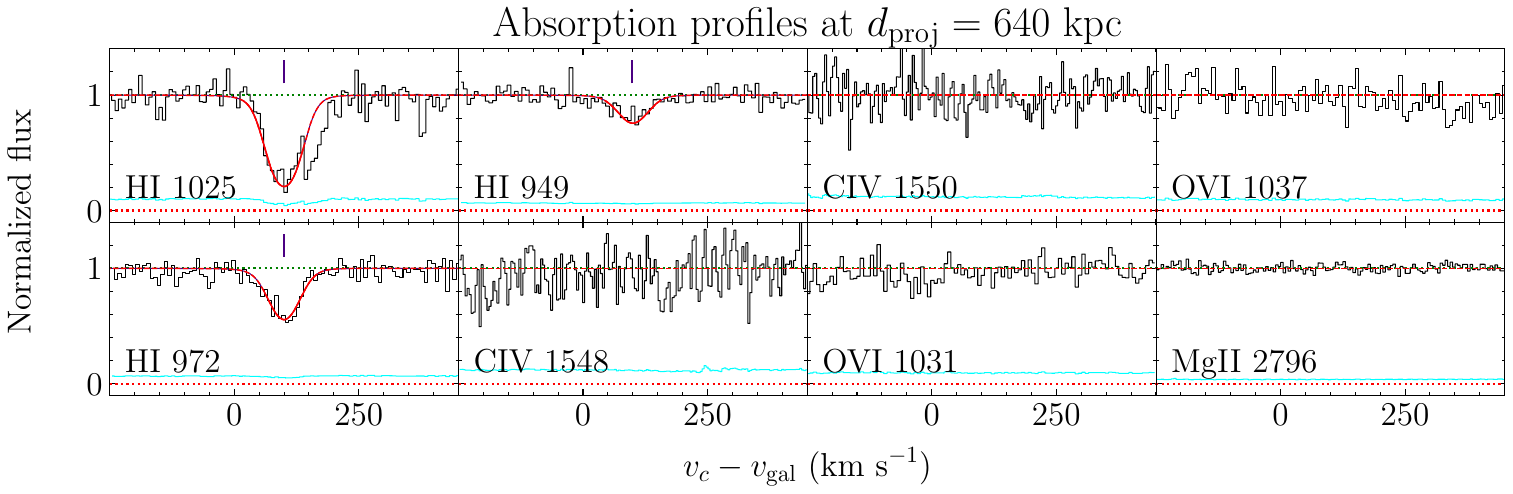}
  \caption{Galaxy image, optical spectrum, and QSO absorption spectra for system s23.} \label{fig:s23}
\end{figure*}

Details for absorption line analysis and photoionization modeling of individual CUBS-midz systems is described below. Note that all available transitions for ionic species within the spectral coverage of the survey are considered during the profile fitting, but only the strongest transitions have been indicated in Figures \ref{fig:s1}--\ref{fig:s23} for brevity.

\subsection{$\mathrm{CUBS \ J230837.40\!-\!525844.94}$ at $d_\mathrm{proj}=34 \ \mathrm{kpc}$} \label{sec:s1}

\begin{figure*}[t]
    \centering
    \includegraphics[width=.8\textwidth]{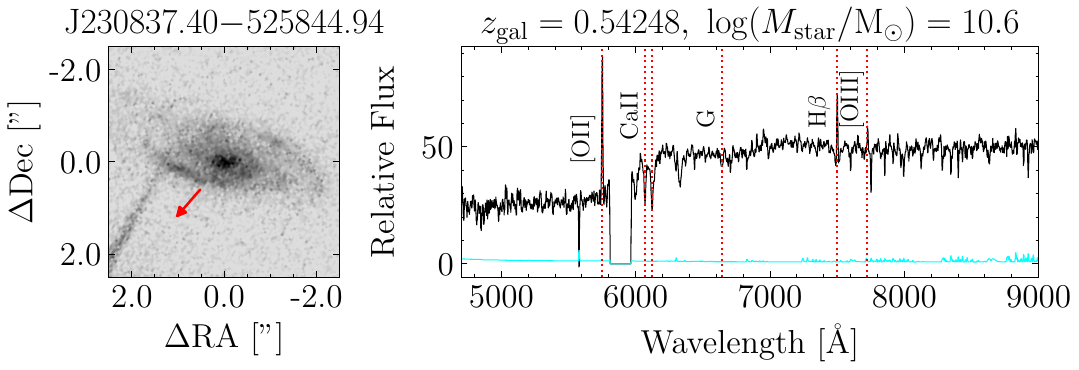}
    \includegraphics[width=.8\textwidth]{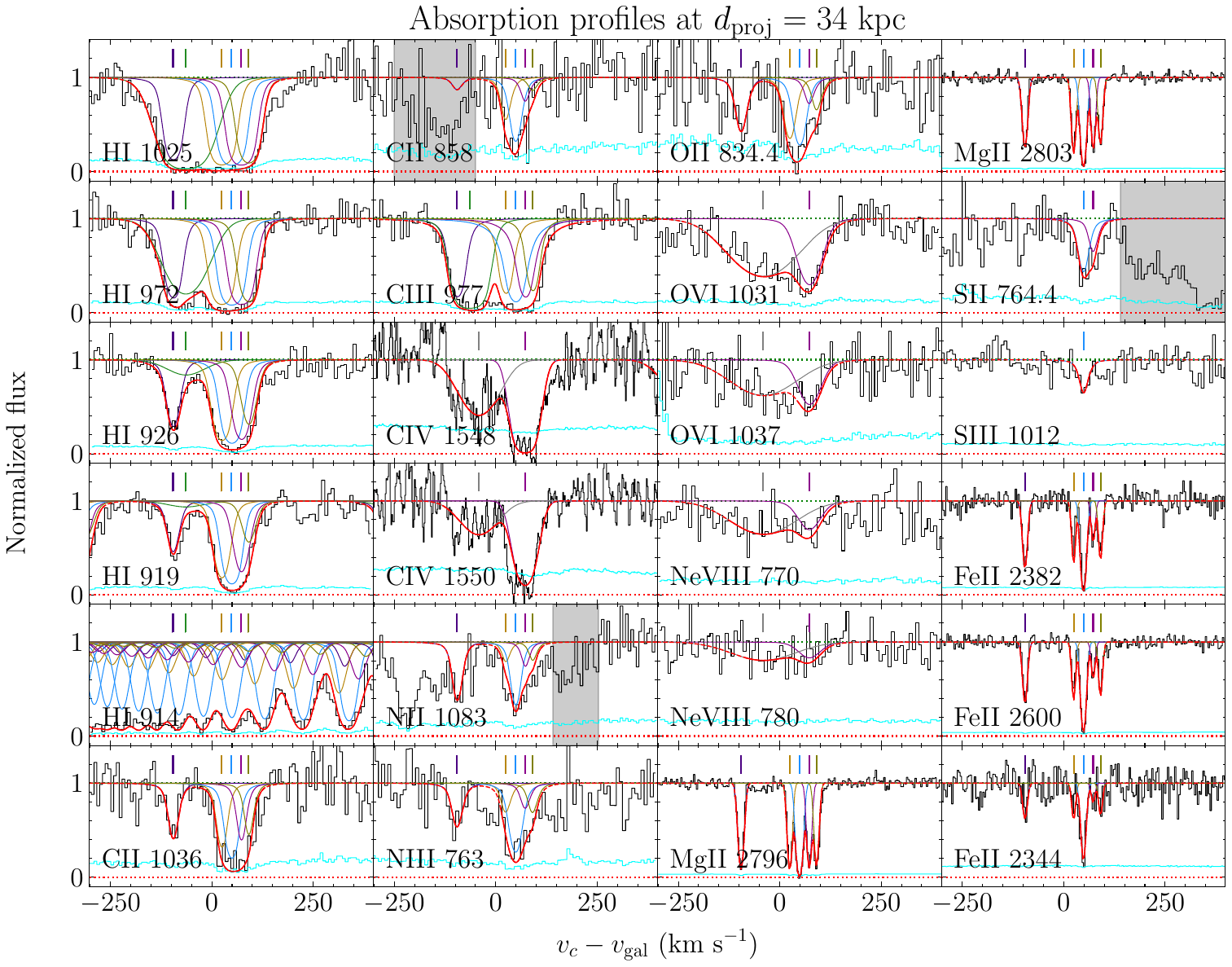}
  \caption{The \textit{HST} F814W image for the host galaxy of system s1 is shown in the \textit{top-left} panel, with the arrow pointing towards the background QSO. The optical MUSE spectrum is shown in the \textit{top-right} panel, with prominent emission and absorption features marked using dotted lines. The associated QSO absorption spectra are shown in the \textit{bottom} panel. The velocity axis is defined relative to the galaxy redshift. The continuum-normalized QSO flux is shown in black, with associated 68\% (1\,$\sigma$) uncertainty indicated in cyan. The best-fit absorption profiles are also included, with individual components distinguished using different colors and the composite profile shown in red. Grayed-out regions of the spectrum are masked during fitting due to significant contamination from interlopers.} \label{fig:s1}
\end{figure*}

This is an isolated, star-forming galaxy at a redshift $z_\mathrm{gal} = 0.5425$ with an estimated stellar mass of $M_\mathrm{star} \approx 4 \times 10^{10} \ \msun$ that was first presented in \cite{Chen:2020,Zahedy:2021}. The associated absorption is a 
Lyman Limit System (LLS) with a total \ion{H}{I} column density of $N(\mathrm{H\ I}) \approx 5 \times 10^{17} \ \mathrm{cm}^{-2}$. This system also exhibits strong absorption in low- and intermediate-ionization species. The component structure is most clearly revealed by the \ion{Mg}{II} and \ion{Fe}{II} transitions covered by available optical echelle spectra from MIKE. The associated \ion{H}{I} can also be resolved into five components kinematically aligned with the \ion{Mg}{II} and \ion{Fe}{II} absorption, but an additional component at $v_c-v_\mathrm{gal} \approx -64 \ \kms$ is required to explain absorption in Ly$\beta$ and Ly$\gamma$. Strong absorption is seen in the \ion{C}{II} $\lambda 858$ and \ion{O}{II} $\lambda 834.4$ transitions, but cannot be decomposed at the COS resolution. The \ion{Mg}{II} line widths are therefore used to guide the \ion{C}{II}, \ion{N}{II}, and \ion{O}{II} line widths. Among intermediate ions, \ion{C}{III} is saturated, but conservative 3\,$\sigma$ lower limits can be placed on the column density. The \ion{N}{III} $\lambda 763$ transition is unsaturated and exhibits a components structure consistent with \ion{N}{II} $\lambda 1083$. Absorption is also identified in \ion{S}{II} and \ion{S}{III} in components c4 and c5 in this system. Among high-ions, strong absorption is identified in \ion{C}{IV}, \ion{O}{VI}, and \ion{Ne}{VIII}. There is absorption coincident with component c5 in these ions, but is distinctly broader, indicating a different physical origin. The blueward absorption in high-ions cannot be resolved into individual components, and is therefore modeled using a single broad component. This component has a large uncertainty on its centroid and may potentially be co-incident with component c2, which has broad \ion{H}{I} and \ion{C}{III} absorption. For the joint Voigt profile analysis involving all available ions, the flux decrement at the Lyman limit is also fit alongside the Lyman series. The best-fit Voigt profiles for this system are presented in the bottom panel of Figure \ref{fig:s1}, and corresponding column density measurements are provided in Table \ref{tab:s1}.

\begin{table}
\caption{Absorption properties along the sightline at $d_\mathrm{proj} = 34$ kpc from system s1.} \label{tab:s1}
\footnotesize\centering
\smallskip
\begin{tabular}{rrr}
\hline
Ion & $\log(N_c/\mathrm{cm}^{-2})$ & $b_c$ (km s$^{-1}$) \\
\hline
\hline
\multicolumn{3}{c}{c1; $v_c - v_\mathrm{gal} = -95.5 \pm 0.1$ km s$^{-1}$} \\
\hline
\ion{H}{I} & $16.12 \pm 0.02$ & $12.2_{-0.3}^{+0.2}$ \\
\ion{C}{II} & $14.3_{-0.2}^{+0.1}$ & $8.4_{-0.6}^{+0.9}$ \\
\ion{C}{III} & $>14.4$ & $8.4$ \footnote{Tied to $b$(\ion{C}{II})} \\
\ion{N}{II} & $14.3 \pm 0.1$ & $8.4$ \\
\ion{N}{III} & $14.3 \pm 0.1$ & $8.4$ \\
\ion{O}{II} & $14.4 \pm 0.2$ & $8.4$ \\
\ion{Mg}{II} & $12.99 \pm 0.01$ & $6.7 \pm 0.1$ \\
\ion{Fe}{II} & $12.99 \pm 0.01$ & $5.7_{-0.4}^{+0.2}$ \\
\hline
\multicolumn{3}{c}{c2; $v_c - v_\mathrm{gal} = -64_{-2}^{+1}$ km s$^{-1}$} \\
\hline
\ion{H}{I} & $15.42_{-0.01}^{+0.01}$ & $58_{-2}^{+1}$ \\
\ion{C}{III} & $>14.3$ & $25_{-1}^{+3}$ \\
\hline
\multicolumn{3}{c}{c3; $v_c - v_\mathrm{gal} = 24.5_{-0.1}^{+0.0}$ km s$^{-1}$} \\
\hline
\ion{H}{I} & $16.72_{-0.02}^{+0.03}$ & $16.1_{-0.2}^{+0.3}$ \\
\ion{C}{II} & $15.1 \pm 0.2$ & $6.7_{-0.4}^{+0.3}$ \\
\ion{C}{III} & $>14.9$ & $6.7$ \\
\ion{N}{II} & $<13.6$\footnote{3\,$\sigma$ upper limit from column density posterior} & $6.7$ \\
\ion{N}{III} & $<14.5$ & $6.7$ \\
\ion{O}{II} & $>14.1$ & $6.7$ \\
\ion{Mg}{II} & $13.07 \pm 0.02$ & $4.9 \pm 0.1$ \\
\ion{Fe}{II} & $12.96_{-0.03}^{+0.04}$ & $3.4_{-0.4}^{+0.2}$ \\
\hline
\multicolumn{3}{c}{c4; $v_c - v_\mathrm{gal} = 48.6_{-0.0}^{+0.1}$ km s$^{-1}$} \\
\hline
\ion{H}{I} & $17.52_{-0.01}^{+0.02}$ & $14 \pm 1$ \\
\ion{C}{II} & $>15.1$ & $[6,10]$ \footnote{3\,$\sigma$ lower and upper limits based on $b$(\ion{Mg}{II})} \\
\ion{C}{III} & $>15.8$ & $[6,10]$ \\
\ion{N}{II} & $14.5 \pm 0.2$ & $[6,10]$ \\
\ion{N}{III} & $>14.3$ & $[6,10]$ \\
\ion{O}{II} & $>11.3$ & $[6,10]$ \\
\ion{Mg}{II} & $13.40_{-0.01}^{+0.02}$ & $6.9 \pm 0.1$ \\
\ion{S}{II} & $13.8 \pm 0.1$ & $6.9$ \\
\ion{S}{III} & $14.21_{-0.06}^{+0.05}$ & $6.9$ \\
\ion{Fe}{II} & $13.60_{-0.03}^{+0.06}$ & $5.8_{-0.3}^{+0.2}$ \\
\hline
\multicolumn{3}{c}{c5; $v_c - v_\mathrm{gal} = 72.6 \pm 0.1$ km s$^{-1}$} \\
\hline
\ion{H}{I} & $16.5 \pm 0.2$ & $17_{-5}^{+6}$ \\
\ion{C}{II} & $14.6_{-0.2}^{+0.1}$ & $[4,7]$ \\
\ion{C}{III} & $>15.8$ & $[4,7]$ \\
\ion{N}{II} & $<13.8$ & $[4,7]$ \\
\ion{N}{III} & $<14.5$ & $[4,7]$ \\
\ion{O}{II} & $<15.2$ & $[4,7]$ \\
\ion{Mg}{II} & $12.99 \pm 0.01$ & $4.8 \pm 0.1$ \\
\ion{S}{II} & $13.26_{-0.08}^{+0.10}$ & $4.8$ \\
\ion{Fe}{II} & $12.59 \pm 0.02$ & $3.3_{-0.6}^{+0.5}$ \\
\hline
\multicolumn{3}{c}{c6; $v_c - v_\mathrm{gal} = 90.7 \pm 0.1$ km s$^{-1}$} \\
\hline
\ion{H}{I} & $16.0 \pm 0.3$ & $16.1_{-0.9}^{+0.8}$ \\
\ion{C}{II} & $14.3 \pm 0.1$ & $6.6_{-0.6}^{+0.5}$ \\
\ion{C}{III} & $>14.8$ & $6.6$ \\
\ion{N}{II} & $13.4_{-0.1}^{+0.2}$ & $6.6$ \\
\ion{N}{III} & $14.0 \pm 0.1$ & $6.6$ \\
\ion{O}{II} & $13.9 \pm 0.2$ & $6.6$ \\
\ion{Mg}{II} & $12.91 \pm 0.01$ & $5.1 \pm 0.1$ \\
\ion{Fe}{II} & $12.87_{-0.03}^{+0.02}$ & $3.7 \pm 0.3$ \\
\hline
\hline
\end{tabular}
\end{table}

For deriving the gas density using photoionization mdoels, the \ion{N}{III}/\ion{N}{II} ratios are well-constrained in components c1 and c6 and suggest an underlying gas density of $\log(n_\mathrm{H}/\mathrm{cm}^{-3}) \approx -2.3$. The \ion{S}{III}/\ion{S}{II} ratio is available for component c4, and indicates a density of $\log(n_\mathrm{H}/\mathrm{cm}^{-3}) \approx -2.1$. For component c2, a density constraint is not available due to non-detections in metal ions with the exception of \ion{C}{III}. For component c3, the absorption in \ion{N}{II} and \ion{N}{III} are weak and are therefore regarded as upper limits. The corresponding \ion{C}{II} column density is well-constrained due to coverage in the \ion{C}{II} $\lambda 1036, \lambda 858$ transitions, and can be used with the \ion{C}{III} lower limit to obtain a corresponding 3\,$\sigma$ upper limit on the gas density, $\log(n_\mathrm{H}/\mathrm{cm}^{-3}) < -1.6$. Meanwhile, the upper limit on the \ion{S}{III} can be used with the \ion{Mg}{II} detection (assuming $\mathrm{[S/Mg]}=0$) to obtain a lower limit on the gas density,  $\log(n_\mathrm{H}/\mathrm{cm}^{-3}) > -2.8$. Fitting all ions together while accommodating non-solar abundance patterns yields a density of $\log(n_\mathrm{H}/\mathrm{cm}^{-3}) \approx -2.3$ with an associated uncertainty of $0.4 \ \mathrm{dex}$. For component c5, available limits on the ion ratios \ion{Mg}{II}/\ion{Mg}{I} and \ion{S}{III}/\ion{S}{II} yield a density constraint of $\log(n_\mathrm{H}/\mathrm{cm}^{-3}) \approx -1.7$ with an associated uncertainty of $0.3 \ \mathrm{dex}$. This solution is unable to reproduce the observed \ion{C}{IV}, \ion{O}{VI}, and \ion{Ne}{VIII} columns in this component. The presence of broad \ion{Ne}{VIII} absorption with $b \approx 30 \ \kms$ likely suggests a collisionally ionized origin for the high-ions. Ionization modeling constraints for individual components are presented in Table \ref{tab:ion_model}.

To infer the underlying gas temperature and non-thermal motions, the measured line widths for \ion{H}{I} and \ion{Mg}{II} are used. Derived gas temperatures range between $\log(T/\mathrm{K}) \approx 3.8-4.2$, with modest non-thermal broadening of $b_\mathrm{NT} \lesssim 6 \ \kms$. Note that the line widths for \ion{C}{II}, \ion{O}{II}, and \ion{N}{II} are not used for this exercise since the broad LSF of COS does not provide resolution for components with $b \lesssim 12 \ \kms$. The \ion{Fe}{II} line widths, on the other hand, are found to be too narrow for the $T,\ b_\mathrm{NT}$ reported in Table \ref{tab:ion_model} that are constrained using \ion{H}{I} and \ion{Mg}{II} absorption.

\subsection{$\mathrm{CUBS \ J213553.27\!-\!531700.70}$ at $d_\mathrm{proj}=35 \ \mathrm{kpc}$} \label{sec:s2}

\begin{table}
\caption{Absorption properties along the sightline at $d_\mathrm{proj} = 35$ kpc from system s2.} \label{tab:s2}
\footnotesize\centering
\smallskip
\begin{tabular}{rrr}
\hline
Ion & $\log(N_c/\mathrm{cm}^{-2})$ & $b_c$ (km s$^{-1}$) \\
\hline
\hline
\multicolumn{3}{c}{c1; $v_c - v_\mathrm{gal} = 5 \pm 3$ km s$^{-1}$} \\
\hline
\ion{H}{I} & $14.47 \pm 0.03$ & $43_{-5}^{+6}$ \\
\ion{C}{IV} & $<13.4$ & $6$ \\
\ion{O}{VI} & $13.89 \pm 0.07$ & $41_{-10}^{+14}$ \\
\ion{Mg}{II} & $<11.0$ & $4$ \\
\hline
\end{tabular}
\end{table}

This is an isolated, star-forming dwarf galaxy at $z_\mathrm{gal}=0.6768$ with a stellar mass of $M_\mathrm{star} \approx 10^8 \ \msun$. Associated absorption is identified in \ion{H}{I} and \ion{O}{VI} but not for any low- or intermediate-ionization species. Although a density estimate cannot be derived due to the lack of sufficient constraints from multiple ionic species, the gas is likely diffuse with $n_\mathrm{H} < 10^{-4} \ \mathrm{cm}^{-3}$ under a photoionization scenario due to the presence of \ion{O}{VI}.

\subsection{$\mathrm{CUBS \ J011139.36\!-\!031605.19}$ at $d_\mathrm{proj}=42 \ \mathrm{kpc}$} \label{sec:s3}

\begin{table}
\caption{Integrated absorption properties along the sightline at $d_\mathrm{proj} = 42$ kpc from system s3.} \label{tab:s3}
\footnotesize\centering
\smallskip
\begin{tabular}{rr}
\hline
Ion & $\log(N_\mathrm{tot}/\mathrm{cm}^{-2})$ \\
\hline
\ion{H}{I} & $20.1\pm0.1$  \\
\ion{C}{IV} & $<12.7$ \\
\ion{O}{VI} & $<13.7$\\
\ion{Mg}{II} & $14.56_{-0.05}^{+0.06}$ \\
\hline
\end{tabular}
\end{table}

This is a quiescent galaxy at $z_\mathrm{gal}=0.5766$ with a stellar mass of $M_\mathrm{star} \approx 10^{11} \msun$ first reported by \cite{Boettcher:2021}. Additional satellite galaxies are identified near this host, extended out to $d_\mathrm{proj} = 220 \ \mathrm{kpc}$. The associated absorption is a Damped Lyman-$\alpha$ system (DLA) with a total \ion{H}{I} column of $N(\mathrm{H \ I}) \approx 10^{20} \ \mathrm{cm}^{-2}$ bearing molecular H$_2$. The \ion{Mg}{II} doublet indicates ten distinct kinematic components. However, these components cannot be resolved in \ion{H}{I} absorption due to saturation. For this reason, only integrated absorption constraints are reported in Table \ref{tab:s3}. Due to the important role of the ISM radiation field in setting the gas ionization state, UVB-only photoionization models used in this work are not suitable for this system, and the reader is referred to \cite{Boettcher:2021} for a detailed examination. 

\subsection{$\mathrm{CUBS \ J015455.18\!-\!071222.15}$ at $d_\mathrm{proj}=43 \ \mathrm{kpc}$} \label{sec:s4}

\begin{table}
\caption{Absorption properties along the sightline at $d_\mathrm{proj} = 43$ kpc from system s4.} \label{tab:s4}
\footnotesize\centering
\smallskip
\begin{tabular}{rrr}
\hline
Ion & $\log(N_c/\mathrm{cm}^{-2})$ & $b_c$ (km s$^{-1}$) \\
\hline
\hline
\multicolumn{3}{c}{c1; $v_c - v_\mathrm{gal} = 467 \pm 3$ km s$^{-1}$} \\
\hline
\ion{H}{I} & $13.90 \pm 0.03$ & $46_{-4}^{+5}$ \\
\ion{C}{IV} & $<12.8$ & $6$ \\
\ion{O}{VI} & $<13.1$ & $30$ \\
\ion{Mg}{II} & $<11.3$ & $4$ \\
\hline
\multicolumn{3}{c}{c2; $v_c - v_\mathrm{gal} = 592 \pm 2$ km s$^{-1}$} \\
\hline
\ion{H}{I} & $13.99 \pm 0.03$ & $33 \pm 3$ \\
\ion{C}{IV} & $<13.1$ & $6$ \\
\ion{O}{VI} & $<13.6$ & $30$ \\
\ion{Mg}{II} & $<11.5$ & $4$ \\
\hline
\end{tabular}
\end{table}

This is an isolated, star-forming dwarf galaxy at $z_\mathrm{gal} = 0.4550$ with stellar mass $M_\mathrm{star} \approx 10^8 \msun$. The associated \ion{H}{I} detection can be ascertained with Ly$\alpha$ and Ly$\beta$ absorption. However, no metal absorption is identified. 

\subsection{$\mathrm{CUBS \ J011035.01\!-\!164833.32}$ at $d_\mathrm{proj}=54 \ \mathrm{kpc}$} \label{sec:s5}

\begin{table}
\caption{Absorption properties along the sightline at $d_\mathrm{proj} = 54$ kpc from system s5.} \label{tab:s5}
\footnotesize\centering
\smallskip
\begin{tabular}{rrr}
\hline
Ion & $\log(N_c/\mathrm{cm}^{-2})$ & $b_c$ (km s$^{-1}$) \\
\hline
\hline
\multicolumn{3}{c}{c1; $v_c - v_\mathrm{gal} = -84 \pm 3$ km s$^{-1}$} \\
\hline
\ion{H}{I} & $15.00 \pm 0.04$ & $39 \pm 2$ \\
\hline
\multicolumn{3}{c}{c2; $v_c - v_\mathrm{gal} = -22.4_{-0.8}^{+0.6}$ km s$^{-1}$} \\
\hline
\ion{H}{I} & $16.04_{-0.03}^{+0.02}$ & $25 \pm 1$ \\
\ion{C}{II} & $13.28 \pm 0.04$ & $14$ \\
\ion{C}{III} & $>13.8$ & $34$ \\
\ion{O}{II} & $13.80_{-0.04}^{+0.05}$ & $14_{-2}^{+1}$ \\
\ion{O}{III} & $14.33_{-0.05}^{+0.03}$ & $34_{-2}^{+1}$ \\
\ion{O}{IV} & $14.06_{-0.03}^{+0.04}$ & $30$ \\
\ion{O}{VI} & $13.89 \pm 0.04$ & $30_{-2}^{+3}$ \\
\ion{Mg}{II} & $12.11_{-0.07}^{+0.06}$ & $12_{-1}^{+2}$ \\
\ion{Si}{III} & $12.81_{-0.07}^{+0.08}$ & $12$ \\
\hline
\multicolumn{3}{c}{c3; $v_c - v_\mathrm{gal} = 31.4 \pm 0.5$ km s$^{-1}$} \\
\hline
\ion{H}{I} & $14.90_{-0.05}^{+0.07}$ & $25 \pm 1$ \\
\ion{C}{II} & $13.22_{-0.07}^{+0.05}$ & $20$ \\
\ion{C}{III} & $>13.7$ & $18$ \\
\ion{C}{IV} & $13.80_{-0.03}^{+0.04}$ & $18_{-2}^{+1}$ \\
\ion{N}{III} & $13.53 \pm 0.04$ & $18_{-2}^{+1}$ \\
\ion{N}{IV} & $13.69 \pm 0.05$ & $18$ \\
\ion{O}{II} & $13.51_{-0.06}^{+0.07}$ & $20 \pm 3$ \\
\ion{O}{III} & $14.22 \pm 0.03$ & $18$ \\
\ion{O}{IV} & $14.50_{-0.04}^{+0.06}$ & $18$ \\
\ion{O}{VI} & $13.98_{-0.03}^{+0.02}$ & $18$ \\
\ion{Si}{III} & $12.90_{-0.08}^{+0.07}$ & $15_{-2}^{+3}$ \\
\hline
\multicolumn{3}{c}{c4; $v_c - v_\mathrm{gal} = 165.9_{-0.5}^{+0.6}$ km s$^{-1}$} \\
\hline
\ion{H}{I} & $15.09_{-0.02}^{+0.03}$ & $17_{-1}^{+2}$ \\
\ion{C}{II} & $13.13_{-0.08}^{+0.06}$ & $6$ \\
\ion{C}{III} & $>13.8$ & $13$ \\
\ion{C}{IV} & $13.88 \pm 0.04$ & $19 \pm 2$ \\
\ion{N}{III} & $13.66_{-0.07}^{+0.06}$ & $13_{-2}^{+3}$ \\
\ion{N}{IV} & $13.59_{-0.06}^{+0.05}$ & $19$ \\
\ion{O}{II} & $13.77_{-0.05}^{+0.06}$ & $6_{-1}^{+2}$ \\
\ion{O}{III} & $14.07 \pm 0.04$ & $13$ \\
\ion{O}{IV} & $14.42_{-0.03}^{+0.04}$ & $19$ \\
\ion{O}{VI} & $14.10 \pm 0.02$ & $19$ \\
\ion{Mg}{II} & $12.08 \pm 0.06$ & $5 \pm 1$ \\
\ion{Si}{III} & $13.0 \pm 0.3$ & $5$ \\
\hline
\multicolumn{3}{c}{c5; $v_c - v_\mathrm{gal} = 251 \pm 2$ km s$^{-1}$} \\
\hline
\ion{H}{I} & $14.09_{-0.03}^{+0.06}$ & $34 \pm 3$ \\
\hline
\end{tabular}
\end{table} 

This is a non-isolated, star-forming galaxy at $z_\mathrm{gal} = 0.4724$ with stellar mass $M_\mathrm{star} \approx 10^{10} \msun$ first identified in \cite{Cooper:2021}. The associated absorption is a partial Lyman Limit System (pLLS) with a neutral hydrogen column density of $N(\mathrm{H \ I}) \approx 10^{16} \ \mathrm{cm}^{-2}$. \ion{O}{II} $\lambda 834$ and \ion{C}{II} $\lambda \lambda 903.9, 903.6$ can be resolved into three components (c2, c3, and c4). Among intermediate ions, \ion{Si}{III} $\lambda 1206$ exhibits kinematically aligned absorption with \ion{O}{II}. \ion{C}{III} $\lambda 977$ also shows absorption along these components, but is saturated. Absorption in these components can be recovered in \ion{O}{III} $\lambda 832$ after removing blended contributions from from the \ion{O}{II} $\lambda 833.3, 832.7$ doublet. \ion{N}{III} $\lambda 989$ shows absorption only along c3 and c4. Among high-ions, the \ion{C}{IV} $\lambda \lambda 1548, 1550$ show narrow ($b \approx 20 \ \kms$), strong absorption in components c3 and c4. A similar component structure is also noted in \ion{N}{IV} $\lambda 765$. \ion{O}{IV} $\lambda 787$ and the \ion{O}{VI} $\lambda \lambda 1031, 1037$ doublet exhibit absorption along c2, in addition to c3 and c4. The absorption in c3 and c4 exhibit consistent line widths with with \ion{C}{IV}, but the absorption in c2 is relatively broad ($b \approx 30 \ \kms$). The \ion{H}{I} absorption can be resolved for components c2, c3, and c4, although non-metal bearing components (c1 and c5) are required to explain the Ly$\alpha$ and Ly$\beta$ absorption at $v_c-v_\mathrm{gal} \approx -84 \ \kms, 251 \ \kms$.

For ionization modeling, the measured \ion{O}{III}/\ion{O}{II} ratios in components c2 through c4 suggest an underlying gas density of $\log(n_\mathrm{H}/\mathrm{cm}^{-3}) \approx -2.6$, reproducing the observed column densities for low- and intermediate-ions. However, this gas phase is unable to reproduce the measured high-ion column densities in these components. An additional, low-density phase with $\log(n_\mathrm{H}/\mathrm{cm}^{-3}) \approx -4.4$ determined by the measured \ion{O}{VI}/\ion{O}{IV} ratio is required alongside the denser phase to fit the measured column densities of \ion{O}{II} through \ion{O}{VI} in components c2, c3, and c4.

\subsection{$\mathrm{CUBS \ J233912.33\!-\!552354.27}$ at $d_\mathrm{proj}=55 \ \mathrm{kpc}$} \label{sec:s6}

\begin{table}
\caption{Absorption properties along the sightline at $d_\mathrm{proj} = 55$ kpc from system s6.} \label{tab:s6}
\footnotesize\centering
\smallskip
\begin{tabular}{rrr}
\hline
Ion & $\log(N_c/\mathrm{cm}^{-2})$ & $b_c$ (km s$^{-1}$) \\
\hline
\hline
\multicolumn{3}{c}{c1; $v_c - v_\mathrm{gal} = -9.9 \pm 0.8$ km s$^{-1}$} \\
\hline
\ion{H}{I} & $14.84 \pm 0.04$ & $36 \pm 4$ \\
\ion{C}{III} & $13.5_{-0.1}^{+0.2}$ & $10$ \\
\ion{C}{IV} & $13.64 \pm 0.05$ & $10 \pm 2$ \\
\ion{N}{IV} & $13.16 \pm 0.07$ & $10$ \\
\ion{O}{III} & $13.85 \pm 0.07$ & $10$ \\
\ion{O}{VI} & $14.43_{-0.03}^{+0.04}$ & $22 \pm 3$ \\
\hline
\multicolumn{3}{c}{c2; $v_c - v_\mathrm{gal} = 47.7 \pm 0.7$ km s$^{-1}$} \\
\hline
\ion{H}{I} & $15.23 \pm 0.02$ & $28 \pm 1$ \\
\ion{C}{III} & $>13.9$ & $27$ \\
\ion{C}{IV} & $13.93 \pm 0.03$ & $27 \pm 2$ \\
\ion{N}{IV} & $13.37 \pm 0.04$ & $27$ \\
\ion{O}{III} & $14.35 \pm 0.03$ & $27$ \\
\ion{O}{VI} & $14.40 \pm 0.03$ & $32_{-3}^{+4}$ \\
\hline
\end{tabular}
\end{table}

This is an isolated, star-forming dwarf galaxy at $z_\mathrm{gal}=0.5726$ with a stellar mass of $M_\mathrm{star} \approx 3 \times 10^8 \msun$. Associated absorption in \ion{C}{IV} $\lambda 1550$, \ion{N}{IV} $\lambda 765$, and \ion{O}{III} $\lambda 832$ exhibit a distinct two-component structure, with component c1 being narrower ($b \approx 10 \ \kms$) than c2 ($b \approx 25 \ \kms$). \ion{C}{IV} $\lambda 1548$ and \ion{O}{IV} $\lambda 787$ are contaminated. No absorption is seen among low-ions. The \ion{O}{VI} $\lambda 1031, 1037$ can also be decomposed into c1 and c2, although the line widths are broader than the other ions ($b \approx 20 \ \kms, 30 \ \kms$), suggesting a distinct physical origin. The \ion{H}{I} absorption in these two components can be resolved using multiple Lyman series lines that are available.

For ionization modeling of c1, the \ion{C}{IV}/\ion{C}{III} ratio suggests a density of $\log(n_\mathrm{H}/\mathrm{cm}^{-3}) \approx -3.9$, but this phase underpredicts the observed \ion{O}{VI}. For c2, the \ion{C}{III} absorption is saturated, but the available limits on the \ion{C}{IV}/\ion{C}{III} and \ion{N}{IV}/\ion{N}{III} ratio (where \ion{N}{III} is a non-detection) constrain a density of $\log(n_\mathrm{H}/\mathrm{cm}^{-3}) \approx -3.6$. The observed \ion{O}{VI} is also not reproduced by this phase. 

\subsection{$\mathrm{CUBS \ J224500.93\!-\!493154.29}$ at $d_\mathrm{proj}=58 \ \mathrm{kpc}$} \label{sec:s7}

This is an isolated, star-forming dwarf galaxy at $z_\mathrm{gal}=0.5338$ with stellar mass $M_\mathrm{star}=10^8 \msun$. The associated absorption is only identified in \ion{H}{I}, with no detections for metal ions.

\begin{table}
\caption{Absorption properties along the sightline at $d_\mathrm{proj} = 58$ kpc from system s7.} \label{tab:s7}
\footnotesize\centering
\smallskip
\begin{tabular}{rrr}
\hline
Ion & $\log(N_c/\mathrm{cm}^{-2})$ & $b_c$ (km s$^{-1}$) \\
\hline
\hline
\multicolumn{3}{c}{c1; $v_c - v_\mathrm{gal} = -129.1 \pm 0.9$ km s$^{-1}$} \\
\hline
\ion{H}{I} & $15.18 \pm 0.04$ & $17 \pm 1$ \\
\ion{C}{IV} & $<12.7$ & $6$ \\
\ion{O}{VI} & $<13.9$ & $30$ \\
\ion{Mg}{II} & $<11.7$ & $4$ \\
\hline
\end{tabular}
\end{table}

\subsection{$\mathrm{CUBS \ J011138.57\!-\!031609.64}$ at $d_\mathrm{proj}=63 \ \mathrm{kpc}$} \label{sec:s8}

This is an isolated, quiescent galaxy at $z_\mathrm{gal}=0.6482$ with stellar mass $M_\mathrm{star} \approx 10^{11} \msun$. The sightline probing this galaxy is transparent, with no associated absorption identified in \ion{H}{I} or metal ions.

\begin{table}[t!]
\caption{Absorption properties along the sightline at $d_\mathrm{proj} = 63$ kpc from system s8.} \label{tab:s8}
\footnotesize\centering
\smallskip
\begin{tabular}{rrr}
\hline
Ion & $\log(N_c/\mathrm{cm}^{-2})$ & $b_c$ (km s$^{-1}$) \\
\hline
\hline
\multicolumn{3}{c}{c1; $v_c - v_\mathrm{gal} = 0$ km s$^{-1}$} \\
\hline
\ion{H}{I} & $<13.6$ & $12$ \\
\ion{C}{IV} & $<12.3$ & $6$ \\
\ion{O}{VI} & $<13.7$ & $30$ \\ 
\ion{Mg}{II} & $<11.1$ & $4$ \\
\hline
\end{tabular}
\end{table}

\subsection{$\mathrm{CUBS \ J213552.98\!-\!531645.51}$ at $d_\mathrm{proj} = 65 \ \mathrm{kpc}$} \label{sec:s9}

This is an isolated, star-forming dwarf galaxy at $z_\mathrm{gal}=0.5082$ with a stellar mass $M_\mathrm{star} \approx 10^8 \msun$, The associated absorption is identified in \ion{H}{I} and \ion{O}{VI}, but no other ions.

\begin{table}
\caption{Absorption properties along the sightline at $d_\mathrm{proj} = 65$ kpc from system s9.} \label{tab:s9}
\footnotesize\centering
\smallskip
\begin{tabular}{rrr}
\hline
Ion & $\log(N_c/\mathrm{cm}^{-2})$ & $b_c$ (km s$^{-1}$) \\
\hline
\hline
\multicolumn{3}{c}{c1; $v_c - v_\mathrm{gal} = 10 \pm 2$ km s$^{-1}$} \\
\hline
\ion{H}{I} & $14.29 \pm 0.03$ & $39_{-4}^{+5}$ \\
\ion{C}{IV} & $<12.9$ & $6$ \\
\ion{O}{VI} & $13.91 \pm 0.05$ & $31_{-6}^{+10}$ \\
\ion{Mg}{II} & $<11.2$ & $4$\\
\hline
\end{tabular}
\end{table}

\subsection{$\mathrm{CUBS \ J035723.09\!-\!481215.82}$ at $d_\mathrm{proj} = 66 \ \mathrm{kpc}$} \label{sec:s10}

\begin{table}[t!]
\caption{Absorption properties along the sightline at $d_\mathrm{proj} = 66$ kpc from system s10.} \label{tab:s10}
\footnotesize\centering
\smallskip
\begin{tabular}{rrr}
\hline
Ion & $\log(N_c/\mathrm{cm}^{-2})$ & $b_c$ (km s$^{-1}$) \\
\hline
\hline
\multicolumn{3}{c}{c1; $v_c - v_\mathrm{gal} = -141.7_{-0.7}^{+0.8}$ km s$^{-1}$} \\
\hline
\ion{H}{I} & $16.10_{-0.03}^{+0.02}$ & $21.6_{-0.4}^{+0.7}$ \\
\ion{C}{II} & $13.34_{-0.04}^{+0.07}$ & $21_{-3}^{+4}$ \\
\ion{C}{III} & $>13.5$ & $15$ \\
\ion{N}{III} & $13.36_{-0.08}^{+0.05}$ & $15 \pm 1$ \\
\ion{O}{II} & $13.80_{-0.03}^{+0.06}$ & $21$ \\
\ion{O}{III} & $13.94_{-0.05}^{+0.04}$ & $15$ \\
\ion{O}{VI} & $13.77_{-0.05}^{+0.07}$ & $36_{-6}^{+11}$ \\
\ion{Si}{III} & $12.79_{-0.06}^{+0.08}$ & $16_{-2}^{+3}$ \\
\hline
\multicolumn{3}{c}{c2; $v_c - v_\mathrm{gal} = -102.4 \pm 0.4$ km s$^{-1}$} \\
\hline
\ion{H}{I} & $17.10 \pm 0.04$ & $17 \pm 1$ \\
\ion{C}{II} & $13.49_{-0.06}^{+0.07}$ & $9_{-2}^{+1}$ \\
\ion{C}{III} & $>13.7$ & $9$ \\
\ion{C}{IV} & $<13.9$ & $19_{-3}^{+4}$ \\
\ion{N}{III} & $13.63_{-0.04}^{+0.05}$ & $9 \pm 1$ \\
\ion{N}{IV} & $<14.2$ & $19$ \\
\ion{O}{II} & $13.92_{-0.05}^{+0.07}$ & $9$ \\
\ion{O}{III} & $>14.6$ & $9$ \\
\ion{O}{IV} & $14.1 \pm 0.1$ & $19$ \\
\ion{O}{VI} & $13.74_{-0.06}^{+0.08}$ & $19$ \\
\ion{Mg}{II} & $12.18 \pm 0.03$ & $7.5 \pm 0.9$ \\
\ion{Si}{III} & $>12.6$ & $7.5$ \\
\hline
\multicolumn{3}{c}{c3; $v_c - v_\mathrm{gal} = -69.6_{-0.4}^{+0.7}$ km s$^{-1}$} \\
\hline
\ion{H}{I} & $16.16_{-0.05}^{+0.03}$ & $17.1_{-0.6}^{+0.8}$ \\
\ion{C}{II} & $13.55_{-0.06}^{+0.04}$ & $9 \pm 1$ \\
\ion{C}{III} & $>13.8$ & $10$ \\
\ion{C}{IV} & $14.01_{-0.08}^{+0.06}$ & $14_{-3}^{+2}$ \\
\ion{N}{III} & $13.97_{-0.04}^{+0.06}$ & $10 \pm 1$ \\
\ion{N}{IV} & $>13.6$ & $14$ \\
\ion{N}{V} & $13.5 \pm 0.1$ & $14$ \\
\ion{O}{II} & $13.78_{-0.07}^{+0.04}$ & $9$ \\
\ion{O}{III} & $>14.4$ & $10$ \\
\ion{O}{IV} & $14.7 \pm 0.1$ & $14$ \\
\ion{O}{VI} & $13.8 \pm 0.1$ & $14$ \\
\ion{Mg}{II} & $12.33_{-0.04}^{+0.03}$ & $7.0_{-0.7}^{+0.5}$ \\
\ion{Si}{III} & $>13.3$ & $7.0$ \\
\ion{Fe}{II} & $12.36_{-0.08}^{+0.10}$ & $5 \pm 1$ \\
\hline
\multicolumn{3}{c}{c4; $v_c - v_\mathrm{gal} = -40 \pm 2$ km s$^{-1}$} \\
\hline
\ion{H}{I} & $14.35_{-0.09}^{+0.07}$ & $28_{-4}^{+6}$ \\
\ion{C}{III} & $>13.4$ & $17$ \\
\ion{C}{IV} & $<13.7$ & $21_{-5}^{+3}$ \\
\ion{N}{III} & $13.64_{-0.04}^{+0.05}$ & $17 \pm 2$ \\
\ion{N}{IV} & $<14.1$ & $21$ \\
\ion{O}{III} & $13.8 \pm 0.1$ & $17$ \\
\ion{O}{IV} & $14.19 \pm 0.09$ & $21$ \\
\ion{O}{VI} & $14.00_{-0.05}^{+0.06}$ & $21$ \\
\ion{Si}{III} & $12.6 \pm 0.1$ & $14_{-2}^{+3}$ \\
\hline
\multicolumn{3}{c}{c5; $v_c - v_\mathrm{gal} = 37_{-4}^{+5}$ km s$^{-1}$} \\
\hline
\ion{H}{I} & $13.73_{-0.08}^{+0.06}$ & $32_{-6}^{+5}$ \\
\hline
\multicolumn{3}{c}{c6; $v_c - v_\mathrm{gal} = 153.8_{-0.5}^{+0.9}$ km s$^{-1}$} \\
\hline
\ion{H}{I} & $13.95_{-0.03}^{+0.04}$ & $21 \pm 2$ \\
\ion{C}{III} & $13.17_{-0.03}^{+0.02}$ & $17 \pm 2$ \\
\ion{O}{III} & $13.80_{-0.04}^{+0.07}$ & $17$ \\
\ion{O}{IV} & $13.9 \pm 0.1$ & $16$ \\
\ion{O}{VI} & $13.48 \pm 0.06$ & $16_{-4}^{+5}$ \\
\hline
\end{tabular}
\end{table}

This is an isolated, star-forming galaxy at $z_\mathrm{gal} = 0.4357$ with stellar mass of $M_\mathrm{star} \approx 5 \times 10^{10} \msun$ first identified in \cite{Chen:2020,Zahedy:2021}. The associated absorption is a LLS with a neutral hydrogen column density of $N(\mathrm{H \ I}) \approx 2 \times 10^{17} \ \mathrm{cm}^{-2}$. The \ion{Mg}{II} doublet absorption exhibits two components, but \ion{O}{II} $\lambda 834$ and \ion{C}{II} $\lambda 1036$ transitions show excess blueward absorption. The non-\ion{Mg}{II}-bearing component can be resolved at $v_c-v_\mathrm{gal} \approx -22 \ \kms$ using the blended \ion{C}{II} $\lambda \lambda 903.9, 903.6$, \ion{C}{II} $\lambda 1036$, and \ion{O}{II} $\lambda 834$. Among intermediate ions, the \ion{C}{III} $\lambda 977$ transition is saturated, but the \ion{N}{III} $\lambda 989$ transition is clean. The \ion{N}{III} absorption exhibits a component structure resembling \ion{C}{II} and \ion{O}{II}, but has excess redward absorption, suggesting the presence of an additional component. The four-component structure in \ion{N}{III} is also reflected in \ion{Si}{III}, and can be accommodated in the \ion{O}{III} $\lambda 832$ absorption after removing contributions from \ion{O}{II} $\lambda \lambda 833.3, 832.7$ based on a fit from \ion{O}{II} $\lambda 834$. Among high-ions, the \ion{C}{IV} $\lambda \lambda 1548, 1550$ doublet and \ion{O}{IV} $\lambda 787$ transitions show absorption in components c2, c3, and c4, with the narrow line widths for \ion{C}{IV} ($b \lesssim 20 \ \kms$) resolved by STIS data for the \ion{C}{IV} doublet. Absorption in the \ion{O}{VI} $\lambda \lambda 1031, 1037$ doublet can be decomposed into components c1 through c4, with line widths in c2 through c4 tied to the \ion{C}{IV} absorption. The \ion{O}{VI} absorption in c1 is broader ($b \approx 35 \ \kms$). The \ion{N}{IV} $\lambda 765$ transition is noisy, but column density lower limits can still be derived tying the absorption line widths to \ion{C}{IV}. \ion{N}{V} $\lambda \lambda 1238, 1242$ absorption is also tentatively identified along component c3, although the spectrum is noisy. \ion{H}{I} absorption in this system can be resolved for components c1 through c4, although excess absorption is observed in Ly$\alpha$ and Ly$\beta$, requiring a non-metal bearing at $v_c-v_\mathrm{gal} \approx 37 \ \kms$. An \ion{H}{I} component, c6, is also noted at $v_c-v_\mathrm{gal} \approx 154 \ \kms$, redward of the absorption complex, with detections in \ion{C}{III}, \ion{O}{III}, \ion{O}{IV}, and \ion{O}{VI}.

The gas in component c1 can by modeled by a single photoionized phase of density $\log(n_\mathrm{H}/\mathrm{cm}^{-3}) \approx -2.4$, according to the measured \ion{O}{III}/\ion{O}{II} ratio. However, components c2, c3, c4, and c6 kinematically aligned low-, intermediate-, and high-ions, making a single phase solution incorrect. For components c2 and c3, the \ion{O}{II} column is well-constrained while \ion{O}{III} is saturated. Combined with the \ion{S}{III} non-detection, limits on the \ion{O}{III}/\ion{O}{II} and \ion{S}{III}/\ion{O}{II} ratios (assuming $\mathrm{[S/O]}=0$, although the \ion{Fe}{III}/\ion{Fe}{II} limit is more constraining for c3) suggest a gas density of $\log(n_\mathrm{H}/\mathrm{cm}^{-3}) \approx -2.7$ to reproduce the low- and intermediate ions. This solution cannot reproduce the observed high-ion columns in these components. Therefore, a low-density phase of $\log(n_\mathrm{H}/\mathrm{cm}^{-3}) \approx -4.3$ constrained by the \ion{O}{VI}/\ion{O}{IV} ratio is required to reproduce the highly ionized species in components c2 and c3. In components c4 and c6, a low-density phase constrained by \ion{O}{IV} and \ion{O}{VI} underpredicts the measured \ion{O}{III} column density, also necessitating a two-phase solution. The high-density phase is poorly constrained for components c4 and c6, with an uncertainty of $0.3 \ \mathrm{dex}$ due to the non-detection of singly-ionized species, implying that the density is largely constrained by the sensitive \ion{O}{III}/\ion{Mg}{II} limit (assuming $\mathrm{[Mg/O]}=0$), in addition to requiring the low-density phase to produce more \ion{O}{IV} relative to the high-density phase (effectively treating the \ion{O}{IV}/\ion{O}{III} ratio as an upper limit for the denser phase).

\subsection{$\mathrm{CUBS \ J035721.44\!-\!481225.86}$ at $d_\mathrm{proj}=67 \ \mathrm{kpc}$} \label{sec:s11}

\begin{table}[t!]
\caption{Absorption properties along the sightline at $d_\mathrm{proj} = 67$ kpc from system s11.} \label{tab:s11}
\footnotesize\centering
\smallskip
\begin{tabular}{rrr}
\hline
Ion & $\log(N_c/\mathrm{cm}^{-2})$ & $b_c$ (km s$^{-1}$) \\
\hline
\hline
\multicolumn{3}{c}{c1; $v_c - v_\mathrm{gal} = -8.2 \pm 0.8$ km s$^{-1}$} \\
\hline
\ion{H}{I} & $14.17_{-0.05}^{+0.04}$ & $46_{-6}^{+7}$ \\
\ion{C}{III} & $13.74_{-0.02}^{+0.03}$ & $22 \pm 1$ \\
\hline
\multicolumn{3}{c}{c2; $v_c - v_\mathrm{gal} = 54_{-2}^{+1}$ km s$^{-1}$} \\
\hline
\ion{H}{I} & $14.32 \pm 0.03$ & $37_{-4}^{+6}$ \\
\ion{C}{III} & $13.12 \pm 0.04$ & $20$ \\
\ion{C}{IV} & $13.1 \pm 0.1$ & $20$ \\
\ion{O}{IV} & $14.15 \pm 0.04$ & $20 \pm 3$ \\
\ion{O}{VI} & $14.00 \pm 0.04$ & $22 \pm 3$ \\
\hline
\end{tabular}
\end{table} 

This is an isolated, star-forming dwarf galaxy at $z_\mathrm{gal} = 0.4489$ with stellar mass $M_\mathrm{star} \approx 2 \times 10^8 \msun$. Absorption in a single component is most clearly identified in \ion{O}{IV} $\lambda 787$, and is also recovered in the \ion{C}{IV} $\lambda \lambda 1548, 1550$ doublet. The corresponding \ion{H}{I} absorption requires an additional blueward component c1 with the stronger component along c2. The \ion{C}{III} absorption can be resolved into two components along c1 and c2. The \ion{O}{VI} doublet is contaminated, although the blue wing of the \ion{O}{VI} $\lambda 1031$ can accommodate a component of line width consistent with \ion{O}{IV} at c2. An ionization model is not available for c1, but for c2 a single phase of density $\log(n_\mathrm{H}/\mathrm{cm}^{-3}) \approx -4.0$ is suggested by the \ion{C}{IV}/\ion{C}{III} ratio. This model underpredicts the observed \ion{O}{VI} column density in c2.

\subsection{$\mathrm{CUBS \ J042052.57\!-\!565038.97}$ at $d_\mathrm{proj}=80 \ \mathrm{kpc}$} \label{sec:s12}

\begin{table}[t!]
\caption{Absorption properties along the sightline at $d_\mathrm{proj} = 80$ kpc from system s12.} \label{tab:s12}
\footnotesize\centering
\smallskip
\begin{tabular}{rrr}
\hline
Ion & $\log(N_c/\mathrm{cm}^{-2})$ & $b_c$ (km s$^{-1}$) \\
\hline
\hline
\multicolumn{3}{c}{c1; $v_c - v_\mathrm{gal} = -112_{-1}^{+1}$ km s$^{-1}$} \\
\hline
\ion{H}{I} & $14.86 \pm 0.02$ & $26 \pm 2$ \\
\ion{C}{III} & $13.33 \pm 0.04$ & $18 \pm 3$ \\
\ion{O}{III} & $13.72 \pm 0.07$ & $18$ \\
\ion{O}{IV} & $14.04 \pm 0.04$ & $36 \pm 4$ \\
\ion{O}{VI} & $13.71_{-0.09}^{+0.07}$ & $36$ \\
\hline
\multicolumn{3}{c}{c2; $v_c - v_\mathrm{gal} = -3.2 \pm 0.6$ km s$^{-1}$} \\
\hline
\ion{H}{I} & $14.42 \pm 0.05$ & $45_{-8}^{+11}$ \\
\ion{C}{III} & $13.05 \pm 0.06$ & $13$ \\
\ion{C}{IV} & $13.45 \pm 0.03$ & $13 \pm 1$ \\
\ion{N}{IV} & $13.12 \pm 0.08$ & $13$ \\
\ion{O}{III} & $13.66_{-0.08}^{+0.07}$ \footnote{Likely contaminated} & $13$ \\
\ion{O}{IV} & $14.22 \pm 0.04$ & $13$ \\
\ion{O}{VI} & $14.33_{-0.03}^{+0.02}$ & $41 \pm 3$ \\
\hline
\end{tabular}
\end{table}

This is a non-isolated, star-forming galaxy at $z_\mathrm{gal}=0.5839$ with stellar mass $M_\mathrm{star} \approx 3 \times 10^9 \msun$. Two absorption components are identified in \ion{C}{III} $\lambda 977$, \ion{O}{III} $\lambda 832$, with component c2 also being identified in the \ion{C}{IV} doublet and \ion{N}{IV} $\lambda 765$. \ion{O}{IV} absorption is identified in c1 and c2, although the absorption along c1 is broader ($b \approx 36 \ \kms$) compared to \ion{C}{III} ($b \approx 18 \ \kms$), indicating different origins. Absorption in \ion{O}{VI} can be constrained in c1 using the line width of \ion{O}{IV}, while the doublet independently suggests broader line width along c2 ($b \approx 40 \ \kms$) compared to \ion{C}{IV} ($b \approx 13 \ \kms$). The \ion{H}{I} absorption is recovered for both c1 and c2. 

For ionization modeling of c1, a low-density phase of $\log(n_\mathrm{H}/\mathrm{cm}^{-3}) \approx -4.3$ constrained using the \ion{O}{VI}/\ion{O}{IV} ratio underpredicts the observed \ion{O}{III} column, necessitating a two-phase solution. For the high-density phase, the \ion{O}{III}/\ion{O}{II} limit (with \ion{O}{II} being a non-detection) with a restriction for \ion{O}{IV} to primarily arise in the \ion{O}{VI}-bearing phase yields a high-density phase of $\log(n_\mathrm{H}/\mathrm{cm}^{-3}) \approx -2.9$, with an uncertainty of $0.3 \ \mathrm{dex}$. For c2, the ratio of \ion{C}{IV}/\ion{C}{III} suggests a gas density of $\log(n_\mathrm{H}/\mathrm{cm}^{-3}) \approx -4.4$, although matching this phase with \ion{O}{IV} is unable to reproduce the measured \ion{O}{III} and \ion{O}{VI} columns. The distinctly broad \ion{O}{VI} line width likely suggests origin in warmer gas, while the \ion{O}{III} $\lambda 702, \lambda 832$ absorption may be contaminated in c2.  

\subsection{$\mathrm{CUBS \ J045416.38\!-\!611640.90}$ at $d_\mathrm{proj}=82 \ \mathrm{kpc}$} \label{sec:s13}

This is a non-isolated, quiescent galaxy at $z_\mathrm{gal}=0.4315$ with stellar mass $M_\mathrm{star} \approx 10^{10} \msun$. The associated absorption includes detections in 
\ion{H}{I} and \ion{O}{VI}. \ion{C}{III} absorption is also identified, although it is contaminated on the red side. 

\begin{table}
\caption{Absorption properties along the sightline at $d_\mathrm{proj} = 82$ kpc from system s13.} \label{tab:s13}
\footnotesize\centering
\smallskip
\begin{tabular}{rrr}
\hline
Ion & $\log(N_c/\mathrm{cm}^{-2})$ & $b_c$ (km s$^{-1}$) \\
\hline
\hline
\multicolumn{3}{c}{c1; $v_c - v_\mathrm{gal} = -73 \pm 1$ km s$^{-1}$} \\
\hline
\ion{H}{I} & $13.67 \pm 0.03$ & $34_{-4}^{+5}$ \\
\ion{C}{III} & $13.36 \pm 0.03$ & $24 \pm 4$ \\
\ion{O}{VI} & $13.96 \pm 0.03$ & $26 \pm 3$ \\
\hline
\end{tabular}
\end{table}

\subsection{$\mathrm{CUBS \ J035723.55\!-\!481215.38}$ at $d_\mathrm{proj}=97 \ \mathrm{kpc}$} \label{sec:s14}

\begin{table}
\caption{Absorption properties along the sightline at $d_\mathrm{proj} = 97$ kpc from system s14.} \label{tab:s14}
\footnotesize\centering
\smallskip
\begin{tabular}{rrr}
\hline
Ion & $\log(N_c/\mathrm{cm}^{-2})$ & $b_c$ (km s$^{-1}$) \\
\hline
\hline
\multicolumn{3}{c}{c1; $v_c - v_\mathrm{gal} = -142 \pm 2$ km s$^{-1}$} \\
\hline
\ion{H}{I} & $14.10 \pm 0.04$ & $24 \pm 7$ \\
\ion{C}{III} & $13.00_{-0.05}^{+0.06}$ & $21_{-6}^{+9}$ \\
\hline
\end{tabular}
\end{table}

This is a non-isolated, star-forming galaxy at $z_\mathrm{gal}=0.4807$ with stellar mass $M_\mathrm{star} \approx 10^{10} \msun$. Associated absorption is identified in \ion{H}{I} and \ion{C}{III}, but no other metal ions.

\subsection{$\mathrm{CUBS \ J015453.61\!-\!071216.79}$ at $d_\mathrm{proj}=99 \ \mathrm{kpc}$} \label{sec:s15}

\begin{table}[t!]
\caption{Absorption properties along the sightline at $d_\mathrm{proj} = 99$ kpc from system s15.} \label{tab:s15}
\footnotesize\centering
\smallskip
\begin{tabular}{rrr}
\hline
Ion & $\log(N_c/\mathrm{cm}^{-2})$ & $b_c$ (km s$^{-1}$) \\
\hline
\hline
\multicolumn{3}{c}{c1; $v_c - v_\mathrm{gal} = 26 \pm 2$ km s$^{-1}$} \\
\hline
\ion{H}{I} & $14.95 \pm 0.03$ & $37 \pm 2$ \\
\ion{C}{III} & $13.21 \pm 0.04$ & $26 \pm 3$ \\
\ion{O}{IV} & $14.27 \pm 0.04$ & $26$ \\
\ion{O}{VI} & $13.83_{-0.07}^{+0.06}$ & $26$ \\
\hline
\multicolumn{3}{c}{c2; $v_c - v_\mathrm{gal} = 77.7 \pm 0.8$ km s$^{-1}$} \\
\hline
\ion{H}{I} & $14.76 \pm 0.07$ & $19 \pm 3$ \\
\ion{C}{III} & $>13.6$ & $16$ \\
\ion{C}{IV} & $13.75 \pm 0.03$ & $16 \pm 2$ \\
\ion{N}{IV} & $13.6 \pm 0.1$ & $16$ \\
\ion{O}{III} & $14.29 \pm 0.05$ & $16$ \\
\ion{O}{IV} & $14.42_{-0.07}^{+0.06}$ & $16$ \\
\ion{O}{VI} & $13.77_{-0.09}^{+0.08}$ & $16$ \\
\hline
\multicolumn{3}{c}{c3; $v_c - v_\mathrm{gal} = 118 \pm 4$ km s$^{-1}$} \\
\hline
\ion{H}{I} & $14.94_{-0.06}^{+0.07}$ & $45_{-6}^{+8}$ \\
\ion{C}{III} & $>13.5$ & $34$ \\
\ion{N}{IV} & $13.2_{-0.2}^{+0.1}$ & $34_{-4}^{+5}$ \\
\ion{O}{III} & $14.08 \pm 0.08$ & $34_{-4}^{+5}$ \\
\ion{O}{IV} & $14.47_{-0.05}^{+0.06}$ & $34$ \\
\ion{O}{VI} & $14.42 \pm 0.03$ & $69$ \\
\ion{Ne}{VIII} & $14.43 \pm 0.03$ & $69 \pm 4$ \\
\hline
\multicolumn{3}{c}{c4; $v_c - v_\mathrm{gal} = 196.5_{-0.9}^{+0.8}$ km s$^{-1}$} \\
\hline
\ion{H}{I} & $14.71 \pm 0.03$ & $14_{-2}^{+1}$ \\
\ion{C}{III} & $13.11 \pm 0.04$ & $14 \pm 3$ \\
\ion{O}{IV} & $13.86_{-0.06}^{+0.05}$ & $14$ \\
\hline
\end{tabular}
\end{table}

This is a non-isolated, star-forming galaxy with stellar mass $M_\mathrm{star} \approx 10^{10} \msun$ first identified in \cite{Qu:2024}. 
Strong, narrow ($b \approx 16 \ \kms$) \ion{C}{IV} absorption is recovered in component c2 at $v_c-v_\mathrm{gal} \approx 78 \ \kms$. This component is also recovered in \ion{O}{III} $\lambda 832$ and \ion{N}{IV} $\lambda 765$, although an additional, broader ($b \approx 34 \ \kms$) component c3 is required to explain absorption in these transitions redward of c2. Components c2 and c3 are recovered in \ion{C}{III} $\lambda 977$ and \ion{O}{IV} $\lambda 787$, although additional components c1 and c4 respectively at $v_c-v_\mathrm{gal} \approx 26 \ \kms, 196 \ \kms$ are required to explain the full absorption in these transitions. The \ion{Ne}{VIII} $\lambda \lambda 770, 780$ doublet exhibits absorption centered at c3 and is modeled using a single broad component with $b \approx 69 \ \kms$. This broad component is also imposed on the \ion{O}{VI} doublet, and additional absorption is accommodated using narrower components at c1 and c2 bearing line widths similar to \ion{O}{IV}. \ion{H}{I} absorption can be recovered along c1 through c4 using the Ly$\beta$ and Ly$\gamma$ transitions. 

For ionization modeling of c1, a single phase of density $\log(n_\mathrm{H}/\mathrm{cm}^{-3}) \approx -4.4$ can be accommodated using the \ion{O}{VI}/\ion{O}{IV} ratio. For c2, the low-density phase of $\log(n_\mathrm{H}/\mathrm{cm}^{-3}) \approx -4.3$ constrained using the \ion{O}{VI}/\ion{O}{IV} ratio underpredicts the observed \ion{O}{III}. The high-density phase is constrained using the \ion{Si}{IV}/\ion{O}{III} limit (assuming $\mathrm{[Si/O]=0}$) and treating \ion{O}{IV}/\ion{O}{III} as an upper bound, resulting in a density of $\log(n_\mathrm{H}/\mathrm{cm}^{-3}) \approx -3.3$. In component c3, a single photoionized phase of density $\log(n_\mathrm{H}/\mathrm{cm}^{-3}) \approx -3.7$ constrained using the \ion{O}{IV}/\ion{O}{III} ratio can explain other the \ion{C}{III} and \ion{N}{IV} measurements, but underpredicts the \ion{O}{VI} and \ion{Ne}{VIII}, which are likely collisionally ionized given their broad line width. An ionization model for c4 can be obtained by using the \ion{C}{III}/\ion{C}{II} and \ion{C}{IV}/\ion{C}{III} limits, resulting in a density of $\log(n_\mathrm{H}/\mathrm{cm}^{-3}) \approx -3.4$ with an uncertainty of $0.3 \ \mathrm{dex}$.

\subsection{$\mathrm{CUBS \ J224458.73\!-\!493158.12}$ at $d_\mathrm{proj} = 114 \ \mathrm{kpc}$} \label{sec:s16}

\begin{table}[t!]
\caption{Absorption properties along the sightline at $d_\mathrm{proj} = 114$ kpc from system s16.} \label{tab:s16}
\footnotesize\centering
\smallskip
\begin{tabular}{rrr}
\hline
Ion & $\log(N_c/\mathrm{cm}^{-2})$ & $b_c$ (km s$^{-1}$) \\
\hline
\hline
\multicolumn{3}{c}{c1; $v_c - v_\mathrm{gal} = -49.2 \pm 0.6$ km s$^{-1}$} \\
\hline
\ion{H}{I} & $15.85 \pm 0.02$ & $23.2_{-0.7}^{+0.6}$ \\
\ion{C}{III} & $13.51_{-0.04}^{+0.05}$ & $22$ \\
\ion{O}{III} & $14.20 \pm 0.06$ & $22 \pm 4$ \\
\ion{O}{IV} & $14.09_{-0.08}^{+0.07}$ & $45_{-14}^{+15}$ \\
\hline
\multicolumn{3}{c}{c2; $v_c - v_\mathrm{gal} = 111.5 \pm 0.8$ km s$^{-1}$} \\
\hline
\ion{H}{I} & $15.47 \pm 0.02$ & $32 \pm 1$ \\
\ion{O}{VI} & $14.05 \pm 0.05$ & $30 \pm 2$ \\
\hline
\end{tabular}
\end{table} 

This is an isolated, star-forming galaxy at $z_\mathrm{gal}=0.5863$ with stellar mass $M_\mathrm{star} \approx 2 \times 10^{10} \msun$. Associated absorption is identified along two distinct components in \ion{H}{I} using multiple Lyman series lines. Absorption in c1 can be identified in \ion{O}{III} $\lambda 832$ and \ion{C}{III} $\lambda 977$, although the former is contaminated on the blue side. \ion{O}{IV} $\lambda 787$ also shows absorption centered at c1, although it is broader than \ion{O}{III}. The \ion{O}{VI} doublet shows absorption alongside c2. For the ionization modeling of c1, the \ion{O}{III}/\ion{O}{II} and \ion{C}{IV}/\ion{C}{III} limits suggest a gas density of $\log(n_\mathrm{H}/\mathrm{cm}^{-3}) \approx -2.7$, with an uncertainty of $0.2 \ \mathrm{dex}$.

\subsection{$\mathrm{CUBS \ J213553.48\!-\!531714.50}$ at $d_\mathrm{proj}=128 \ \mathrm{kpc}$} \label{sec:s17}

\begin{table}[t!]
\caption{Absorption properties along the sightline at $d_\mathrm{proj} = 128$ kpc from system s17.} \label{tab:s17}
\footnotesize\centering
\smallskip
\begin{tabular}{rrr}
\hline
Ion & $\log(N_c/\mathrm{cm}^{-2})$ & $b_c$ (km s$^{-1}$) \\
\hline
\hline
\multicolumn{3}{c}{c1; $v_c - v_\mathrm{gal} = -141 \pm 2$ km s$^{-1}$} \\
\hline
\ion{H}{I} & $14.50_{-0.07}^{+0.04}$ & $31_{-4}^{+3}$ \\
\hline
\multicolumn{3}{c}{c2; $v_c - v_\mathrm{gal} = 8_{-1}^{+2}$ km s$^{-1}$} \\
\hline
\ion{H}{I} & $15.40_{-0.03}^{+0.02}$ & $26 \pm 1$ \\
\ion{C}{III} & $13.94_{-0.06}^{+0.07}$ & $21$ \\
\ion{C}{IV} & $14.36 \pm 0.04$ & $21 \pm 1$ \\
\ion{N}{III} & $13.5_{-0.2}^{+0.1}$ & $21$ \\
\ion{O}{VI} & $14.65_{-0.04}^{+0.06}$ & $24 \pm 1$ \\
\hline
\multicolumn{3}{c}{c3; $v_c - v_\mathrm{gal} = 123.9_{-0.1}^{+0.2}$ km s$^{-1}$} \\
\hline
\ion{H}{I} & $18.19 \pm 0.06$ & $16.7_{-0.6}^{+0.8}$ \\
\ion{C}{II} & $13.7 \pm 0.1$ & $11 \pm 1$ \\
\ion{C}{III} & $>14.1$ & $11$ \\
\ion{Mg}{I} & $11.09_{-0.07}^{+0.08}$ & $9.4$ \\
\ion{Mg}{II} & $12.71 \pm 0.03$ & $9.4_{-0.4}^{+0.2}$ \\
\ion{Si}{II} & $13.2_{-0.4}^{+0.3}$ & $9.4$ \\
\ion{Fe}{II} & $12.48 \pm 0.03$ & $7.6 \pm 0.8$ \\
\hline
\multicolumn{3}{c}{c4; $v_c - v_\mathrm{gal} = 150.7_{-0.2}^{+0.9}$ km s$^{-1}$} \\
\hline
\ion{H}{I} & $16.2_{-0.1}^{+0.3}$ & $24_{-9}^{+4}$ \\
\ion{C}{II} & $13.80_{-0.08}^{+0.07}$ & $8.8 \pm 0.8$ \\
\ion{C}{III} & $>14.1$ & $8.8$ \\
\ion{Mg}{II} & $12.53_{-0.02}^{+0.01}$ & $6.5_{-0.3}^{+0.6}$ \\
\ion{Si}{II} & $13.1_{-0.4}^{+0.2}$ & $6.5$ \\
\ion{Fe}{II} & $12.47_{-0.03}^{+0.04}$ & $7 \pm 1$ \\
\hline
\multicolumn{3}{c}{c5; $v_c - v_\mathrm{gal} = 174_{-5}^{+7}$ km s$^{-1}$} \\
\hline
\ion{H}{I} & $15.1 \pm 0.1$ & $42_{-9}^{+7}$ \\
\ion{C}{III} & $>13.9$ & $36$ \\
\ion{C}{IV} & $13.57 \pm 0.05$ & $36_{-5}^{+3}$ \\
\ion{O}{VI} & $13.73 \pm 0.09$ & $36$ \\
\hline
\end{tabular}
\end{table} 

This is a non-isolated, quiescent galaxy at $z_\mathrm{gal}=0.6219$ with stellar mass $M_\mathrm{star} \approx 5 \times 10^{10} \msun$ first reported in \cite{Chen:2020,Zahedy:2021}. The corresponding absorption is yet another LLS with $N(\mathrm{H \ I}) \approx 2 \times 10^{18} \ \mathrm{cm}^{-2}$. The \ion{Mg}{II} and \ion{Fe}{II} transitions indicate a two-component structure that is also observed in \ion{C}{II} $\lambda 1334$ and \ion{Si}{II} $\lambda 1260$ covered by STIS. \ion{C}{III} $\lambda 977$ suggests the presence of additional components both blueward and redward of the \ion{Mg}{II}-bearing components. The blueward component at $v_c-v_\mathrm{gal} \approx 8 \ \kms$ is not observed in low-ions, but is recovered in \ion{C}{IV}, \ion{N}{III}, and \ion{O}{VI}, indicating the presence of lower density gas. The redward absorption in \ion{C}{III} is also confirmed by isolated broad components in \ion{C}{IV} and \ion{O}{VI} centered at $v_c-v_\mathrm{gal} \approx 174 \ \kms$. \ion{H}{I} absorption can be recovered for all four metal-bearing components, with an additional non-metal bearing component at $v_c-v_\mathrm{gal} \approx -141 \ \kms$ indicated by Ly$\beta$. During the joint Voigt profile fitting of all ions, it is necessary to impose specific restrictions on the component structure of \ion{C}{III} to handle saturation and blending between components c3, c4, and c5. Adopting the line widths of $b \approx 10 \ \kms$ from \ion{C}{II} for components c3 and c4, the \ion{C}{III} in these components is likely saturated and stronger than component c5, which has a broader line width of $b \approx 35 \ \kms$ indicated by the associated \ion{C}{IV} absorption. For this reason, $N_{c3}$(\ion{C}{III}) $>N_{c5}$(\ion{C}{III}) and $N_{c4}$(\ion{C}{III}) $>N_{c5}$(\ion{C}{III}) are imposed during the fitting procedure, and corresponding 3\,$\sigma$ lower limits on the \ion{C}{III} column density are reported in Table \ref{tab:s17}.

Ionization modeling results are available for components c2, c3, and c4. For component c2, the \ion{C}{IV}/\ion{C}{III} ratio suggests low-density gas of $\log(n_\mathrm{H}/\mathrm{cm}^{-3}) \approx -4.4$. For components c3 and c4, available limits on the \ion{C}{III}/\ion{CII} and \ion{Si}{IV}/\ion{Si}{II} ratios are stringent and suggest a gas density of $\log(n_\mathrm{H}/\mathrm{cm}^{-3}) \approx -2.1$, with associated uncertainties of $0.3 \ \mathrm{dex}$.

\subsection{$\mathrm{CUBS \ J224458.93\!-\!493129.29}$ at $d_\mathrm{proj}=138 \ \mathrm{kpc}$} \label{sec:s18}

\begin{table}
\caption{Absorption properties along the sightline at $d_\mathrm{proj} = 138$ kpc from system s18.} \label{tab:s18}
\footnotesize\centering
\smallskip
\begin{tabular}{rrr}
\hline
Ion & $\log(N_c/\mathrm{cm}^{-2})$ & $b_c$ (km s$^{-1}$) \\
\hline
\hline
\multicolumn{3}{c}{c1; $v_c - v_\mathrm{gal} = 0$ km s$^{-1}$} \\
\hline
\ion{H}{I} & $<14.0$ & $12$ \\
\ion{C}{IV} & $<13.0$ & $6$ \\
\ion{O}{VI} & $<14.1$ & $30$ \\
\ion{Mg}{II} & $<11.6$ & $4$ \\
\hline
\end{tabular}
\end{table}

This is an isolated, star-forming galaxy at $z_\mathrm{gal} = 0.4893$ with stellar mass $M_\mathrm{star} \approx 10^{10} \ \msun$. No associated absorption is identified in \ion{H}{I} or metal ions for this system. 

\subsection{$\mathrm{CUBS \ J011957.16\!-\!201003.55}$ at $d_\mathrm{proj}=141 \ \mathrm{kpc}$} \label{sec:s19}

\begin{table}[t!]
\caption{Absorption properties along the sightline at $d_\mathrm{proj} = 141$ kpc from system s19.} \label{tab:s19}
\footnotesize\centering
\smallskip
\begin{tabular}{rrr}
\hline
Ion & $\log(N_c/\mathrm{cm}^{-2})$ & $b_c$ (km s$^{-1}$) \\
\hline
\hline
\multicolumn{3}{c}{c1; $v_c - v_\mathrm{gal} = -56.1 \pm 0.5$ km s$^{-1}$} \\
\hline
\ion{H}{I} & $15.44_{-0.04}^{+0.03}$ & $15.2_{-0.6}^{+0.5}$ \\
\hline
\multicolumn{3}{c}{c2; $v_c - v_\mathrm{gal} = 20.7_{-0.3}^{+0.4}$ km s$^{-1}$} \\
\hline
\ion{H}{I} & $15.81 \pm 0.02$ & $18 \pm 1$ \\
\ion{C}{II} & $13.93_{-0.09}^{+0.10}$ & $8 \pm 1$ \\
\ion{C}{III} & $>14.6$ & $12 \pm 1$ \\
\ion{C}{IV} & $14.03_{-0.08}^{+0.09}$ & $16_{-2}^{+3}$ \\
\ion{N}{II} & $13.76 \pm 0.08$ & $8$ \\
\ion{N}{III} & $14.20 \pm 0.06$ & $12 \pm 1$ \\
\ion{N}{IV} & $13.7 \pm 0.2$ & $16_{-2}^{+3}$ \\
\ion{O}{II} & $14.0 \pm 0.1$ & $8$ \\
\ion{O}{III} & $14.7 \pm 0.1$ & $12 \pm 1$ \\
\ion{O}{IV} & $14.56 \pm 0.08$ & $16_{-2}^{+3}$ \\
\ion{O}{VI} & $13.90 \pm 0.08$ & $16_{-2}^{+3}$ \\
\ion{Mg}{II} & $12.37 \pm 0.02$ & $6.4 \pm 0.7$ \\
\ion{Si}{IV} & $13.33 \pm 0.08$ & $14_{-2}^{+3}$ \\
\hline
\multicolumn{3}{c}{c3; $v_c - v_\mathrm{gal} = 45 \pm 3$ km s$^{-1}$} \\
\hline
\ion{H}{I} & $14.53_{-0.07}^{+0.06}$ & $39_{-3}^{+4}$ \\
\ion{C}{III} & $<13.9$ & $14$ \\
\ion{C}{IV} & $13.6 \pm 0.1$ & $14_{-2}^{+3}$ \\
\ion{N}{III} & $<13.9$ & $14$ \\
\ion{N}{IV} & $<13.5$ & $14$ \\
\ion{O}{III} & $13.9 \pm 0.1$ & $14$ \\
\ion{O}{IV} & $14.2 \pm 0.1$ & $14$ \\
\ion{O}{VI} & $<13.8$ & $14$ \\
\hline
\multicolumn{3}{c}{c4; $v_c - v_\mathrm{gal} = 277.1_{-0.9}^{+0.7}$ km s$^{-1}$} \\
\hline
\ion{H}{I} & $14.27 \pm 0.03$ & $20_{-3}^{+2}$ \\
\hline
\end{tabular}
\end{table}

This is a non-isolated, star-forming galaxy at $z_\mathrm{gal}=0.4517$ with stellar mass $M_\mathrm{star} \approx 4 \times 10^{10} \msun$. Associated absorption can be identified in low-ions along a single component c2 at $v_c-v_\mathrm{gal} \approx 21 \ \kms$. The \ion{C}{IV} doublet exhibits an asymmetric profile on the red side, suggesting the presence of an additional component c3, which can also be accommodated in the \ion{O}{III} $\lambda 832$ and \ion{O}{IV} $\lambda 787$ transitions. A joint fit to these transitions suggests that the offset of c3 from c2 is $v_{c3}-v_\mathrm{c2} \approx 24 \ \kms$. Although the presence of c3 is not as significant in \ion{N}{III} and \ion{N}{IV}, this component is included during the joint fitting to account for blending with c2. \ion{O}{VI} $\lambda 1031$ is contaminated, but the red side can accommodate absorption from c2 and c3. However, \ion{O}{VI} $\lambda 1037$ is heavily contaminated and is therefore not included during the fitting. The \ion{H}{I} absorption indicates the presence of additional non-metal-bearing components c1 at $v_c-v_\mathrm{gal} \approx -56 \ \kms$ in the higher-order Lyman series, and component c4 at $v_c-v_\mathrm{gal} \approx 277 \ \kms$ in Ly$\alpha$ and Ly$\beta$. 

For the ionization modeling of c2, a two-phase solution is necessary. The high-density phase is constrained at $\log(n_\mathrm{H}/\mathrm{cm}^{-3}) \approx -2.6$ using the \ion{N}{III}/\ion{N}{II} and \ion{O}{III}/\ion{O}{II} ratios. The low-density phase is estimated to be at a density of $\log(n_\mathrm{H}/\mathrm{cm}^{-3}) \approx -4.3$ using the \ion{O}{IV} and \ion{O}{VI} detections as well as the \ion{S}{V} and \ion{S}{VI} upper limits (assuming $\mathrm{[S/O]} = 0$). Component c3 can be described using a single phase of density $\log(n_\mathrm{H}/\mathrm{cm}^{-3}) \approx -3.7$ using the \ion{O}{IV}/\ion{O}{III} ratio.

\subsection{$\mathrm{CUBS \ J045415.97\!-\!611602.32}$ at $d_\mathrm{proj}=162 \ \mathrm{kpc}$} \label{sec:s20}

\begin{table}
\caption{Absorption properties along the sightline at $d_\mathrm{proj} = 162$ kpc from system s20.} \label{tab:s20}
\footnotesize\centering
\smallskip
\begin{tabular}{rrr}
\hline
Ion & $\log(N_c/\mathrm{cm}^{-2})$ & $b_c$ (km s$^{-1}$) \\
\hline
\hline
\multicolumn{3}{c}{c1; $v_c - v_\mathrm{gal} = 0$ km s$^{-1}$} \\
\hline
\ion{H}{I} & $<13.7$ & $12$ \\
\ion{C}{IV} & $<12.9$ & $6$ \\
\ion{O}{VI} & $<13.6$ & $30$ \\
\ion{Mg}{II} & $<11.5$ & $4$ \\
\hline
\end{tabular}
\end{table}

This is an isolated quiescent galaxy at $z_\mathrm{gal}=0.6011$ with stellar mass $M_\mathrm{star}\approx 3 \times 10^{10}\msun$. No associated absorption is identified in \ion{H}{I} or metal ions. 

\subsection{$\mathrm{CUBS \ J224457.76\!-\!493159.13}$ at $d_\mathrm{proj}=177 \ \mathrm{kpc}$} \label{sec:s21}

\begin{table}[t!]
\caption{Absorption properties along the sightline at $d_\mathrm{proj} = 177$ kpc from system s21.} \label{tab:s21}
\footnotesize\centering
\smallskip
\begin{tabular}{rrr}
\hline
Ion & $\log(N_c/\mathrm{cm}^{-2})$ & $b_c$ (km s$^{-1}$) \\
\hline
\hline
\multicolumn{3}{c}{c1; $v_c - v_\mathrm{gal} = -394 \pm 1$ km s$^{-1}$} \\
\hline
\ion{H}{I} & $14.78 \pm 0.05$ & $12 \pm 1$ \\
\ion{C}{III} & $13.0 \pm 0.1$ & $34_{-12}^{+14}$ \\
\ion{Ne}{VIII} & $14.00 \pm 0.09$ & $52_{-11}^{+16}$ \\
\hline
\multicolumn{3}{c}{c2; $v_c - v_\mathrm{gal} = -309.2 \pm 0.3$ km s$^{-1}$} \\
\hline
\ion{H}{I} & $15.82_{-0.02}^{+0.03}$ & $28 \pm 1$ \\
\ion{C}{II} & $13.6 \pm 0.1$ & $8_{-2}^{+3}$ \\
\ion{C}{III} & $>13.9$ & $11$ \\
\ion{C}{IV} & $13.54 \pm 0.05$ & $30_{-3}^{+4}$ \\
\ion{O}{II} & $14.4_{-0.1}^{+0.3}$ & $8$ \\
\ion{O}{III} & $14.8_{-0.2}^{+0.3}$ & $11_{-1}^{+3}$ \\
\ion{O}{IV} & $14.22 \pm 0.08$ & $30$ \\
\ion{Mg}{II} & $12.30_{-0.06}^{+0.08}$ & $2.1_{-0.4}^{+0.5}$ \\
\hline
\end{tabular}
\end{table} 

This is a non-isolated, quiescent galaxy at $z_\mathrm{gal}=0.6203$ with stellar mass $M_\mathrm{star} \approx 5 \times 10^{10} M_\odot$ first reported by \cite{Qu:2024}. Associated absorption is seen in low-ions along a single component, c2 at $v_c-v_\mathrm{gal} \approx -309 \ \kms$. The single-component structure persists in \ion{O}{III} $\lambda 832$ and \ion{O}{IV} $\lambda 787$, although the \ion{O}{IV} absorption is broader. The \ion{H}{I} absorption reveals an additional narrow component c1 at $v_c - v_\mathrm{gal} \approx -394 \ \kms$. Broad absorption is identified in c1 for \ion{C}{III} $\lambda 977$ and \ion{Ne}{VIII} $\lambda \lambda 770, 780$, but not in \ion{O}{VI} $\lambda \lambda 1031, 1037$.

For the ionization modeling of c2, a high-density phase is constrained at $\log(n_\mathrm{H}/\mathrm{cm}^{-3}) \approx -2.5$ using the \ion{O}{III}/\ion{O}{II} ratio. This phase underpredicts the observed \ion{O}{IV}, so a low-density phase is required, constrained using the \ion{O}{VI}/\ion{O}{IV} limit and requiring most of the \ion{O}{IV} to come from this phase. The constrained density is $\log(n_\mathrm{H}/\mathrm{cm}^{-3}) \approx -3.8$, with an associated uncertainty of $0.4 \ \mathrm{dex}$.

\subsection{$\mathrm{CUBS \ J011037.69\!-\!164915.63}$ at $d_\mathrm{proj} = 364 \ \mathrm{kpc}$} \label{sec:s22}

\begin{table}[t!]
\caption{Absorption properties along the sightline at $d_\mathrm{proj} = 364$ kpc from system s22.} \label{tab:s22}
\footnotesize\centering
\smallskip
\begin{tabular}{rrr}
\hline
Ion & $\log(N_c/\mathrm{cm}^{-2})$ & $b_c$ (km s$^{-1}$) \\
\hline
\hline
\multicolumn{3}{c}{c1; $v_c - v_\mathrm{gal} = 167.5 \pm 0.2$ km s$^{-1}$} \\
\hline
\ion{H}{I} & $16.38 \pm 0.03$ & $13.8_{-0.4}^{+0.3}$ \\
\ion{C}{II} & $14.3_{-0.1}^{+0.2}$ & $8 \pm 1$ \\
\ion{C}{III} & $14.3_{-0.2}^{+0.3}$ & $8.9$ \\
\ion{C}{IV} & $13.97_{-0.08}^{+0.09}$ & $8 \pm 1$ \\
\ion{N}{II} & $14.0_{-0.2}^{+0.1}$ & $8$ \\
\ion{N}{III} & $14.28 \pm 0.06$ & $8.9 \pm 0.7$ \\
\ion{N}{IV} & $13.7 \pm 0.1$ & $8$ \\
\ion{O}{II} & $14.5 \pm 0.1$ & $8$ \\
\ion{O}{III} & $14.4 \pm 0.1$ & $8.9$ \\
\ion{O}{IV} & $14.3_{-0.2}^{+0.3}$ & $8$ \\
\ion{Mg}{I} & $11.55_{-0.05}^{+0.06}$ & $6.2$ \\
\ion{Mg}{II} & $>13.2$ & $6.2_{-0.6}^{+0.5}$ \\
\ion{Si}{II} & $13.54_{-0.08}^{+0.09}$ & $5_{-1}^{+2}$ \\
\ion{Si}{IV} & $13.05_{-0.04}^{+0.05}$ & $6.7 \pm 0.9$ \\
\ion{S}{II} & $13.3 \pm 0.1$ & $5$ \\
\ion{S}{III} & $14.0 \pm 0.1$ & $5$ \\
\ion{S}{IV} & $13.01_{-0.09}^{+0.08}$ & $6.7$ \\
\ion{S}{V} & $12.8 \pm 0.1$ & $6.7$ \\
\ion{Fe}{II} & $12.75_{-0.07}^{+0.08}$ & $3_{-1}^{+2}$ \\
\hline
\multicolumn{3}{c}{c2; $v_c - v_\mathrm{gal} = 187_{-1}^{+1}$ km s$^{-1}$} \\
\hline
\ion{H}{I} & $13.9_{-0.5}^{+0.3}$ & $16_{-5}^{+7}$ \\
\ion{C}{III} & $13.1 \pm 0.1$ & $7$ \\
\ion{C}{IV} & $<13.2$ & $7_{-2}^{+1}$ \\
\ion{N}{III} & $<14.0$ & $7$ \\
\ion{N}{IV} & $13.0 \pm 0.1$ & $7$ \\
\ion{O}{III} & $13.7_{-0.2}^{+0.1}$ & $7$ \\
\ion{O}{IV} & $<14.3$ & $7$ \\
\ion{O}{VI} & $13.9 \pm 0.1$ & $7$ \\
\hline
\multicolumn{3}{c}{c3; $v_c - v_\mathrm{gal} = 246 \pm 4$ km s$^{-1}$} \\
\hline
\ion{H}{I} & $14.53_{-0.04}^{+0.03}$ & $42 \pm 4$ \\
\hline
\end{tabular}
\end{table}

This is a non-isolated, quiescent galaxy at $z_\mathrm{gal}=0.5404$ with stellar mass $M_\mathrm{star} \approx 2 \times 10^{11} \msun$ reported by \cite{Cooper:2021}. The associated absorption corresponds to a pLLS with $N(\mathrm{H \ I}) \approx 3 \times 10^{16} \ \mathrm{cm}^{-2}$. Narrow metal absorption is seen in a single component, c1, in several low-ions (e.g., \ion{Mg}{II}, \ion{Si}{II}, and \ion{Fe}{II}) and high-ions (e.g., \ion{Si}{IV}). This system has successive stages detected for sulfur (\ion{S}{II} through \ion{S}{V}), providing a complementary anchor to carbon, nitrogen, and oxygen ions for ionization modeling. The sulfur lines in COS are fit using line widths measured from the high-resolution \ion{Si}{II} and \ion{Si}{IV} transitions in STIS. Among low-ions, \ion{C}{II}, \ion{N}{II}, and \ion{O}{II} also exhibit absorption in c1. A single component model underpredicts the absorption for \ion{C}{III}, \ion{O}{III}, and \ion{N}{IV}. Since the location of the satellite component c2 is not very precisely determined by a single transition due to blending with the primary component, a joint fit is performed using \ion{C}{III}, \ion{O}{III}, and \ion{N}{III} to find the offset of c2 from c1, $v_\mathrm{c2}-v_\mathrm{c1} \approx 20 \ \kms$. The presence of such c2 is not clearly suggested by the profiles of  \ion{N}{III}, \ion{C}{IV}, and \ion{O}{IV}, but a second component is still included during the joint fitting to account for potential blending with the primary component. The \ion{O}{VI} $\lambda 1031$ transition is contaminated, but the \ion{O}{VI} $\lambda1037$ exhibits weak, narrow absorption that is better aligned with the location of c2. For \ion{H}{I}, the absorption in c1 is well-constrained by the higher-order Lyman series lines, and an additional component c3 at $v_c-v_\mathrm{gal} \approx 246 \ \kms$ is suggested by Ly$\beta$ and Ly$\gamma$. The corresponding constraint for \ion{H}{I} in c2 is relatively poor due to blending with c1 and c3 on either side. 

For the ionization modeling of c1, the \ion{C}{III}/\ion{C}{II}, \ion{N}{III}/\ion{N}{II}, \ion{O}{III}/\ion{O}{II}, and \ion{S}{III}/\ion{S}{II} ratios suggest a consistent density of $\log(n_\mathrm{H}/\mathrm{cm}^{-3}) \approx -2.2$. However, this phase is unable to reproduce the observed column densities for high-ions. The measured \ion{S}{V}/\ion{S}{IV}, alongside the stringent \ion{S}{VI} upper limit constrains the density of the high-ionization phase to be $\log(n_\mathrm{H}/\mathrm{cm}^{-3}) \approx -4.0$. For the satellite component c2, the \ion{C}{III}, \ion{N}{IV}, and \ion{O}{III} columns are robustly measured. The corresponding absorption for \ion{N}{III}, \ion{C}{IV}, and \ion{O}{IV} is weak, so a column density upper limit is evaluated from the posterior distribution for these ions. A single gas phase is density $\log(n_\mathrm{H}/\mathrm{cm}^{-3}) \approx -3.4$ is constrained using the \ion{C}{IV}/\ion{C}{III}, \ion{N}{IV}/\ion{N}{III}, and \ion{O}{IV}/\ion{O}{III} limits, albeit with a large uncertainty of $0.4 \ \mathrm{dex}$. This phase is unable to reproduce the measured \ion{O}{VI} column density in c2, despite the comparably narrow width of the \ion{O}{VI} absorption. 

\subsection{$\mathrm{CUBS \ J045406.34\!-\!611518.54}$ at $d_\mathrm{proj}=640 \ \mathrm{kpc}$} \label{sec:s23}

This is a non-isolated, quiescent galaxy at $z_\mathrm{gal} = 0.5811$ with stellar mass $M_\mathrm{star} \approx 4 \times 10^{11} \msun$. Associated absorption is identified in \ion{H}{I} but no metal ions.

\begin{table}
\caption{Absorption properties along the sightline at $d_\mathrm{proj} = 640$ kpc from system s23.} \label{tab:s23}
\footnotesize\centering
\smallskip
\begin{tabular}{rrr}
\hline
Ion & $\log(N_c/\mathrm{cm}^{-2})$ & $b_c$ (km s$^{-1}$) \\
\hline
\hline
\multicolumn{3}{c}{c1; $v_c - v_\mathrm{gal} = 99 \pm 1$ km s$^{-1}$} \\
\hline
\ion{H}{I} & $14.79 \pm 0.02$ & $35 \pm 2$ \\
\ion{C}{IV} & $<12.8$ & $6$ \\
\ion{O}{VI} & $<13.7$ & $30$ \\
\ion{Mg}{II} & $<11.5$ & $4$ \\
\hline
\end{tabular}
\end{table}




\end{document}